\documentclass[sigconf, nonacm]{acmart}

\usepackage{tikz}
\usepackage[utf8]{inputenc}
\usepackage{enumitem}
\usepackage{multirow}
\usepackage[linesnumbered,ruled,vlined,algo2e]{algorithm2e}
\SetKw{Break}{break}

\newcommand{\eat}[1]{}
\title{Bipartite Graph Matching Algorithms for Clean-Clean \\Entity Resolution: An Empirical Evaluation}
\author{George Papadakis$^1$, Vasilis Efthymiou$^2$, Emmanouil Thanos$^3$, Oktie Hassanzadeh$^4$}
\affiliation{%
  \institution{$^1$National and Kapodistrian University of Athens, Greece $\>\>$ \texttt{gpapadis@di.uoa.gr}\\
  $^2$Foundation for Research and Technology - Hellas, Greece $\>\>$ \texttt{vefthym@ics.forth.gr}\\
  $^3$KU Leuven, Belgium $\>\>$ \texttt{emmanouil.thanos@kuleuven.be}\\
  $^4$IBM Research, USA $\>\>$ \texttt{hassanzadeh@us.ibm.com}
  \country{}
  }
}

\begin{document}

\begin{abstract}
Entity Resolution (ER) is the task of finding records that refer to the same real-world entities. A common scenario is when entities across two clean sources need to be resolved, which we refer to as Clean-Clean ER. In this paper, we perform an extensive empirical evaluation of 8 bipartite graph matching algorithms that take in as input a bipartite similarity graph and provide as output a set of matched entities. We consider a wide range of matching algorithms, including algorithms that have not previously been applied to ER, or have been evaluated only in other ER settings. We assess the relative performance of the algorithms with respect to accuracy and time efficiency
over 10 established, real datasets, from which we extract $>$700 different similarity graphs. Our results provide insights into the relative performance of these algorithms and guidelines for choosing the best one, depending on the data at hand.
\end{abstract}

\eat{
\begin{abstract}
Record Linkage (RL) is the task of finding records that refer to the same real-world entities among two individually clean, but overlapping data sources. Its output is a bipartite similarity graph, whose edge weights capture the matching likelihood of the adjacent nodes/entities.
In this paper,  we perform an extensive empirical evaluation of the state-of-the-art bipartite graph partitioning algorithms that convert the bipartite similarity graph into 
a set of matched records. We consider a wide range of matching algorithms, including algorithms that have not previously been applied to RL, or have been evaluated only in other RL settings. We assess the relative performance of the algorithms with respect to effectiveness and time efficiency. Our results provide insights into the relative performance of these algorithms as well as guidelines for 
choosing the best one, depending on the RL application at hand.
\end{abstract}
}

\maketitle

\section{Introduction}

Entity Resolution is a challenging, yet well-studied problem in data integration \cite{DBLP:books/daglib/0030287,DBLP:journals/pvldb/KopckeTR10}. A common scenario is 
Clean-Clean ER (CCER) \cite{DBLP:journals/csur/ChristophidesEP21}, where the two data sources 
to be integrated are both clean (i.e., duplicate-free), or are cleaned using single-source entity resolution frameworks. Example applications include Master Data Management~\cite{OttoMdmSurvey10}, where a new clean source needs to be integrated into the clean reference data, and Knowledge Graph matching and completion~\cite{Gutierrez20,DBLP:journals/csimq/SaeediNPR18}, where an existing clean knowledge base needs to be augmented with an external source. 


We focus 
on methods that take advantage of a large body of work on blocking and matching algorithms, which 
efficiently compare the entities across two sources and provide as output pairs of entities along with a confidence or similarity score \cite{DBLP:series/synthesis/2015Dong,DBLP:journals/csur/ChristophidesEP21}. The output can then be used to decide which pairs should be matched. The simplest approach is 
specifying as duplicates all the pairs with 
a score higher than
a given threshold.
Choosing a single threshold fails to address the issue that in most cases the similarity scores vary significantly depending on the characteristics of the entities. More importantly, for CCER,
this approach does not guarantee that each source entity can be matched with at most one other entity. If we view the output as a bipartite {\em similarity graph}, where the nodes are 
entity profiles
and the edge weights are the matching scores between the candidate duplicates, what we need is finding a {\em matching} (or independent edge set~\cite{matchingTheoryBook}) so that each entity from one source is matched to at most one entity in the other source.

In this paper, we present the results of our thorough evaluation of efficient bipartite graph matching algorithms for CCER.
To the best of our knowledge, our study is the first to primarily focus on bipartite graph matching algorithms, examining the relative performance of the algorithms in a variety of data sets and methods of creating the input similarity graph. 
Our goal is to answer the following questions: \textit{Which bipartite graph matching algorithm is the most accurate one, which is the most robust one, and which offers the best balance between effectiveness and time efficiency? How well do the main algorithms scale? 
Which characteristics of the input graphs determine the absolute and the relative performance of the algorithms?}
By answering these questions we intend to facilitate the selection of the best algorithm for the data at hand.

In summary, we make the following contributions:

$\bullet$ In Section \ref{sec:algorithms}, we present an overview of eight efficient bipartite graph1 matching algorithms along with an analysis of their behavior and complexity. Some of the algorithms are adaptations of efficient graph clustering algorithms that have not been applied

$\bullet$ In Section \ref{sec:inputs}, we organize the input of bipartite graph partitioning algorithms into a taxonomy that is based on 
    the learning-free source of similarity scores/edge weights. 

$\bullet$ We perform an extensive experimental analysis that involves 739 different similarity graphs from 10 established real-world CCER datasets, whose sizes range from several thousands to hundreds of million edges, as explained in Section \ref{sec:expSetup}.
    
$\bullet$ In Section \ref{sec:expAnalysis}, we assess the relative performance of the matching algorithms with respect to effectiveness and time efficiency.

$\bullet$ We have publicly released the implementation of all algorithms as well as our experimental results.\footnote{See \url{https://github.com/scify/JedAIToolkit} for more details.}

\section{Preliminaries}
\label{sec:preliminaries}

We assume that an \emph{entity profile} or simply \textit{entity} is the description of a real-world object, provided as a set of attribute-value pairs in some \emph{entity collection} $V$.  The problem of Entity Resolution (ER) is to identify such entity profiles (called \emph{matches} or \emph{duplicates}) that correspond to the same real-world object, and place them in a common \emph{cluster} $c$. In other words, the output of ER, ideally, is a set of clusters $C$, each containing all the matching profiles that correspond to a single real-world entity. 

In this paper, we focus on the case of \emph{Clean-Clean ER} (CCER), in which we want to match profiles coming from two clean (i.e., duplicate-free) entity collections $V_1$ and $V_2$. This means that the resulting clusters should contain at most two profiles, one from each collection. \emph{Singular clusters}, corresponding to profiles for which no match has been found, are also acceptable. 

To generate this clustering, a typical CCER pipeline~\cite{DBLP:journals/csur/ChristophidesEP21} involves the steps of \textit{(i)} \emph{(meta-)blocking},
i.e., indexing steps that generate \emph{candidate matching pairs}, this way reducing the otherwise quadratic search space of matches, \textit{(ii)} \emph{matching}, assigning a similarity score to each candidate pair, and \textit{(iii)} \emph{bipartite graph matching}, which receives the scored candidate pairs and decides which pairs will be placed together in a cluster. In this work, we evaluate how different methods for the last step perform, when the previous ones are fixed.

\textit{Problem Definition.} The task of \textbf{Bipartite Graph Matching}
receives as input a bipartite \emph{similarity graph} $G = (V_1, V_2, E)$, where $V_1$ and $V_2$ are two clean entity collections, and $E \subseteq V_1 \times V_2$ is the set of edges with weights in [0,1], corresponding to the similarity scores between entity profiles of the two collections. 
Its output comprises a set of partitions/clusters $C$, with each one containing one node $v_i \in V_1 \cup V_2$ or two nodes $v_i \in V_1$ and $v_j \in V_2$ that represent the same real-world object. 



\begin{figure*}[th]\centering
	\center {\includegraphics[width=\linewidth]{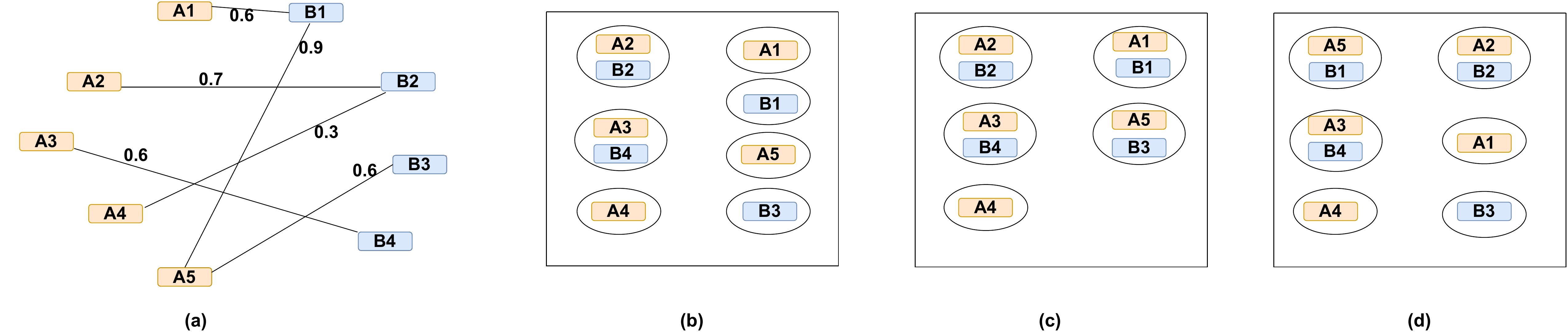}}
    \vspace{-15pt}
	\caption{Example of processing a similarity graph: (a) the similarity graph constructed for a pair of clean entity collections ($V_1$ in orange and $V_2$ in blue), (b) the resulting clusters after applying \textsf{CNC}, (c) the resulting partitions/clusters assuming that the approximation algorithms \textsf{RCA} or \textsf{BAH} 
	retrieved the optimal solution for the maximum weight bipartite matching or the assignment problem, and (d) the resulting clusters after applying \textsf{UMC}, \textsf{BMC} or \textsf{EXC}.
	}
	\label{fig:exampleCCERclustering}
    \vspace{-10pt}
\end{figure*}

Figure~\ref{fig:exampleCCERclustering}(a) shows an example of a bipartite similarity graph, in which node partitions (entity collections) are labeled as $A$ (in orange) and $B$ (in blue). The edges connect only nodes from $A$ to $B$ and are associated with a weight that reflects the similarity (matching likelihood) of the adjacent nodes. Figures~\ref{fig:exampleCCERclustering}(b)--1(d) show three different outputs of CCER, in which nodes within the same oval (cluster) correspond to matching entities.

\textbf{Related work.}
There is a rich  body of 
literature on 
ER \cite{DBLP:journals/csur/ChristophidesEP21,DBLP:books/daglib/0030287}. Following the seminal Fellegi-Sunter model for record linkage~\cite{FellegiS69}, a major focus of prior work has been on classifying pairs of input records as {\em match}, {\em non-match}, or {\em potential match}. While even some of the early work on record linkage incorporated a 1-1 matching constraint~\cite{Winkler06}, the primary focus of prior work, especially the most recent one, has been on the 
effectiveness of the classification task, mainly by leveraging Machine~\cite{DBLP:journals/pvldb/KondaDCDABLPZNP16} and Deep Learning \cite{DBLP:conf/edbt/BrunnerS20,DBLP:journals/jdiq/LiLSWHT21,DBLP:conf/sigmod/MudgalLRDPKDAR18}.


Inspired by the recent progress and success of prior work on improving the efficiency of ER with blocking and filtering~\cite{DBLP:journals/csur/PapadakisSTP20}, we target ER frameworks where the output of the matching is used to construct a similarity graph that needs to be partitioned for the final step of entity resolution. Hassanzadeh et al.~\cite{DBLP:journals/pvldb/HassanzadehCML09} also target such a framework, and perform an evaluation of various graph clustering algorithms for entity resolution. However, they target a scenario where input data sets are not clean or more than two clean sources are merged into a dirty source that contains duplicates in itself; as a result, each cluster could contain more than two records. We refer to this variation of ER as {\em Dirty ER} \cite{DBLP:journals/csur/ChristophidesEP21}. Some of the bipartite matching algorithms we use in this paper are adaptations of the graph clustering algorithms used 
in \cite{DBLP:journals/pvldb/HassanzadehCML09}
for Dirty ER.



More recent clustering methods for Dirty ER were proposed in \cite{DBLP:journals/jdiq/DraisbachCN20}. After estimating the connected components, \textit{Global Edge Consistency Gain} iteratively switches the label of edges so as to maximize the overall consistency, i.e,. the number of triangles with the same label in all edges. \textit{Maximum Clique Clustering} ignores edge weights and iteratively removes the maximum clique along with its vertices until all nodes have been assigned to an equivalence cluster. This approach is generalized by \textit{Extended Maximum Clique Clustering}, which removes maximal cliques from the similarity graph and enlarges them by adding edges that are incident to a minimum portion of their nodes.

Gemmel et al.~\cite{DBLP:journals/corr/abs-1108-6016} present two algorithms for CCER
as well as more algorithms for different ER settings (e.g., one-to-many and many-to-many). Both algorithms are covered by the clustering algorithms that are included in our study: the MutualFirstChoice 
is equivalent to our Exact clustering, while 
the Greedy algorithm 
is equivalent to UniqueMappingClustering.
Finally, the MaxWeight method~\cite{DBLP:journals/corr/abs-1108-6016} utilises the exact solution of the maximum weight bipartite matching, for which an efficient heuristic approach is considered in our {Best Assignment Heuristic Clustering}.

FAMER~\cite{DBLP:journals/csimq/SaeediNPR18} is a framework that supports multiple matching and clustering algorithms for multi-source ER. Although it studies some common clustering algorithms with those explored in this paper (e.g., Connected Components),
our focus on bipartite graphs, which do not support multi-source settings,
makes the direct comparison inapplicable. Note, though, that adapting FAMER's top-performing algorithm, i.e., CLIP clustering,
to work in a CCER setting yields an algorithm equivalent to Unique Mapping Clustering.

Wang et al.~\cite{DBLP:conf/icde/WangTLXXL19} follow a reinforcement learning approach, based on a Q-learning~\cite{DBLP:journals/ml/WatkinsD92} algorithm, for which a state is represented by the pair ($|L|, |R|$), where $L \subseteq V_1, R \subseteq V_2$ are the nodes matched from the two partitions, and the reward is computed as the sum of the weights of the selected matches. We leave this algorithm outside the scope of this study, 
as we consider only 
learning-free methods, but we plan to further explore it in our future works.

Kriege et al.~\cite{DBLP:conf/icdm/KriegeGB019} present a linear approximation to the weighted graph matching problem, but for that, they require that the edge weights are assigned by a tree metric, i.e. a similarity measure that satisfies a looser version of the triangle inequality. In this work, we investigate algorithms that are agnostic to such similarity measure properties, assuming only that the weights are in [0,1], as is the case of most existing algorithms.

\section{Algorithms}
\label{sec:algorithms}




We consider algorithms that satisfy the following selection criteria:
\begin{enumerate}
    \item They are crafted for bipartite similarity graphs, which apply exclusively to CCER. Algorithms for the types of graphs that correspond to Dirty and Multi-source ER have been examined in \cite{DBLP:journals/pvldb/HassanzadehCML09} and \cite{DBLP:journals/csimq/SaeediNPR18},~respectively.
    \item Their functionality is learning-free in the sense that they do not learn a pruning model over a set of labelled instances. We only use the ground-truth of real matches to optimize their internal parameter configuration.
    \item Their time complexity is not worse than the brute-force approach of ER, $O(n^2)$, where $n = |V_1 \cup V_2|$ is the number of nodes in the bipartite similarity graph $G=(V_1, V_2, E)$. 
    \item Their space complexity is $O(n+m)$, where $m = |E|$ is the number of edges in the given similarity graph.
\end{enumerate}

\begin{table}[t]\centering
    \caption{Configuration parameters per algorithm.}
    \vspace{-10pt}
    {\footnotesize
    	\begin{tabular}{ | l | c | c |}
		\hline
		\multicolumn{1}{|c|}{\textbf{Algor.}} &
		\multicolumn{1}{c|}{\textbf{Similarity Threshold} $t$} &
		\multicolumn{1}{c|}{\textbf{Other}}  \\
		\hline
        \hline
        $CNC$ & \checkmark & $\times$ \\
        \hline
        $RSR$ & \checkmark & $\times$\\
        \hline
        $RCA$ & \checkmark & $\times$\\
        \hline
        \multirow{ 2}{*}{$BAH$} & \multirow{ 2}{*}{\checkmark} & maximum search steps (10,000) \\
        & & maximum run-time per search step (2 min.)\\
        \hline
        $BMC$ & \checkmark & node partition used as basis\\
        \hline
        $EXC$ & \checkmark & $\times$\\
        \hline
        $KRC$ & \checkmark & $\times$\\
        \hline
        $UMC$ & \checkmark & $\times$\\
		\hline
	\end{tabular}
	}
	\label{tb:confParameters}
	\vspace{-10pt}
\end{table}

Due to the third criterion, we exclude the classic Hungarian algorithm, also known as the Kuhn-Munkres algorithm \cite{Kuhn55thehungarian}, 
whose time complexity is cubic, $O(n^3)$. For the same reason, we exclude
the work of
Schwartz et al.~\cite{DBLP:conf/wea/SchwartzSW05} 
on
1-1 bipartite graph matching with minimum cumulative weights, which
reduces the problem to a minimum cost flow problem and uses the matching algorithm of Fredman \& Tarjan~\cite{DBLP:journals/jacm/FredmanT87} to provide an approximate solution~in~O($n^2logn$). Note that most of the considered algorithms depend on the number of edges $m$ in the similarity graph, which is equal to $n^2$ in the worst case. In practice, though, its value is determined by the similarity threshold $t$, which is used by each algorithm to prune all edges with a lower weight. For reasonable thresholds, $O(n) \leq m \ll O(n^2)$.

Below, we describe the selected algorithms.
Table \ref{tb:confParameters} summarizes their configuration parameters. 
Their implementation 
(in Java) is publicly available through the JedAI toolkit~\cite{DBLP:journals/is/PapadakisMGSTGB20}.


\textbf{Connected Components (\textsf{CNC}).}
This is the simplest algorithm: it discards all edges with a weight lower than the similarity threshold and then computes the transitive closure of the pruned similarity graph. In the output, it solely retains the partitions/clusters that contain two entities -- one from each entity collection. Using a simple depth-first approach, its time complexity is $O(m)$  \cite{DBLP:books/daglib/0017733}.

\textbf{Ricochet Sequential Rippling Clustering (\textsf{RSR}).}
This algorithm is an adaptation of the homonymous 
method for Dirty ER in \cite{DBLP:journals/pvldb/HassanzadehCML09} such that it exclusively considers clusters with just one entity from each entity collection. After pruning the edges weighted lower than $t$, \textsf{RSR} sorts all nodes from both $V_1$ and $V_2$ in descending order of the average weight of their adjacent edges. Whenever a new seed is chosen from the sorted list, the first adjacent vertex that is currently unassigned or is closer to the new seed than it is to the seed of its current partition is re-assigned to the new cluster. If a partition is reduced to a singleton after a re-assignment, it is placed in its nearest single-node cluster. The algorithm stops when all nodes have been considered. Its time complexity is $O(n \; m)$~\cite{wijaya2009ricochet}. 

\textbf{Row Column Assignment Clustering (\textsf{RCA}).}
This approach is based on the Row-Column Scan approximation method in
\cite{kurtzberg1962approximation} that solves the assignment problem. It requires two passes of the similarity graph, with each pass generating a candidate solution.
In the first pass, each entity from $V_1$ creates a new partition, to which the most similar, currently unassigned entity from $V_2$ is assigned. Note that, in principle, any pair of entities can be assigned to the same partition at this step even if their similarity is lower than $t$, since the assignment problem assumes that each vertex from $V_1$ is connected to all vertices from $V_2$ (any ``job'' can be performed by all ``men''). The clusters of pairs with similarity less than $t$ are then discarded.
In the second pass, the same procedure is applied to the entities/nodes of $V_2$. The value of each solution is the sum of the edge weights between the nodes assigned to the same (2-node) partition. The solution with the highest value is returned as output.
Its time complexity 
is $O(|V_1| \; |V_2|)$ \cite{paperExtendedVersion}. 

\textbf{Best Assignment Heuristic (\textsf{BAH}).}
This algorithm applies a simple swap-based random-search algorithm to heuristically solve the maximum weight bipartite matching problem and uses the resulting solution to create the output partitions.
Initially, each entity from the smaller entity collection is connected to an entity from the larger one. In each iteration of the search process, two entities from the larger entity collection are randomly selected in order to swap their current connections. 
If the sum of the edge weights of the new pairs is higher than the previous pairs, the swap is accepted. The algorithm stops when a maximum number of search steps is reached or when a maximum run-time has been exceeded. In our case, the run-time limit has been set to 2 minutes.

\textbf{Best Match Clustering (\textsf{BMC}).}
This algorithm is inspired from the Best Match strategy of~\cite{DBLP:conf/icde/MelnikGR02}, which solves the stable marriage problem~\cite{DBLP:journals/tamm/GaleS62}, as simplified in BigMat~\cite{DBLP:conf/bigdataconf/0002MD19}. 
For each entity 
of the one entity collection, this algorithm creates a new partition,
in which the most similar, not-yet-clustered entity from the other entity collection is also placed -- provided that the corresponding edge weight is higher than $t$.
Note that the greedy heuristic for \textsf{BMC}
introduced in~\cite{DBLP:conf/icde/MelnikGR02} is the same, in principle, to Unique Mapping Clustering (see below). Note also that 
an additional configuration parameter
is the entity collection that is used as the basis for creating partitions, which can be set to $V_1$ or $V_2$. In our experiments, we examine both options and retain the best one.
Its time complexity is $O(m)$~\cite{paperExtendedVersion}.

\textbf{Exact Clustering (\textsf{EXC}).}
This algorithm is inspired from the Exact strategy of~\cite{DBLP:conf/icde/MelnikGR02}. 
\textsf{EXC} places two entities in the same partition 
only if they are mutually the best matches, i.e., 
the most similar candidates of each other, and their edge weight exceeds $t$.
This approach is basically a stricter, symmetric version of \textsf{BMC} and could also be conceived as a strict version of the reciprocity filter that was employed in \cite{DBLP:conf/edbt/Efthymiou0SC19}.
Its time complexity 
is $O(n \; m)$.

\textbf{Kir\'{a}ly's Clustering (\textsf{KRC}).}
This algorithm is an adaptation of the linear time 3/2 approximation to the maximum stable marriage problem, called ``New Algorithm'' in~\cite{DBLP:journals/algorithms/Kiraly13}. 
Intuitively, the entities of $V_1$ (``men''~\cite{DBLP:journals/algorithms/Kiraly13}) propose to the entities 
from $V_2$ with an edge weight higher than $t$ (``women''~\cite{DBLP:journals/algorithms/Kiraly13}) to form a partition (``get engaged''~\cite{DBLP:journals/algorithms/Kiraly13}). 
The entities of $V_2$ accept a proposal under certain conditions (e.g., if it's the first proposal they receive),
and the partitions and preferences are updated accordingly.
Entities from $V_1$ get a second chance to make proposals 
and the algorithm terminates when all entities of $V_1$ are in a partition,
or no more proposal chances are left.
We omit some of the details (e.g., the rare case of ``uncertain man''), due to space restrictions, and refer the reader to~\cite{paperExtendedVersion,DBLP:journals/algorithms/Kiraly13} for more information (e.g., the acceptance criteria for proposals).
Its time complexity 
is $O(n + m \; logm)$ \cite{DBLP:journals/algorithms/Kiraly13}.

\textbf{Unique Mapping Clustering (\textsf{UMC}).}
This algorithm prunes all edges with a weight lower than $t$, sorts the remaining ones 
in decreasing weight/similarity
and iteratively forms a partition for the top-weighted pair
as long as none of its entities has already been matched to some other.
This comes from the \textit{unique mapping constraint} of CCER, i.e., the restriction that 
each entity from the one entity collection
matches with at most one entity from the other.
Note that the \textit{CLIP Clustering algorithm}, introduced for the multi-source 
ER problem in \cite{DBLP:conf/esws/SaeediPR18}, is equivalent to \textsf{UMC} 
in the CCER case that we study. 
Its time complexity 
is $O(m \; logm)$ \cite{paperExtendedVersion}. 


\textbf{Example.}
Figure \ref{fig:exampleCCERclustering} demonstrates an example of applying the above algorithms to the similarity graph in Figure \ref{fig:exampleCCERclustering}(a). For all algorithms, we assume a weight threshold of $0.5$. 

\textsf{CNC} completely discards the 4-node connected component $(A1,\\ B1, A5, B3)$ and considers exclusively the valid partitions $(A2, B2)$ and $(A3, B4)$, 
as demonstrated in Figure \ref{fig:exampleCCERclustering}(b). 

Algorithms that aim to maximize the total sum of edge weights 
between the matched entities,
such as \textsf{RCA} and \textsf{BAH}, will cluster $A1$ with $B1$ and $A5$ with $B3$, as shown in Figure \ref{fig:exampleCCERclustering}(c), if they manage to find the optimal solution for the given graph. The reason is that this combination of edge weights yields a sum of $0.6+0.6=1.2$, which is higher than $0.9$, i.e., the sum resulting from clustering $A5$ with $B1$ and leaving $A1$ and $B3$ as singletons.

\textsf{UMC} starts from the top-weighted edges, 
matching 
$A5$ with $B1$, $A2$ with $B2$ and $A3$ with $B4$; 
$A1$ and $B3$ are left as singletons, as shown in Figure \ref{fig:exampleCCERclustering}(d), because their candidates have already been matched to other entities.  
The same output is produced by \textsf{EXC}, as
the entities in each partition consider each other as their most similar candidate. For this reason, \textsf{BMC} also yields the same results assuming that $V_2$ (blue) is used as the basis entity collection.

The partitions generated by \textsf{RSR} and \textsf{KRL} depend on the sequence of adjacent vertices and proposals, respectively. Given, though, that higher similarities are generally more preferred than increasing total sum by both of these algorithms, the outcome in Figure \ref{fig:exampleCCERclustering}(d) is the most possible one for these algorithms, too.



\vspace{-10pt}
\section{Similarity Graphs}
\label{sec:inputs}


Two types of methods can be used for the generation of the similarity graphs that constitute the input to the above algorithms~{\color{blue}\cite{DBLP:books/daglib/0030287}}:

(i) \textit{learning-free} methods, which produce similarity scores in an unsupervised manner based on the content of the input entities,~and 

(ii) \textit{learning-based} methods, which produce probabilistic similarities based on a training set.

In this work, we exclude the latter, focusing exclusively on learning-free methods. Thus, we 
make the most of the selected datasets, without sacrificing valuable parts for the construction of the training (and perhaps the validation) set. We also avoid depending on the fine-tuning of numerous configuration parameters, especially in the case of Deep Learning-based methods~\cite{DBLP:conf/icdm/WangSWDJ20}. Besides, our goal is not to optimize the performance of the CCER process, but to investigate how the main graph matching algorithms perform under a large variety of real settings. For this reason, we produce a large number of similarity graphs per dataset, rather than generating synthetic data.

In this context, we do not apply any blocking method when producing these inputs. Instead, we consider all pairs of entities from different datasets with a similarity higher than 0. This allows for experimenting with a large variety of similarity graph sizes, which range from several thousand to hundreds of million edges. Besides, the role of blocking, i.e., the pruning of the entity pairs with very low similarity scores, is performed by the similarity threshold $t$ that is employed by all algorithms. 

The resulting similarity graphs differ in the number of edges and the corresponding weights, which were produced using different \textit{similarity functions}.
Each similarity function consists of two parts: 
(i) the \textit{representation model}, and
(ii) the \textit{similarity measure}.

The \textbf{representation model} transforms a textual value into a model that is suitable for applying the selected similarity measure. 
Depending on the \underline{\textit{scope}} of these representations, we distinguish them into (i) \textit{schema-agnostic} and (ii) \textit{schema-based}. The former consider all attribute values in an entity description, while the latter consider only the value of a specific attribute. 
Depending on their \underline{\textit{form}}, we also distinguish the representations into (i) \textit{syntactic} and (ii) \textit{semantic}. The former operate on the original text of the entities, while the latter operate on vector transformations (embeddings) of the original text that aim to capture its actual connotation, leveraging external information that has been extracted from large and generic corpora through unsupervised learning. 

The \textit{schema-based syntactic representations} 
process each value as a sequence of characters or words and apply to mostly short textual values.
For example, the attribute value ``Joe Biden'' can be represented as the set of tokens \{`Joe', `Biden'\}, or the set of character 3-grams \{`Joe', `oe\_', `e\_B', `\_Bi', `Bid', `ide', `den'\}.

The \textit{schema-agnostic syntactic representations} process 
the set of all individual attribute values. 
We use two types of 
models that have been widely applied to document classification tasks {\color{blue}\cite{DBLP:journals/www/0001GP16}}:

\vspace{3pt}
(i) an n-gram vector \cite{DBLP:books/daglib/0021593}, whose dimensions correspond to character or token n-grams and are weighted according to their frequency (TF or TF-IDF score). This approach does not consider the order of n-gram appearances in each~value. 

\vspace{3pt}
(ii) an n-gram graph \cite{DBLP:journals/tslp/GiannakopoulosKVS08}, which transforms each value into a graph, where the nodes correspond to character or token n-grams, the edges connect those co-occurring in a window of size $n$ and the edge weights denote the n-gram's co-occurrence frequency. Thus, the order of n-grams in a value~is~preserved.

\vspace{2.5pt}
Following the previous example, the character 3-gram vector of ``Joe Biden'' would be a sparse vector with as many dimensions as all the 3-grams appearing in the entity collection and with zeros in all other places except the ones corresponding to the seven character 3-grams of ``Joe Biden'' listed above. For the places corresponding to those seven 3-grams, the value would be the TF or TF-IDF of each 3-gram. 
Similarly, a token 2-gram vector of ``Joe Biden'' would be all zeros, for each token 2-gram appearing in all the values, except for the place corresponding to the 2-gram `Joe Biden', where its value would be 1. 
A character 3-gram graph would be a graph with seven nodes, one for each 3-gram listed above, connecting the node `Joe' to the nodes `oe\_' and `e\_B', each with an edge of weight 1, `oe\_' to `e\_B' and `\_Bi', etc. See \cite{paperExtendedVersion} for more details.

Both approaches build an aggregate representation per entity: the n-gram vectors treat each entity as a ``document'' and adjust their weights accordingly, while the individual n-gram graphs of each value are merged 
into a larger ``entity graph'' through the update operator discussed in \cite{DBLP:journals/tslp/GiannakopoulosKVS08}. For both approaches, we consider $n \in \{2, 3, 4\}$ for character and $n \in \{1, 2, 3\}$ for token n-grams.


The \textit{semantic representations} treat every text as a sequence of items (words or character n-grams) of arbitrary length and convert it into a dense numeric vector based on learned external patterns. The closer the connotation of two texts is, the closer are their vectors. These representations come in two main forms, which apply uniformly to schema-agnostic and schema-based settings:

(i) The pre-trained embeddings of word- or character-level. Due to the highly specialized content of ER tasks (e.g., arbitrary alphanumerics in product names), the former, which include \textit{word2vec} \cite{DBLP:conf/nips/MikolovSCCD13} and \textit{GloVe} \cite{DBLP:conf/emnlp/PenningtonSM14}, suffer from a high portion of out-of-vocabulary tokens -- these are words that cannot be transformed into a vector because they are not included in the training corpora \cite{DBLP:conf/sigmod/MudgalLRDPKDAR18}. This drawback is addressed by the character-level embeddings: \textit{fastText} vectorizes a token by summing the embeddings of all its character n-grams \cite{DBLP:journals/tacl/BojanowskiGJM17}. For this reason, we exclusively consider the 300-dimensional fastText in the following.

(ii) Transformer-based language models \cite{DBLP:conf/naacl/DevlinCLT19} go beyond the shallow, context-agnostic pre-trained embeddings by vectorizing an item based on its context. In this way, they assign different vectors to homonyms, which share the same form, but different meaning (e.g., ``bank'' as a financial institution or as the border of a river). They also assign similar vectors to synonyms, which have different form, but almost the same meaning (e.g., ``enormous'' and ``vast''). Several BERT-based language models have been applied to ER in \cite{DBLP:conf/edbt/BrunnerS20,DBLP:journals/jdiq/LiLSWHT21}. 
They do suffer from out-of-vocabulary tokens, but to the best of our knowledge, there is no established character-level language model that could address this issue, as fastText does for pre-trained embeddings.
Among them, we exclusively consider the 768-dimensional ALBERT, due to its higher efficiency \cite{DBLP:conf/iclr/LanCGGSS20}.


Every \textbf{similarity measure} receives as input two representation models and produces a score that is proportional to the likelihood that the respective entities correspond to the same real world object: the higher the score, the more similar are the input models and their textual values and the higher is the matching likelihood.

For each type of representation models, we considered a large variety of established similarity measures. The following are combined with the character-level schema-based representation models: Damerau-Levenshtein, Levenshtein and q-grams distance,  Jaro Similarity, Needleman Wunch, Longest Common Subsequence and Longest Common Subsequence. To the token-level, schema-based models we apply: Cosine, Dice and (Generalized) Jaccard similarity as well as Monge-Elkan, Overlap Coefficient, Block and Euclidean distance. The schema-agnostic n-gram vectors are coupled with Arcs and Jaccard similarity as well as with Cosine and Generalized Jaccard similarity with TF or TF-IDF weights. For the n-gram graphs, we consider Containment, (Normalized) Value and Overall similarity. Finally, the semantic similarity models are combined with Cosine, Euclidean and World Mover's similarity.
We formally define these measures in the Appendix of~\cite{paperExtendedVersion}.

\section{Experimental Setup}
\label{sec:expSetup}


\begin{table*}[t]\centering
    \caption{Technical characteristics of the real datasets for Clean-Clean ER in increasing number of computational cost. }
    \vspace{-8pt}
	\begin{tabular}{ | l | r | r | r | r | r | r | r | r | r | r | }
		\cline{2-11}
		\multicolumn{1}{c|}{}&
		\multicolumn{1}{c|}{$\mathbf{D_{1}}$} &
		\multicolumn{1}{c|}{$\mathbf{D_{2}}$} &
		\multicolumn{1}{c|}{$\mathbf{D_{3}}$} &
		\multicolumn{1}{c|}{$\mathbf{D_{4}}$} &
		\multicolumn{1}{c|}{$\mathbf{D_{5}}$} &
        \multicolumn{1}{c|}{$\mathbf{D_{6}}$} &
        \multicolumn{1}{c|}{$\mathbf{D_{7}}$} &
        \multicolumn{1}{c|}{$\mathbf{D_{8}}$} &
        \multicolumn{1}{c|}{$\mathbf{D_{9}}$} &
        \multicolumn{1}{c|}{$\mathbf{D_{10}}$} \\
		\hline
        \hline
        Dataset${_1}$ & Rest.1 & Abt & Amazon & DBLP & IMDb & IMDb & TMDb & Walmart &  DBLP & IMDb \\ 
        Dataset${_2}$ & Rest.2 & Buy & Google Pr. & ACM & TMDb &  TVDB & TVDB & Amazon & Scholar & DBpedia \\
        $|V_1|$ & 339 & 1,076 & 1,354 & 2,616 & 5,118 & 5,118 & 6,056 & 2,554 & 2,516 & 27,615 \\
		$|V_2|$ & 2,256 & 1,076 & 3,039 & 2,294 & 6,056 & 7,810 & 7,810 & 22,074 & 61,353 & 23,182 \\
        NVP${_1}$ & 1,130 & 2,568 & 5,302 & 10,464 & 21,294 & 21,294 & 23,761 & 14,143 &  10,064 & 1.6$\cdot10^5$ \\
        NVP${_2}$ & 7,519 & 2,308 & 9,110 & 9,162 & 23,761 & 20,902 & 20,902 & 1.14$\cdot10^5$ & 1.98$\cdot10^5$ & 8.2$\cdot10^5$ \\
        $|A_1|$ & 7 & 3 & 4 & 4 & 13 & 13 & 30 & 6 & 4 & 4 \\
        $|A_2|$ & 7 & 3 & 4 & 4 & 30 & 9 & 9 & 6 & 4 & 7 \\
		$|\bar{p}_1|$ & 3.33 & 2.39 & 3.92 & 4.00 & 4.16 & 4.16 & 3.92 & 5.54 & 4.00 & 5.63 \\
		$|\bar{p}_2|$ & 3.33 & 2.14 & 3.00 & 3.99 & 3.92 & 2.68 & 2.68 & 5.18 & 3.24 & 35.20 \\
		$|D(V_1$$\cap$$V_2)|$ & 89 & 1,076 & 1,104 & 2,224 & 1,968 & 1,072 & 1,095 & 853 & 2,308 & 22,863 \\
		$||V_1 \times V_2||$ & 7.65$\cdot10^5$ & 1.16$\cdot10^6$ & 4.11$\cdot10^6$ &  6.00$\cdot10^6$ & 3.10$\cdot10^7$ & 4.00$\cdot10^7$ & 4.73$\cdot10^7$ & 5.64$\cdot10^7$ &  1.54$\cdot10^8$ & 6.40$\cdot10^8$ \\
		\hline
	\end{tabular}
	\label{tb:ccerDatasets}
	\vspace{-8pt}
\end{table*}

All experiments were carried out on a server running Ubuntu 18.04.5 LTS with a 32-core Intel Xeon CPU E5-4603 v2 (2.20GHz), 128 GB of RAM and 1.7 TB HDD. All time experiments were executed on a single core. For the implementation of the schema-based syntactic similarity functions, we used the Simmetrics Java package\footnote{\url{https://github.com/Simmetrics/simmetrics}}.  For the schema-agnostic syntactic similarity functions, we used the implementation provided by the JedAI toolkit (the implementation of n-gram graphs and the corresponding graph similarities is based on the JIinsect toolkit\footnote{\url{https://github.com/ggianna/JInsect}}). For the semantic representation models, we employed the Python sister package\footnote{\url{https://pypi.org/project/sister}}, which supports both fastText and ALBERT. For the computation of the semantic similarities, we used the Python scipy package.\footnote{\url{https://www.scipy.org}}






\textbf{Datasets.}
In our experiments, we use 10 real-world, established datasets for ER, whose technical characteristics appear in Table~\ref{tb:ccerDatasets}, where $|V_x|$ stands for the number of input entities, $|NVP_x|$ for the total number of name-value pairs, $|A_x|$ 
    for the number of attributes and $|\bar{p}_x|$ for the average number of name-value pairs per entity profile in Dataset$_x$. $|D(V_1 \cap V_2)|$ denotes the number of duplicates in the ground-truth and $||V_1 \times V_2||$ the number of pairwise comparisons executed by the brute-force approach.
    $D_{1}$, which was introduced in OAEI 2010\footnote{\url{http://oaei.ontologymatching.org/2010/im}}, contains data about restaurants.
    $D_{2}$ matches products from the online retailers Abt.com and Buy.com
    \cite{DBLP:journals/pvldb/KopckeTR10}.
    $D_{3}$ interlinks products from Amazon and the Google Base data API (Google Pr.)
    \cite{DBLP:journals/pvldb/KopckeTR10}.
    $D_{4}$ contains data about publications from DBLP and ACM
    \cite{DBLP:journals/pvldb/KopckeTR10}.
    $D_{5} - D_{7}$ contain data about television shows from TheTVDB.com (TVDB) and movies from IMDb and themoviedb.org (TMDb)
    \cite{DBLP:journals/corr/abs-2101-06126}.
    $D_{8}$ contains data about products from Walmart and Amazon
    \cite{DBLP:conf/sigmod/MudgalLRDPKDAR18}.
    $D_{9}$ contains data about scientific publications from DBLP and Google Scholar 
    \cite{DBLP:journals/pvldb/KopckeTR10}.
    $D_{10}$ matches movies from IMDb and DBpedia \cite{DBLP:journals/is/PapadakisMGSTGB20} (note that $D_{10}$ contains a different snapshot of IMDb movies than $D_5$ and $D_6$).
    %
All datasets are publicly available through the JedAI data repository.\footnote{\url{https://github.com/scify/JedAIToolkit/tree/master/data}}

Note that for the schema-based settings (both the syntactic and semantic ones), we used only the attributes that combine high coverage with high distinctiveness. That is, they appear in the majority of entities, while conveying a rich diversity 
of values, thus yielding 
high effectiveness. These attributes are
    ``name'' and ``phone'' for $D_1$,
    ``name'' for $D_2$,
    ``title'' for $D_3$,
    ``title'' and ``authors'' for $D_4$,
    ``modelno'' and ``title'' for $D_5$,
    ``title'' and ``authors'' for $D_6$,
    ``name'' and ``title'' for $D_7$,
    ``title'' and ``name'' for $D_8$,
    ``title'' and ``abstract'' for $D_9$, and 
    ``title'' for $D_{10}$.






\textbf{Evaluation Measures.}
In order 
to assess the relative performance of the above graph matching algorithms, we evaluate both their effectiveness and their time efficiency (and scalability). We measure their \textit{effectiveness}, with respect to a ground truth of known matches, in terms of three measures:

$\bullet$ \emph{Precision} 
    denotes the portion of output partitions that involve two matching entities.

$\bullet$ \emph{Recall} 
    denotes that portion of partitions with two matching entities that are included in the output.

$\bullet$ \emph{F-Measure} ($F1$) 
    is the harmonic mean of Precision and Recall.

\noindent
All are defined in $[0, 1]$. Higher values show higher effectiveness.

For \textit{time efficiency}, we measure the average run-time of an algorithm for each setting, i.e., the time that intervenes between receiving the weighted similarity graph as input and returning the partitions as output, over 10 repeated executions. 

\textbf{Generation Process.} To generate a large variety of input similarity graphs, we apply every similarity function described in Section \ref{sec:inputs} to all datasets in Table \ref{tb:ccerDatasets}. We actually apply all combinations of representation models and similarity measures,
thus yielding 60 schema-agnostic syntactic similarity graphs per dataset,
16 schema-based similarity graphs per attribute in each dataset, and 12 semantic similarity graphs per dataset. Note that we did not apply any fine-tuning to ALBERT, as our goal is not optimize ER performance, but rather to produce diverse inputs.

To estimate the algorithms' performance, we first apply min-max normalization to the edge weights of all similarity graphs, regardless of the similarity function that produced them, to ensure that they are restricted to $[0, 1]$. Next, we apply every algorithm to every input similarity graph by varying its similarity threshold from 0.05 to 1.0 with a step of 0.05 (preliminary experiments showed that there is no significant difference in the experimental results when using a smaller step size like 0.01. Thus, we set it to 0.05 to reduce the effort for the experiments, due to the large number of algorithms, inputs and datasets they involve). The largest threshold that achieves the highest F-Measure is selected as the optimal one, determining the performance of the algorithm for the particular~input. 

\begin{table}[t]\centering
 \setlength{\tabcolsep}{3.5pt}
    \caption{The number of similarity graphs $|G|$ as well as their size, in terms of the average number of edges $|\bar{E}|$, per dataset. In parenthesis, the ratio of $|\bar{E}|$ to $||V_1 \times V_2||$ (cf. Table \ref{tb:ccerDatasets}).}
    \vspace{-10pt}
    {\scriptsize
    	\begin{tabular}{ | l | r | r | r | r | r | r | r | r | }
		\cline{2-9}
		\multicolumn{1}{c|}{} &
		\multicolumn{4}{c|}{\underline{~~~~~Syntactic Similarities~~~~~}} &
		\multicolumn{4}{c|}{\underline{~~~~~Semantic Similarities~~~~~}} \\
		\multicolumn{1}{c|}{} &
		\multicolumn{2}{c|}{\underline{Schema-based}} &
		\multicolumn{2}{c|}{\underline{Schema-ag.}} &
		\multicolumn{2}{c|}{\underline{Schema-based}} &
		\multicolumn{2}{c|}{\underline{Schema-ag.}} \\
		\multicolumn{1}{c|}{} &
		\multicolumn{1}{c|}{$\mathbf{|G|}$} &
		\multicolumn{1}{c|}{$\mathbf{|\bar{E}|}$$\cdot$10$^6$} &
		\multicolumn{1}{c|}{$\mathbf{|G|}$} &
		\multicolumn{1}{c|}{$\mathbf{|\bar{E}|}$$\cdot$10$^6$} &
		\multicolumn{1}{c|}{$\mathbf{|G|}$} &
		\multicolumn{1}{c|}{$\mathbf{|\bar{E}|}$$\cdot$10$^6$} &
        \multicolumn{1}{c|}{$\mathbf{|G|}$} &
		\multicolumn{1}{c|}{$\mathbf{|\bar{E}|}$$\cdot$10$^6$}  \\
		\hline
        \hline
        $D_1$ & 20 & 0.16 (21.2\%) & 46 & 0.72 (93.5\%) & 8 & 0.26 (34.3\%) & 2 & 0.76 (100\%)\\
        $D_2$ & 12 & 1.05 (90.5\%) & 47 & 0.64 (55.1\%) & 2 & 1.16 (100\%) & 2 & 1.16 (100\%)\\
        $D_3$ & 14 & 2.89 (70.5\%) & 53 & 2.65 (64.5\%) & 2 & 4.11 (100\%) & 2 & 4.11 (100\%)\\
        $D_4$ & 27 & 4.49 (74.8\%) & 24 & 3.84 (64.0\%) & 12 & 5.99 (99.8\%) & 4 & 6.00 (100\%)\\
        $D_5$ & 24 & 5.81 (18.7\%) & 48 & 11.92 (38.5\%) & 12 & 8.22 (26.7\%) & 2 & 30.64 (98.8\%)\\
        $D_6$ & 25 & 8.39 (21.0\%) & 45 & 10.99 (27.5\%) & 12 & 12.31 (30.8\%) & 2 & 39.81 (99.6\%)\\
        $D_7$ & 26 & 2.80 (05.9\%) & 42 & 12.21 (25.8\%) & 12 & 36.44 (07.8\%) & 5 & 46.99 (99.3\%)\\
        $D_8$ & 26 & 28.10 (49.8\%) & 47 & 37.31 (66.2\%) & 2 & 37.70 (67.0\%) & - & - \\
        $D_9$ & 20 & 119.18 (77.2\%) & 46 & 77.56 (50.2\%) & 6 & 154.26 (100\%) & 2 & 154.36 (100\%) \\
        $D_{10}$ & 13 & 250.73 (39.2\%) & 43 & 317.17 (49.5\%) & 2 & 378.51 (59.0\%)& - & -\\
		\hline
		\hline
		$\Sigma$ & 207 & - & 441 & - & 70 & - & 21 & -\\
		\hline
	\end{tabular}
	}
	\label{tb:inputs}
	\vspace{-5pt}
\end{table}

Special care was taken to clean the experimental results from noise. We removed all similarity graphs where all matching entities had a zero edge weight.
We also removed all noisy graphs, where all 
algorithms achieve an F-Measure lower than 0.25.
Finally, we cleaned our data from duplicate inputs
, i.e., similarity graphs that emanate from the same dataset but different similarity functions and have the same number of edges, while at least two different algorithms achieve their best performance with the same similarity threshold, exhibiting almost identical effectiveness, i.e., the difference in F-Measure and precision or recall~is less than 0.2\%. 

The characteristics of the retained similarity graphs appear in Table~\ref{tb:inputs}. Overall, there are 739 different similarity graphs, most of which rely on syntactic similarity functions and the schema-agnostic settings, in particular. The reason is that the semantic similarities assign relatively high similarity scores to most pairs of entities, thus resulting in poor performance for all considered algorithms -- especially in the schema-agnostic settings. Every dataset is represented by at least 58 similarity graphs, in total, while the average number of edges ranges from 160K to 379M. This large set of real-world similarity graphs allows for a rigorous testing of the graph matching algorithms under diverse conditions.




\section{Experimental Analysis}
\label{sec:expAnalysis}






\textbf{Effectiveness Measures.} The most important performance aspect of clustering algorithms is their ability to effectively distinguish the matching from the non-matching pairs. In this section, we examine this aspect, addressing the following questions:
\begin{enumerate}[
    leftmargin=*,
    label={QE(\arabic*):},
    ref={QE(\arabic*)}]
    \item What is the trade-off between precision and recall that is achieved by each algorithm?
    \item Which algorithm is the most/least effective?
    \item How does the type of input affect the effectiveness of the evaluated algorithms?
    \item Which other factors affect their effectiveness?
\end{enumerate}

To answer QE(1) and QE(2), we consider the macro-average performance ($\mu$) of all algorithms across all input similarity graphs, which is reported in Table \ref{tb:avPer}. We observe that all algorithms emphasize on precision at the cost of lower recall. The most balanced algorithm is \textsf{UMC}, as it yields the smallest difference between the two measures (just 0.017). In contrast, \textsf{CNC} constitutes the most imbalanced algorithm, as its precision is almost double its recall. The former achieves the second best F-Measure, being very close to the top performer \textsf{KRC}, while the latter achieves the second worst F-Measure, surpassing only \textsf{BAH}. Note that \textsf{BAH} is the least robust with respect to all measures, as indicated by their standard deviation ($\sigma$), due to its stochastic functionality, while \textsf{CNC}, \textsf{UMC} and \textsf{KRC} are the most robust with respect to precision, recall and F-Measure, respectively. Among the other algorithms, \textsf{EXC} and \textsf{BMC} are closer to \textsf{KRC} and \textsf{UMC}, with the former 
achieving the third highest F1, while \textsf{RSR} and \textsf{RCA} lie closer to \textsf{CNC}, with \textsf{RSR} exhibiting the third lowest~F1.

\begin{table}[t]\centering
    \caption{{\small Macro-average performance across all similarity graphs.}}
    	\begin{tabular}{ | l | r | r | r | r | r | r | }
		\cline{2-7}
		\multicolumn{1}{c|}{} &
		\multicolumn{2}{c|}{\underline{Precision}} &
		\multicolumn{2}{c|}{\underline{Recall}} &
		\multicolumn{2}{c|}{\underline{F-Measure}} \\
		\multicolumn{1}{c|}{} &
		\multicolumn{1}{c|}{$\mu$} &
		\multicolumn{1}{c|}{$\sigma$} &
		\multicolumn{1}{c|}{$\mu$} &
		\multicolumn{1}{c|}{$\sigma$} & 
		\multicolumn{1}{c|}{$\mu$} &
		\multicolumn{1}{c|}{$\sigma$} \\
		\hline
        \hline
        $CNC$ & \textbf{0.801} & 0.185 & 0.403 & 0.257 & 0.490 & 0.237 \\
        $RSR$ & 0.615 & 0.228 & 0.455 & 0.239 & 0.499 & 0.216 \\
        $RCA$ & 0.590 & 0.224 & 0.502 & 0.238 & 0.518 & 0.211 \\
        $BAH$ & 0.548 & 0.236 & 0.383 & 0.282 & 0.408 & 0.246 \\
        $BMC$ & 0.631 & 0.212 & 0.582 & 0.221 & 0.586 & 0.196 \\
        $EXC$ & 0.735 & 0.197 & 0.544 & 0.242 & 0.591 & 0.199 \\
        $KRC$ & 0.696 & 0.200 & 0.597 & 0.223 & \textbf{0.619} & 0.187 \\
        $UMC$ & 0.645 & 0.212 & \textbf{0.628} & 0.212 & 0.618 & 0.193 \\
		\hline
	\end{tabular}
	\label{tb:avPer}
	\vspace{-8pt}
\end{table}

\begin{figure}
    \centering
    \includegraphics[width=0.4\textwidth]{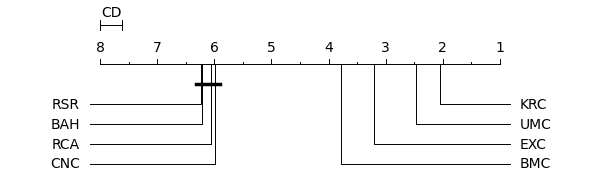}
    \vspace{-14pt}
    \caption{Nemenyi diagram based on F-Measure.}
    \vspace{-18pt}
    \label{fig:nemenyi}
\end{figure}

To assess the statistical significance of these patterns, we perform an analysis~\cite{Herbold2020} 
based on their F-Measure over the 739 paired samples. In more detail, we first perform the non-parametric Friedman test~\cite{DBLP:journals/jmlr/Demsar06} and reject the null hypothesis (with $\alpha$ = 0.05) that the differences between the evaluated methods are statistically insignificant. Then, we perform a post-hoc Nemenyi test~\cite{nemenyi1963distribution} to identify the critical distance (CD = 0.37) between the methods. The Nemenyi diagram based on F-Measure, which appears in Figure~\ref{fig:nemenyi}, shows that there are no significant differences among the methods with the worst ranks with respect to F-Measure (\textsf{CNC}, \textsf{RCA}, \textsf{BAH}, and \textsf{RSR}). All other differences are significant, with \textsf{KRC}, \textsf{UMC}, \textsf{EXC}, and \textsf{BMC}, ranking first (in that order). 

We have also performed the same analysis for Precision and for Recall \cite{paperExtendedVersion} (the same CD of 0.37 applies). 
The Precision-based ranking (reporting Mean Rank, MR) of the methods is: \textsf{CNC} (MR=1.28), \textsf{EXC} (MR=2.5), \textsf{KRC} (MR=3.7), \textsf{UMC} (MR=4.81), \textsf{BMC} (MR=5.3), \textsf{RSR} (MR=5.66), \textsf{BAH} (MR=6.12), and \textsf{RCA} (MR=6.64), so there is no significant difference between \textsf{BMC} and \textsf{RSR}, while all other differences are significant.
The Recall-based ranking of the methods is: 
\textsf{UMC} (MR=1.77), \textsf{KRC} (MR=2.44), \textsf{BMC} (MR=3.15), \textsf{EXC} (MR=4.34), \textsf{RCA} (MR=5.46), \textsf{RSR} (MR=5.92), \textsf{BAH} (MR=5.93), and \textsf{CNC} (MR=7), so there is no significant difference between \textsf{RSR} and \textsf{BAH}, while all other differences are significant.


An interesting observation drawn from these patterns is that \textsf{EXC} consistently achieves higher precision and lower recall than \textsf{BMC}. This should be expected, given that \textsf{EXC} requires an additional reciprocity check before declaring that two entities match.
We notice, however, that the gain in precision is greater than the loss in recall and, thus, \textsf{EXC} yields a higher F-Measure than \textsf{BMC}, on average. Note also that in the vast majority of cases, \textsf{BMC} works best when choosing the smallest entity collection as the basis for creating clusters.



\begin{figure*}[ht!]
\centering
\includegraphics[width=0.3\textwidth]{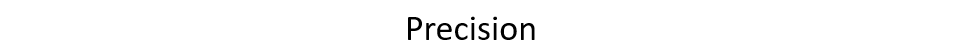}
\includegraphics[width=0.3\textwidth]{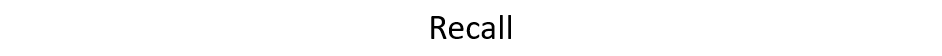}
\includegraphics[width=0.3\textwidth]{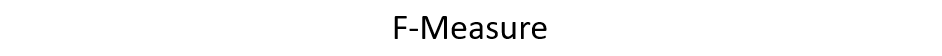}
\includegraphics[width=0.32\textwidth]{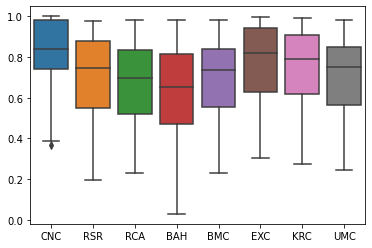}
\includegraphics[width=0.32\textwidth]{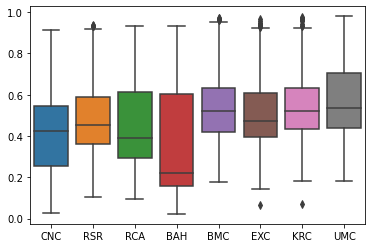}
\includegraphics[width=0.32\textwidth]{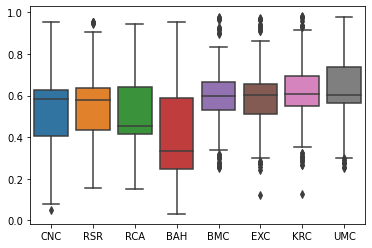}
\includegraphics[width=0.67\textwidth]{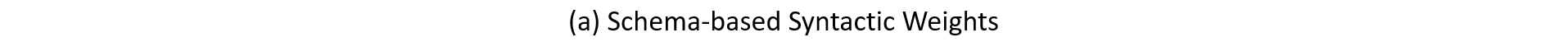}\\
\includegraphics[width=0.32\textwidth]{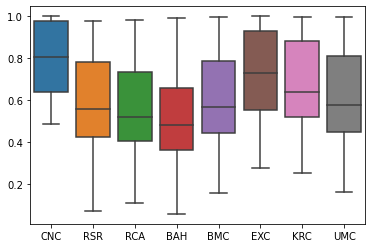}
\includegraphics[width=0.32\textwidth]{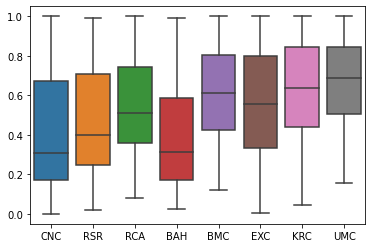}
\includegraphics[width=0.32\textwidth]{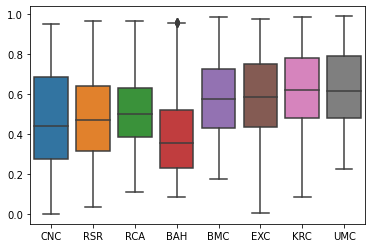}
\includegraphics[width=0.67\textwidth]{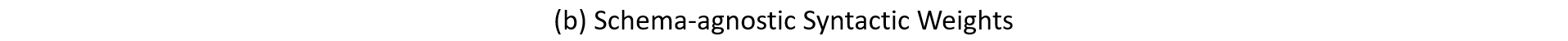}\\
\includegraphics[width=0.32\textwidth]{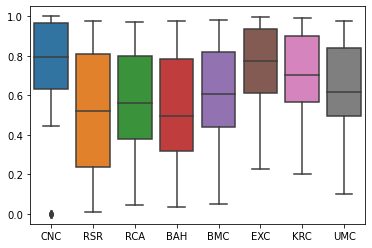}
\includegraphics[width=0.32\textwidth]{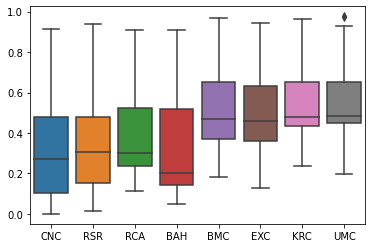}
\includegraphics[width=0.32\textwidth]{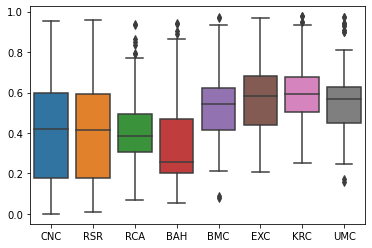}
\includegraphics[width=0.67\textwidth]{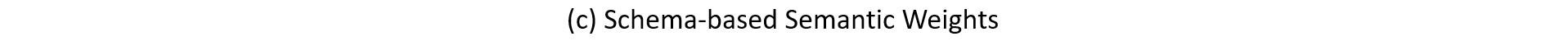}\\
\includegraphics[width=0.32\textwidth]{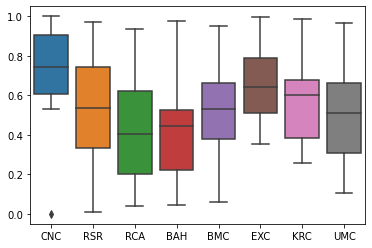}
\includegraphics[width=0.32\textwidth]{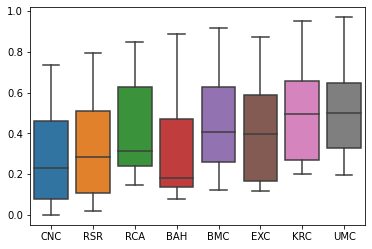}
\includegraphics[width=0.32\textwidth]{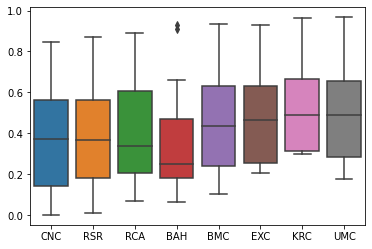}
\includegraphics[width=0.67\textwidth]{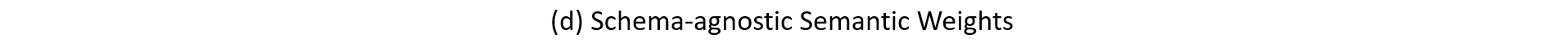}
\vspace{-10pt}
\caption{Precision (left), recall (center) and F-Measure (right) of all algorithms over similarity graphs with (a) schema-based syntactic, (b) schema-agnostic syntactic, (c) schema-based semantic, and (d) schema-agnostic semantic edge weights.}
\vspace{-12pt}
\label{fig:effectiveness}
\end{figure*}

To answer QE(3), Figure \ref{fig:effectiveness} presents the distribution of precision, recall and F-Measure of all algorithms across the four types of similarity graphs' origin. For the schema-based syntactic weights, we observe in Figure~\ref{fig:effectiveness}(a) that the average precision of all algorithms increases significantly in comparison to the one in Table \ref{tb:avPer} - from 4.0\% (\textsf{CNC}) to 16.8\% (\textsf{BMC}). For \textsf{CNC} and \textsf{RSR}, this is accompanied by an increase in average recall (by 7.7\% and 5.1\%, respectively), while for all other algorithms, the average recall drops between 4.8\% (\textsf{EXC}) and 7.5\% (\textsf{KRC}). This means that the schema-based syntactic similarities reinforce the imbalance between precision and recall in Table \ref{tb:avPer} in favor of the former for all algorithms except \textsf{CNC} and \textsf{RSR}. The average F-Measure drops only for \textsf{KRC} (by 0.2\%), which is now outperformed by \textsf{UMC}. Similarly, \textsf{BMC} exceeds \textsf{EXC} in terms of average F1 (0.603 vs 0.599), because the increase in its mean precision is much higher than the decrease in its mean recall. Finally, it is worth noting that this type of input increases significantly the robustness of all algorithms, as the standard deviation of F1 drops by more than 12\% for all algorithms, but the stochastic~\textsf{BAH}.

The opposite patterns are observed for schema-agnostic syntactic weights in Figure~\ref{fig:effectiveness}(b): the imbalance between precision and recall is reduced, as on average, the former drops from 1.9\% (\textsf{EXC}) to 5.3\% (\textsf{BAH}), while the latter raises from 2.4\% (\textsf{RSR}) to 7.6\% (\textsf{RCA}). The imbalance is actually reversed for \textsf{BMC} and \textsf{UMC}, whose average recall (0.613 and 0.664, resp.) 
exceeds their average precision (0.606 and 0.622, resp.). The only exception is \textsf{CNC}, which increases both its average and average precision. Overall, there are minor, positive changes in the average F-Measure of most algorithms, with \textsf{KRC} and \textsf{EXC} retaining a minor edge over \textsf{UMC} and \textsf{BMC}, respectively.

Regarding the semantic similarity weights, we observe in Figure~\ref{fig:effectiveness}(c) similar patterns with the schema-based syntactic ones: average precision increases for all algorithms except \textsf{CNC} and \textsf{RSR}, while average recall drops in all cases. In this case, though, the latter change is stronger than the former, leading to lower average F-Measures than those in Table \ref{tb:avPer}. In the case of schema-agnostic semantic weights, all measures in Figure \ref{fig:effectiveness}(d) drop to a significant extend (>15\% in most cases) when compared to Table \ref{tb:avPer}. It is also remarkable that the standard deviation of all measures increases to a significant extent for both schema-based and schema-agnostic weights in relation to their syntactic counterparts, despite the fewer similarity graphs. As a result, the robustness of all algorithms over semantic weights is limited. Nevertheless, \textsf{KRC} and \textsf{EXC} maintain a clear lead over \textsf{UMC} and \textsf{MBC}, respectively.



\begin{table*}[t]\centering
    \caption{The number of times each algorithm achieves the highest and second highest F1 for a particular similarity graph, $\#Top1$ and $\#Top2$, resp., as well as the average difference $\Delta~(\%)$ with the second highest F1 across all types of edge weights for balanced (BLC), one-sided (OSD) and scarce (SCR) entity collections. OVL stands for the overall sums or averages across all similarity graphs per category. Note that there are ties for both $\#Top1$ and $\#Top2$: 16 and 40, resp., over schema-based syntactic weights, 17 and 11, resp., over schema-agnostic syntactic weights, 9 and 2, resp., over schema-based semantic weights. 
    }
    \vspace{-10pt}
    	\begin{tabular}{ | l | l |  r | r | r | r || r | r | r | r || r | r | r | r || r | r | r | r |}
		\cline{3-18}
		\multicolumn{2}{c|}{} &
		\multicolumn{8}{c||}{\underline{~~~~~Syntactic Similarities~~~~~}} &
		\multicolumn{8}{c|}{\underline{~~~~~Semantic Similarities~~~~~}} \\
		\multicolumn{2}{c|}{} &
		\multicolumn{4}{c||}{\underline{Schema-based}} &
		\multicolumn{4}{c||}{\underline{Schema-agnostic}} &
		\multicolumn{4}{c||}{\underline{Schema-based}} &
		\multicolumn{4}{c|}{\underline{Schema-agnostic}} \\
		\multicolumn{2}{c|}{} &
		\multicolumn{1}{c|}{BLC} &
		\multicolumn{1}{c|}{OSD} &
		\multicolumn{1}{c|}{SCR} &
		\multicolumn{1}{c||}{OVL} &
	    \multicolumn{1}{c|}{BLC} &
		\multicolumn{1}{c|}{OSD} &
		\multicolumn{1}{c|}{SCR} &
		\multicolumn{1}{c||}{OVL} &
		\multicolumn{1}{c|}{BLC} &
		\multicolumn{1}{c|}{OSD} &
		\multicolumn{1}{c|}{SCR} &
		\multicolumn{1}{c||}{OVL} &
		\multicolumn{1}{c|}{BLC} &
		\multicolumn{1}{c|}{OSD} &
		\multicolumn{1}{c|}{SCR}  &
		\multicolumn{1}{c|}{OVL} \\
		\hline
        \hline
        \multirow{2}*{$CNC$} & $\#Top1$ & - & - & 18 & 18 & - & - & 48 & 48 & - & - & 1 & 1 & - & - & - & - \\
        & $\Delta~(\%)$ & - & - & 0.41 & 0.41 & - & - & \textbf{7.59} & \textbf{7.59} & - & - & 0.33 & 0.33 & - & - & - & - \\
        & $\#Top2$ & - & - & 8 & 8 & - & - & 8 & 8 & - & - & 4 & 4 & - & - & - & - \\
        \hline
        \multirow{2}*{$RSR$} & $\#Top1$ & - & - & 4 & 4 & - & - & 1 & 1 & - & - & 1 & 1 & - & - & - & - \\
        & $\Delta~(\%)$ & - & - & \textbf{1.90} & 1.90 & - & - & 0.51 & 0.51 & - & - & 0.33 & 0.33 & - & - & - & - \\
        & $\#Top2$ & - & - & 7 & 7 & - & - & 5 & 5 & - & - & 1 & 1 & - & - & 1 & 1 \\
        \hline
        \multirow{2}*{$RCA$} & $\#Top1$ & - & - & - & - & - & - & - & - & - & - & - & - & - & - & - & - \\
        & $\Delta~(\%)$ & - & - & - & - & - & - & - & - & - & - & - & - & - & - & - & - \\
        & $\#Top2$ & - & - & - & - & - & - & 1 & 1 & - & - & - & - & - & - & - & - \\
        \hline
        \multirow{2}*{$BAH$} & $\#Top1$ & 8 & - & 1 & 9 & 40 & - & 2 & 42 & 2 & - & 1 & 3 & 2 & - & 1 & 3 \\
        & $\Delta~(\%)$ & 3.69 & - & \textbf{1.90} & \textbf{3.49} & \textbf{5.55} & - & 0.46 & 5.31 & \textbf{12.72} & - & 0.67 & \textbf{8.70} & \textbf{13.96} & - & 0.52 & \textbf{9.48} \\
        & $\#Top2$ & 8 & - & 3 & 11 & 7 & 5 & 6 & 18 & - & - & 2 & 2 & - & - & - & - \\
        \hline
        \multirow{2}*{$BMC$} & $\#Top1$ & - & - & 6 & 6 & - & - & 7 & 7 & - & - & 2 & 2 & - & - & - & - \\
        & $\Delta~(\%)$ & - & - & 0.57 & 0.57 & - & - & 1.41 & 1.41 & - & - & 0.21 & 0.21 & - & - & - & - \\
        & $\#Top2$ & 2 & 2 & 27 & 31 & 12 & 5 & 18 & 35 & - & - & 2 & 2 & - & - & - & - \\
        \hline
        \multirow{2}*{$EXC$} & $\#Top1$ & - & 4 & 35 & 39 & - & 11 & \textbf{79} & 90 & - & 3 & 14 & 17 & - & - & \textbf{5} & 5 \\
        & $\Delta~(\%)$ & - & 0.53 & 0.87 & 0.84 & - & 0.18 & 1.18 & 1.67 & - & \textbf{1.62} & 2.54 & 2.38 & - & - & 4.18 & 4.18 \\
        & $\#Top2$ & - & 10 & 31 & 41 & - & 19 & 72 & 91 & - & \textbf{3} & \textbf{19} & 22 & - & \textbf{2} & 2 & 4 \\
        \hline
        \multirow{2}*{$KRC$} & $\#Top1$ & \textbf{22} & \textbf{15} & \textbf{43} & \textbf{80} & 16 & \textbf{47} & \textbf{79} & \textbf{142} & \textbf{11} & \textbf{5} & \textbf{30} & \textbf{46} & \textbf{3} & \textbf{4} & 4 & \textbf{11} \\
        & $\Delta~(\%)$ & 1.36 & 1.60 & 0.35 & 0.87 & 4.15 & 2.54 & 4.05 & 3.56 & 2.92 & 1.47 & \textbf{4.92} & 4.07 & 4.06 & \textbf{10.32} & \textbf{5.01} & 6.68 \\
        & $\#Top2$ & 12 & \textbf{17} & \textbf{57} & \textbf{86} & 39 & \textbf{40} & \textbf{87} & \textbf{166} & 3 & \textbf{3} & 11 & 17 & 1 & - & \textbf{5} & 6 \\
        \hline
        \multirow{2}*{$UMC$} & $\#Top1$ & \textbf{22} & \textbf{15} & 30 & 67 & \textbf{58} & 41 & 29 & 128 & 3 & - & 6 & 9 & 1 & - & 1 & 2 \\
        & $\Delta~(\%)$ & \textbf{4.99} & \textbf{1.75} & 1.19 & 2.56 & 4.51 & \textbf{3.21} & 2.51 & 3.64 & 2.11 & - & 0.34 & 0.93 & 0.22 & - & 1.00 & 0.61 \\
        & $\#Top2$ & \textbf{30} & 5 & 28 & 63 & \textbf{56} & 30 & 42 & 128 & \textbf{13} & 2 & 9 & \textbf{24} & \textbf{5} & \textbf{2} & 3 & \textbf{10} \\
        \hline
    \end{tabular}
    \vspace{-10pt}
	\label{tb:duplicatesTypes}
\end{table*}

To answer QE(4), we distinguish the similarity graphs into three categories according to the portion of duplicates in their ground truth with respect to the size of $|V_1|$ and $|V_2|$:

(i) \textit{Balanced} (BLC) are the entity collections where the vast majority of entities in $V_i$ are matched with an entity in $V_j$ (i=1 $\wedge$ j=2 or i=2 $\wedge$ j=1). This category includes all similarity graphs generated from $D_2$, $D_4$ and $D_{10}$.

(ii) \textit{One-sided} (OSD) are the entity collections, where only the vast majority of entities in $V_1$ are matched with an entity from $V_2$, or vice versa. OSD includes all graphs stemming from $D_3$ and $D_9$.

(iii) \textit{Scarce} (SCR) are the entity collections, where a small portion of entities in $V_i$ are matched with an entity in $V_j$ (i=1 $\wedge$ j=2 or i=2 $\wedge$ j=1). This category includes all graphs generated from $D_1$, $D_5$-$D_8$.

We apply this categorization to the four main types of similarity graphs defined in Section \ref{sec:inputs} and for each subcategory, we consider three new effectiveness measures:

(i) $\#Top1$ denotes the number of times an algorithm 
achieves the maximum F-Measure for a particular category of similarity graphs,

(ii) $\Delta~(\%)$ stands for the average difference (expressed as a percentage) between the highest and the second highest F1 across all similarity graphs of the same category, and 

(iii) $\#Top2$ denotes the number of times an algorithm scores 
the second highest F1 for a particular category of similarity graphs.

Note that in case of ties, we increment $\#Top1$ and $\#Top2$ for all involved algorithms. Note also that these three effectiveness measures also allow for answering QE(2) in more detail.
 
The results for these measures are reported in Table \ref{tb:duplicatesTypes}. For schema-based syntactic weights, there is a strong competition between \textsf{KRC} and \textsf{UMC} for the highest effectiveness.
Both algorithms achieve the maximum F1 for the same number of similarity graphs in the case of balanced and one-sided entity collections. For the former inputs, though, \textsf{UMC} exhibits consistently high performance, as it ranks second in almost all cases that it is not the top performer, unlike \textsf{KRC}, which comes second in 1/3 of these cases. Additionally, \textsf{UMC} achieves significantly higher $\Delta$ than \textsf{KRC}. For one-sided entity collections, \textsf{KRC} takes a minor lead over \textsf{UMC}: even though its $\Delta$ is slightly lower, it comes second three times more often than \textsf{UMC}. For scarce entity collections, \textsf{KRC} takes a clear lead over \textsf{UMC}, outperforming it with respect to both $\#Top1$ and $\#Top2$ to a large extent. \textsf{UMC} excels only with respect to $\Delta$.

Among the remaining algorithms, \textsf{CNC}, \textsf{RSR}, \textsf{BMC} and $\textsf{EXC}$ seem suitable only for scarce entity collections. \textsf{RSR} actually achieves the highest $\Delta$, while \textsf{EXC} achieves the second highest $\#Top1$ and $\#Top2$, outperforming \textsf{UMC}. Regarding \textsf{BAH}, we observe that for balanced entity collections, it outperforms all algorithms for 15\% of the similarity graphs, achieving the highest $\Delta$ and comes second for an equal number of inputs. This is in contrast to the poor average performance reported in Table \ref{tb:avPer}, but is explained by its stochastic nature, which gives rise to an unstable performance, as indicated by the significantly higher $\sigma$ than all other algorithms for all effectiveness measures.

In the case of schema-agnostic syntactic edge weights, \textsf{UMC} verifies its superiority over \textsf{KRC} for balanced entity collections with respect to all measures. \textsf{KRC} is actually outperformed by \textsf{BAH}, which achieves the top F1 2.5 times more often, while exhibiting the highest $\Delta$ among all algorithms. For one-sided entity collections, \textsf{KRC} excels with respect to $\#Top1$ and $\#Top2$, but \textsf{UMC} achieves significantly higher $\Delta$, while \textsf{EXC} constitutes the third best algorithm overall, as for the schema-based syntactic edge weights. In the case of scarce entity collections, the two competing algorithms are \textsf{KRC} and \textsf{EXC}, as they share the highest $\#Top1$. Yet, the former achieves three times higher $\Delta$ and slightly higher $\#Top2$. Suprisingly, \textsf{CNC} ranks second  in terms of $\#Top1$, while achieving the highest $\Delta$ by far, among all algorithms. As a result, \textsf{UMC} is left at the fourth place, followed by \textsf{BMC}.

For the semantic edge weights, we observe the following patterns: for the balanced entity collections, only \textsf{KRC}, \textsf{BAH} and \textsf{UMC} exhibit the highest performance for both schema-based and schema-agnostic weights. They excel in $\#Top1$, $\Delta$ and $\#Top2$, respectively. For one-sided entity collections, \textsf{KRC} is the dominant algorithm, especially in the case of schema-agnostic weights. For the schema-based ones, \textsf{EXC} consistently achieves very high performance, too. For scarce entity collections, there is a strong competition between \textsf{KRC} and \textsf{EXC}; the former consistently outperforms the latter with respect to $\Delta$, while \textsf{EXC} excels in $\#Top1$ for schema-agnostic weights and in $\#Top2$ for schema-based ones.

We examined other patterns with respect to additional characteristics of the entity collections, such as the distribution of positive and negative weights (i.e., between matching and non-matching entities, respectively) and the domain (e-commerce for $D_2$, $D_3$ and $D_8$, bibliographic data for $D_4$ and $D_9$ as well as movies for $D_5$-$D_8$ and $D_{10}$). Yet, no clear patterns emerged in these cases.

\begin{figure*}[t]
\centering
\includegraphics[width=0.24\textwidth]{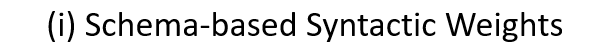}
\includegraphics[width=0.24\textwidth]{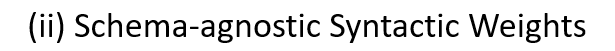}
\includegraphics[width=0.24\textwidth]{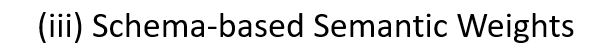}
\includegraphics[width=0.24\textwidth]{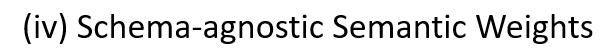}
\includegraphics[width=0.95\textwidth]{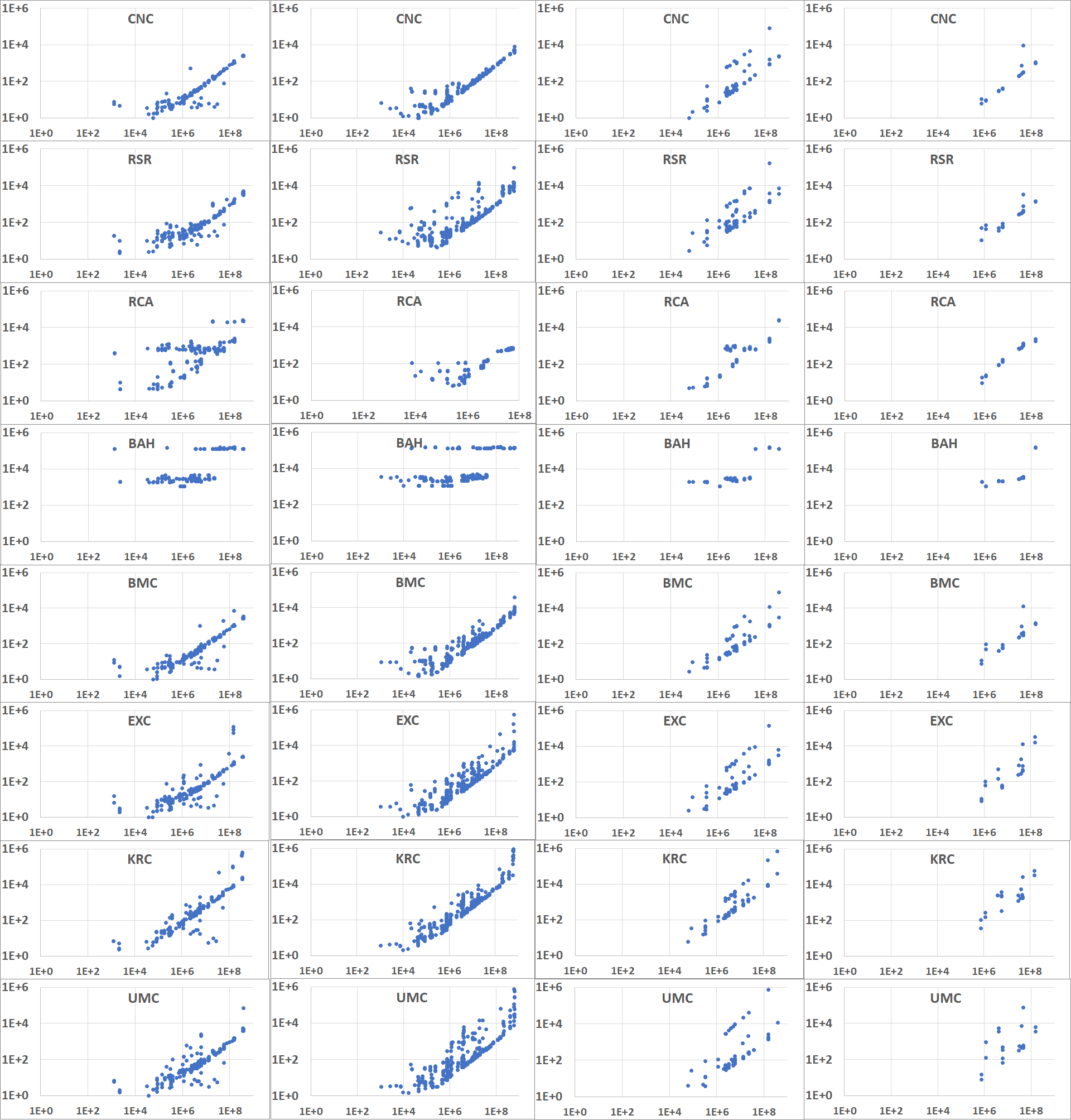}
\vspace{-10pt}
\caption{Scalability analysis of all algorithms over all similarity graphs with (i) schema-based syntactic, (ii) schema-agnostic syntactic, (iii) schema-based semantic and (iv) schema-agnostic semantic edge weights.
The horizontal axis corresponds to the number of edges in the similarity graphs and the vertical one to the run-time in milliseconds (maximum value=16.7 min).}
\vspace{-10pt}
\label{fig:scalability}
\end{figure*}


\textbf{Time Efficiency.} The (relative) run-time of the evaluated algorithms is a crucial aspect for the task of Entity Resolution, due to
the very large similarity graphs, which comprise thousands of entities/nodes and (hundreds of) millions of edges/entity pairs, as reported in Table \ref{tb:inputs}. Below, we study this aspect 
along with the scalability of the considered algorithms
over the 739 different similarity graphs. More specifically, we examine the following questions:
\begin{enumerate}[
    leftmargin=*,
    label={QT(\arabic*):},
    ref={QT(\arabic*)}]
    \item Which algorithm is the fastest one?
    \item Which factors affect the run-time of the algorithms?
    \item How scalable are the algorithms to large input sizes?
    \item Which algorithms offer the best trade-off between F-Measure and run-time?
\end{enumerate}
The average run-times over 10 executions of the evaluated algorithms per dataset and type of edge weights are reported in Table~\ref{tb:runTime}. 

Regarding QT(1), we observe that all algorithms are quite fast, as they are all able to process even the largest similarity graphs (i.e., those of $D_9$ and $D_{10}$) within minutes or even seconds. \textsf{CNC} is the fastest one almost in all cases, due to the simplicity of its approach. It is followed in close distance by \textsf{BMC} and \textsf{RSR}, with the former consistently outperforming the latter. \textsf{EXC} is also very efficient, but as expected, it is usually slower than \textsf{BMC}, due to the additional reciprocity check it involves. On the other extreme lies \textsf{BAH}, which constitutes by far the slowest method, yielding in many cases 2 or even 3 orders of magnitude longer run-times. The reason is the large number (10,000) of search steps we allow per dataset. For the largest datasets, its maximum run-time actually equals the run-time limit of 2 minutes, except for $D_9$, where the very large number of entities in $V_2$ delays the activation of the time-out. The rest of the algorithms lie between these two extremes: \textsf{KRC} is the slowest one, on average, while \textsf{UMC} and \textsf{RCA} exhibit significantly lower run-times.
Among the most effective algorithms, \textsf{EXC} is significantly faster than \textsf{UMC}, which is significantly faster than \textsf{KRC}.


\begin{table}[t]\centering
  \setlength{\tabcolsep}{2.5pt}
    \caption{Mean run-time per algorithm, dataset and type of input. Milliseconds are reported, except for BAH, $D_{10}$ and cases followed by s, which are measured in seconds.}
    \vspace{-8pt}
    {\scriptsize
	\begin{tabular}{ | l | r | r | r | r | r | r | r | r | }
		\cline{2-9}
		\multicolumn{1}{c|}{}&
		\multicolumn{1}{c|}{\textbf{CNC}} &
		\multicolumn{1}{c|}{\textbf{RSR}} &
		\multicolumn{1}{c|}{\textbf{RCA}} &
		\multicolumn{1}{c|}{\textbf{BAH} (sec)} &
		\multicolumn{1}{c|}{\textbf{BMC}} &
		\multicolumn{1}{c|}{\textbf{EXC}} &
		\multicolumn{1}{c|}{\textbf{KRC}} &
		\multicolumn{1}{c|}{\textbf{UMC}} \\
		\hline
        $\mathbf{D_1}$ & 3$\pm$1 & 8$\pm$5 & 8$\pm$4 & 1.9$\pm$.0 & 4$\pm$5 & 4$\pm$3 & 14$\pm$9 & 4$\pm$3 \\
        $\mathbf{D_2}$ & 8$\pm$1 & 15$\pm$4 & 20$\pm$2 & 1.1$\pm$.0 & 13$\pm$5 & 59$\pm$73 & 66$\pm$19 & 39$\pm$50 \\
        $\mathbf{D_3}$ & 21$\pm$13 & 36$\pm$16 & 57$\pm$15 & 2.2$\pm$.1 & 24$\pm$16 & 34$\pm$37 & 200$\pm$91 & 61$\pm$64 \\
        $\mathbf{D_4}$ & 32$\pm$18 & 57$\pm$23 & 136$\pm$20 & 2.0$\pm$.0 & 43$\pm$25 & 69$\pm$164 & 384$\pm$381 & 254$\pm$601 \\
        $\mathbf{D_5}$ & 44$\pm$41 & 56$\pm$38 & 662$\pm$66 & 3.6$\pm$.8 & 49$\pm$49 & 43$\pm$35 & 286$\pm$254 & 53$\pm$45 \\
        $\mathbf{D_6}$ & 52$\pm$60 & 86$\pm$79 & 706$\pm$97 & 3.0$\pm$.2 & 63$\pm$75 & 58$\pm$64 & 422$\pm$508 & 79$\pm$93 \\
        $\mathbf{D_7}$ & 41$\pm$95 & 43$\pm$17 & 980$\pm$251 & 3.1$\pm$.3 & 28$\pm$16 & 27$\pm$13 & 204$\pm$138 & 50$\pm$81 \\
        $\mathbf{D_8}$ & 196$\pm$168 & 209$\pm$167 & 557$\pm$136 & 123.5$\pm$.6 & 282$\pm$402 & 182$\pm$153 & 1.2s$\pm$1.0s & 213$\pm$194 \\
        $\mathbf{D_9}$ & 946$\pm$421 & 994$\pm$408 & 1.8s$\pm$.3s & 147.5$\pm$2.7 & 1.1s$\pm$1.4s & 18s$\pm$36s & 17s$\pm$29s & 1.1s$\pm$.5s\\
        $\mathbf{D_{10}}$ & 1.6$\pm$1.1 & 2.8$\pm$1.4 & 21.9$\pm$1.2 & 127$\pm$2 & 1.9$\pm$1.2 & 1.6$\pm$1.1 & 164$\pm$243 & 7.9$\pm$18.4\\
        \hline
        \multicolumn{9}{c}{(a) \textbf{Schema-based, syntactic inputs}} \\
        \hline
        $\mathbf{D_1}$ & 11$\pm$8 & 19$\pm$27 & 10$\pm$2 & 1.9$\pm$.1& 9$\pm$5 & 22$\pm$34 & 52$\pm$0.2 & 25$\pm$47 \\
        $\mathbf{D_2}$ & 6$\pm$3 & 18$\pm$8 & 18$\pm$4 & 1.1$\pm$.0 & 10$\pm$10 & 14$\pm$16 & 51$\pm$40 & 74$\pm$166\\
        $\mathbf{D_3}$ & 21$\pm$13 & 39$\pm$21 & 57$\pm$13 & 2.3$\pm$.4 & 34$\pm$48 & 81$\pm$177 & 582$\pm$686 & 599$\pm$1.2s\\
        $\mathbf{D_4}$ & 31$\pm$19 & 56$\pm$22 & 130$\pm$17 & 2.2$\pm$.3 & 40$\pm$26 & 44$\pm$74 & 334$\pm$705 & 125$\pm$356\\
        $\mathbf{D_5}$ & 93$\pm$58 & 206$\pm$393 & 616$\pm$55 & 2.9$\pm$.4 & 172$\pm$201 & 203$\pm$404 & 858$\pm$821 & 672$\pm$2.0s\\
        $\mathbf{D_6}$ & 86$\pm$83 & 124$\pm$84 & 676$\pm$70 & 3.3$\pm$.4 & 99$\pm$92 & 151$\pm$216 & 635$\pm$578 & 116$\pm$107\\
        $\mathbf{D_7}$ & 102$\pm$109 & 129$\pm$127 & 912$\pm$91 & 3.5$\pm$.4 & 104$\pm$96 & 112$\pm$168 & 637$\pm$713
        & 232$\pm$716\\
        $\mathbf{D_8}$ & 280$\pm$121 & 319$\pm$132 & 581$\pm$91 & 122.8$\pm$.4 & 245$\pm$103 & 443$\pm$1,264 & 1.8s$\pm$747 & 360$\pm$155\\
        $\mathbf{D_9}$ & 611$\pm$492 & 728$\pm$515 & 1.5s$\pm$340 & 144.8$\pm$2 & 627$\pm$499 & 1.7s$\pm$6.5s & 5.3s$\pm$9.7s & 2.1s$\pm$8.8s\\
        $\mathbf{D_{10}}$ & 2.5$\pm$2.1s & 8.4$\pm$14.3 & 24.6$\pm$2.8 & 128.4$\pm$2.1 & 4.1$\pm$6.0 & 21.1$\pm$86.4 & 205$\pm$354 & 157$\pm$600\\
        \hline
        \multicolumn{9}{c}{(b) \textbf{Schema-agnostic, syntactic inputs}} \\
        \hline
        $\mathbf{D_1}$ & 11$\pm$18 & 32$\pm$43 & 9$\pm$5 & 1.9$\pm$.0 & 9$\pm$7 & 16$\pm$19 & 33$\pm$26 & 20$\pm$29\\
        $\mathbf{D_2}$ & 7$\pm$0 & 91$\pm$51 & 23$\pm$3 & 1.0$\pm$.0 & 15$\pm$2 & 30$\pm$26 & 118$\pm$42 & 79$\pm$48\\
        $\mathbf{D_3}$ & 28$\pm$4 & 136$\pm$124 & 88$\pm$16 & 2.2$\pm$.0 & 282$\pm$11 & 602$\pm$612 & 2.4s$\pm$.2 & 5.6s$\pm$.0s\\
        $\mathbf{D_4}$ & 210$\pm$388 & 392$\pm$514 & 152$\pm$14 & 2.1$\pm$.1 & 217$\pm$366 & 295$\pm$569 & 1.5s$\pm$1.4s & 1.6$\pm$3.6s\\
        $\mathbf{D_5}$ & 384$\pm$869 & 883$\pm$1.7s & 629$\pm$49 & 2.8$\pm$.2 & 373$\pm$947 & 445$\pm$1s & 1.4s$\pm$2.8s & 2.3s$\pm$6.3\\
        $\mathbf{D_6}$ & 550$\pm$1.3s & 1.4s$\pm$2.8s & 730$\pm$70 & 3.0$\pm$.2 & 250$\pm$482 & 754$\pm$2s & 2.1s$\pm$4.4s & 4s$\pm$12s\\
        $\mathbf{D_7}$ & 181$\pm$372 & 264$\pm$415 & 917$\pm$77 & 3.0$\pm$.1 & 124$\pm$247 & 168$\pm$301 & 547$\pm$741 & 882$\pm$2.2s\\
        $\mathbf{D_8}$ & 216$\pm$0 & 381$\pm$78 & 654$\pm$31 & 123.6$\pm$.3 & 245$\pm$2 & 4.8s$\pm$6.4s & 1.8s$\pm$.0s & 364$\pm$10\\
        $\mathbf{D_9}$ & 1s$\pm$317 & 1.8s$\pm$1.1s & 2.2s$\pm$.3s & 147.8$\pm$2.7 & 1.1s$\pm$.1s & 1.2s$\pm$.2s & 8.2s$\pm$.6s & 1.7s$\pm$.5s\\
        $\mathbf{D_{10}}$ & 2.3$\pm$.1 & 5.2$\pm$2.4 & 24.9$\pm$1.3 & 128.8$\pm$1.2 & 39.4$\pm$51.6 & 4.8$\pm$2.4 & 365$\pm$463 & 137$\pm$191\\
        \hline
        \multicolumn{9}{c}{(c) \textbf{Schema-based, semantic inputs}}\\
        \hline
        $\mathbf{D_1}$ & 8$\pm$3 & 31$\pm$28 & 14$\pm$6 & 1.9$\pm$.1 & 9$\pm$3 & 9$\pm$2 & 72$\pm$50 & 11$\pm$5\\
        $\mathbf{D_2}$ & 9$\pm$0 & 58$\pm$22 & 23$\pm$1 & 1.1$\pm$.0 & 70$\pm$30 & 80$\pm$30 & 215$\pm$89 & 521$\pm$561\\
        $\mathbf{D_3}$ & 30$\pm$2 & 43$\pm$9 & 92$\pm$3 & 2.1$\pm$.0 & 39$\pm$0 & 320$\pm$255 & 2.5s$\pm$.0s & 4.5s$\pm$1.5s\\
        $\mathbf{D_4}$ & 40$\pm$1 & 65$\pm$13 & 147$\pm$16 & 2.0$\pm$.0 & 65$\pm$14 & 54$\pm$7 & 2.2s$\pm$1.4s & 250$\pm$194\\
        $\mathbf{D_5}$ & 199$\pm$5 & 280$\pm$16 & 700$\pm$22 & 2.8$\pm$.0 & 230$\pm$19 & 517$\pm$401 & 1.8s$\pm$.8s & 441$\pm$180\\
        $\mathbf{D_6}$ & 486$\pm$336 & 372$\pm$12 & 816$\pm$54 & 3.2$\pm$.2 & 660$\pm$412 & 1.1s$\pm$1.1s & 3.6s$\pm$2.5s & 3.9s$\pm$4.9s\\
        $\mathbf{D_7}$ & 2.1s$\pm$4.1s & 1.1s$\pm$1.3s & 1.2s$\pm$.1s & 3.3$\pm$.2 & 2.8s$\pm$5.6 & 3.0s$\pm$5.6s & 7.0s$\pm$10.9s & 15s$\pm$33s\\
        $\mathbf{D_8}$ & - & - & - & - & - & - & - & -\\
        $\mathbf{D_9}$ & 1.1s$\pm$.1 & 1.3s$\pm$.1s & 2.0s$\pm$.3s & 148.6$\pm$3.3 & 1.4s$\pm$.1 & 24s$\pm$11s & 46s$\pm$18s & 4.8s$\pm$1.8s \\
        $\mathbf{D_{10}}$ & - & - & - & - & - & - & - & -\\
        \hline
        \multicolumn{9}{c}{(d) \textbf{Schema-agnostic, semantic inputs}}
	\end{tabular}
	\vspace{-15pt}
	}
	\label{tb:runTime}
\end{table}

Regarding QT(2), there are two main factors that affect the reported run-times: (i) the 
time complexity of the algorithms, and (ii) the similarity threshold used for pruning the search space. Regarding the first factor, we observe that the run-times in Table~\ref{tb:runTime} verify the time complexities described in Section~\ref{sec:algorithms}. With $O(m)$, \textsf{CNC} and \textsf{BMC} are the fastest ones, followed by \textsf{RSR} and \textsf{EXC} with $O(n \; m)$, \textsf{RCA} with $O(|V_1|\;|V_2|)$, \textsf{UMC} with $O(m \; \log m)$ and \textsf{KRC} with $O(n + m \; \log m)$. \textsf{BAH}'s run-time is determined by the number of search steps and the run-time limit. 

Equally important is the effect of the similarity thresholds: the higher their optimal value (i.e., the one maximizing F1) is, the fewer edges are retained in the similarity graph and the faster is its processing. The optimal similarity threshold depends on 
the type of edge weights and the size of the similarity graph, as we explain in the threshold analysis in \cite{paperExtendedVersion}. This means that the relative time efficiency of algorithms with the same theoretical complexity should be attributed to their different similarity threshold. For example, the average optimal thresholds for \textsf{CNC} and \textsf{BMC} over all schema-based syntactic weights are 0.755 and 0.669, respectively, while over schema-agnostic syntactic weights they are 0.409 and 0.327, respectively. These large differences account for the significantly lower run-time of \textsf{CNC} in almost all datasets for both cases. The larger the difference in the similarity threshold, the larger is the difference in the run-times. 

Note that the similarity threshold typically accounts for the relative run-times that conflict with the relative time complexities, too: in case an algorithm runs faster than another one with lower time complexity, this is typically caused by the higher similarity threshold it employs. For example, \textsf{KRC} exhibits a lower average run-time than \textsf{EXC} over $D_9$ for schema-based syntactic weights, because their mean optimal similarity thresholds amount to 0.550 and 0.490, respectively. Similarly, \textsf{UMC} runs much faster than \textsf{EXC} over $D_8$ with schema-agnostic syntactic weights, because their mean optimal similarity thresholds amount to 0.427 and 0.387, respectively.

Finally, the similarity threshold 
accounts for the relative run-times between the same algorithm over different types of edge weights. For example, 
\textsf{EXC} is 13 times slower over the schema-agnostic syntactic weights of $D_{10}$ than their schema-based counterparts, even though the former involve just 25\% more edges than the latter, as reported in Table \ref{tb:inputs}. This significant difference should be attributed to the large deviation in the mean optimal thresholds:
0.153 for the former weights and 0.535 for the latter ones. The same applies to \textsf{UMC}, whose average run-time increases by 20 times when comparing the schema-based with the schema-agnostic weights, because its average optimal threshold drops from 0.481~to~0.110. 

To answer QT(3), Figure~\ref{fig:scalability} presents the scalability analysis of every algorithm over all similarity graphs for each type of edge weights. In each diagram, every point corresponds to the run-time of a different similarity graph. We observe that for all algorithms, the run-time increases linearly with the size of the similarity graphs: as the number of edges increases by four orders of magnitude, from $10^4$ to $10^8$, the run-times increase to a similar extent in most cases. For all algorithms, though, there are outlier points that deviate from the ``central'' curve. The larger the number of outliers is, the less robust is the time efficiency of the corresponding algorithm, due to its sensitivity to the size of the graph and the similarity threshold. In this respect, the least robust algorithms are \textsf{RSR} over schema-based syntactic weights along with \textsf{UMC} and \textsf{EXC} over schema-agnostic syntactic weights. These patterns seem to apply to the semantic weights, too,
despite the limited number of the similarity graphs, especially in the case of schema-agnostic settings.

Note that there are two exceptions to these patterns, namely \textsf{RCA} and \textsf{BAH}.
The diagrams of the former algorithm seem to involve a much lower number of points, as its time complexity depends exclusively on the number of entities in the input entity collections, i.e., $O(|V_1|\;|V_2|)$. As a result, different similarity graphs from the same dataset yield similar run-times that coincide in the diagrams of Figure~\ref{fig:scalability}. Regarding \textsf{BAH}, it exhibits a step-resembling scalability graph, because its processing terminates after a pre-defined timeout or a fixed number of iterations (whichever comes first), independently of the size of the similarity graph.




\begin{figure}[t]
\centering
\includegraphics[width=0.34\textwidth]{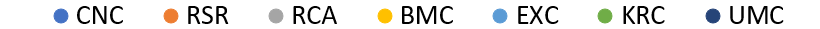}
\includegraphics[width=0.4\textwidth]{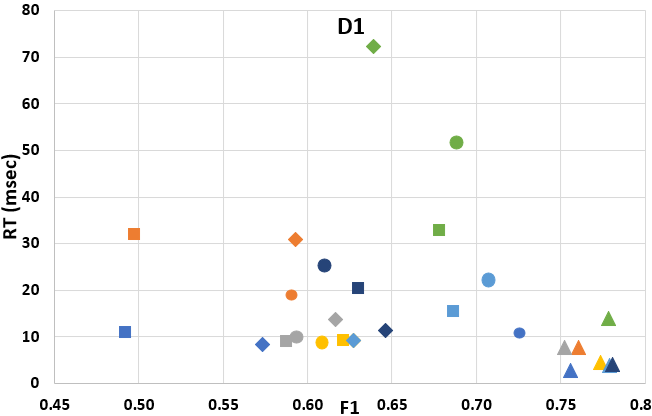}
\vspace{-7pt}
\caption{{\small F1-runtime diagram for all algorithms, but \textsf{BAH} over~$D_1$.}}
\vspace{-14pt}
\label{fig:d1F1RT}
\end{figure}

Finally, to answer QT(4), Figure \ref{fig:d1F1RT} depicts the trade-off between macro-averaged F1 and run-time (RT) per algorithm and type of edge weights in $D_1$. Note that every type corresponds to a different shape: circle stands for the schema-agnostic syntactic weights, triangle for the schema-based syntactic ones, rhombus for the schema-agnostic semantic ones and rectangle for the schema-based semantic ones. We observe that the schema-based syntactic similarity graphs dominate the other types of input, as they exhibit very high F1 in combination with the lowest run-times. The former should be attributed to the relatively clean values of names and phones for the duplicate entities and the latter to the lack of attribute values for most non-matching entities. As a result, the average size of these graphs is significantly lower than the other types of graphs, especially the schema-agnostic ones, as shown in Table~\ref{tb:inputs}. Among the schema-based syntactic inputs, the differences in F1 are lower than 4\%, with \textsf{UMC} achieving the best trade-off between the two measures (average F1=0.781 for average run-time=4 msec). This combination practically dominates all others. Only \textsf{CNC} is significantly faster (RT=3 msec), but its F1 (0.756) is also significantly lower. For the patterns pertaining to rest of the datasets, please refer to the extended version of the paper \cite{paperExtendedVersion}.


\textbf{Comparison with the best matching methods.} We now compare the performance achieved by bipartite graph matching algorithms with the recent state-of-the-art matching methods: ZeroER \cite{DBLP:conf/sigmod/WuCSCT20}, which leverages unsupervised learning, and DITTO \cite{DBLP:journals/pvldb/0001LSDT20}, which is based on deep learning. We consider the four common datasets, namely $D_2$-$D_5$ ($D_1$ is a larger and noisier version of FZ in \cite{DBLP:conf/sigmod/WuCSCT20,DBLP:journals/pvldb/0001LSDT20}, and thus not directly comparable). Table \ref{tb:bestF1} reports the relative performance in terms of maximum F-Measure for ZeroER and DITTO, as it was reported reported in \cite{DBLP:conf/sigmod/WuCSCT20} and \cite{DBLP:journals/pvldb/0001LSDT20}, respectively.

Bipartite matching is represented by \textsf{UMC} in combination with cosine similarity over schema-agnostic vector models with TF-IDF weights; the best representation model and the corresponding similarity threshold depend on the dataset. These settings do not necessarily 
correspond to the highest F-Measure across all algorithms and similarity graphs we have considered, but
demonstrate the capabilities of bipartite matching when varying just two configuration parameters. The results appear in Table \ref{tb:bestF1}.

Compared to ZeroER, we observe that \textsf{UMC} consistently achieves higher performance: its F1 is higher by 3\%, 9\%, 25\% and 83\% over $D_4$, $D_5$, $D_3$ and $D_2$, respectively. Compared to DITTO, \textsf{UMC} achieves identical performance over $D_4$ and 2\% lower F1 over $D_5$. The difference is much larger over $D_3$, where \textsf{UMC} underperforms by 21\%, due to the contextual evidence captured by the training set and the RoBERTa language model \cite{DBLP:journals/corr/abs-1907-11692} that lies at the core of DITTO. Note, though, that the difference is reduced to 12\%, when considering the best schema-based configuration of \textsf{UMC} (the Overlap Coefficient of the tokens of the ``Title'' attribute with a similarity threshold of 0.3). For $D_2$, \textsf{UMC} outperforms DITTO~by~7\%.

Overall, we can conclude that bipartite graph matching underperforms the best (deep learning-based) performance in the literature in just one out of four benchmark datasets.

\begin{table}[t]\centering
    \caption{Comparison to state-of-the-art matching methods.}
    \vspace{-10pt}
    {\footnotesize
    	\begin{tabular}{ | c | c | c | l |}
		\cline{2-4}
		\multicolumn{1}{c}{} &
		\multicolumn{1}{|c|}{ZeroER} &
		\multicolumn{1}{c|}{DITTO} &
		\multicolumn{1}{l|}{UMC (schema-agnostic TF-IDF weights, cosine sim.)} \\
		\hline
        \hline
        $D_2$ & 0.52 & 0.89 &~~0.95 (character bi-grams, $t=0.35$) \\
        $D_3$ & 0.48 & 0.76 &~~0.60 (token bi-grams, $t=0.05$)\\
        $D_4$ & 0.96 & 0.99 &~~0.99 (token uni-grams, $t=0.40$)\\
        $D_5$ & 0.86 & 0.96 &~~0.94 (character four-grams, $t=0.35$)\\ 
		\hline
	\end{tabular}
	}
	\label{tb:bestF1}
	\vspace{-16pt}
\end{table}


\vspace{-6pt}
\section{Conclusions}

We draw the following important patterns from our experiments:

(i) 
The best performing algorithm for a particular similarity graph mainly depends
on the type of edge weights and the portion of duplicates with respect to the total number of nodes/entities.

(ii) \textsf{CNC} constitutes the fastest algorithm, due to its simplicity and the high similarity thresholds it employs, achieving the highest precision at the cost of low recall. It frequently outperforms all other algorithms with respect to F1 in the case of scarce entity collections with syntactic weights, especially the schema-agnostic~ones.

(iii) \textsf{RSR} is a fast algorithm that rarely achieves high effectiveness,
in the case of scarce entity collections.

(iv) \textsf{RCA} is an efficient method that never excels in effectiveness.

(v) \textsf{BAH} constitutes a slow, stochastic approach that is capable of the best and the worst. It frequently achieves, by far, the highest F1 over balanced entity collections (and rarely over scarce ones), but in most cases, it yields the lowest scores with respect to all effectiveness measures.

(vi) \textsf{BMC} is the second fastest algorithm that tries to balance precision and recall, being particularly effective in the case of scarce entity collections, 
especially in combination with syntactic weights.

(vii) \textsf{EXC} improves \textsf{BMC} by boosting precision at the cost of lower recall and higher run-time. It consistently achieves (close to) the maximum F1 over scarce and one-sided entity collections, losing only to \textsf{KRC} and (rarely) to \textsf{UMC}. Given, though, that it outperforms both algorithms to a significant extent with respect to run-time, it constitutes the best choice for applications requiring both high effectiveness and efficiency/scalability.

(viii) \textsf{KRC} achieves very high or the highest effectiveness in most cases, especially over one-sided and scarce entity collections. This comes, though, at the cost of higher (yet stable) run-times than its top-performing counterparts.

(ix) \textsf{UMC} is the best choice for balanced entity collections, especially when coupled with syntactic weights, exhibiting a much more robust performance than \textsf{BAH}. It achieves very high (and frequently the highest) effectiveness in the rest of the cases, too. Its run-time, though, is rather unstable, depending largely on the optimal similarity threshold.



\begin{acks} 
This project has received funding from the Hellenic Foundation for Research and Innovation (HFRI) and the General Secretariat for Research and Technology (GSRT), under grant agreement No~969.
\end{acks}

\balance

\bibliographystyle{ACM-Reference-Format}
\bibliography{refs}


\begin{thebibliography}{54}


\ifx \showCODEN    \undefined \def \showCODEN     #1{\unskip}     \fi
\ifx \showDOI      \undefined \def \showDOI       #1{#1}\fi
\ifx \showISBNx    \undefined \def \showISBNx     #1{\unskip}     \fi
\ifx \showISBNxiii \undefined \def \showISBNxiii  #1{\unskip}     \fi
\ifx \showISSN     \undefined \def \showISSN      #1{\unskip}     \fi
\ifx \showLCCN     \undefined \def \showLCCN      #1{\unskip}     \fi
\ifx \shownote     \undefined \def \shownote      #1{#1}          \fi
\ifx \showarticletitle \undefined \def \showarticletitle #1{#1}   \fi
\ifx \showURL      \undefined \def \showURL       {\relax}        \fi
\providecommand\bibfield[2]{#2}
\providecommand\bibinfo[2]{#2}
\providecommand\natexlab[1]{#1}
\providecommand\showeprint[2][]{arXiv:#2}

\bibitem[\protect\citeauthoryear{Assi, Mcheick, and Dhifli}{Assi
  et~al\mbox{.}}{2019}]%
        {DBLP:conf/bigdataconf/0002MD19}
\bibfield{author}{\bibinfo{person}{Ali Assi}, \bibinfo{person}{Hamid Mcheick},
  {and} \bibinfo{person}{Wajdi Dhifli}.} \bibinfo{year}{2019}\natexlab{}.
\newblock \showarticletitle{{BIGMAT:} {A} Distributed Affinity-Preserving
  Random Walk Strategy for Instance Matching on Knowledge Graphs}. In
  \bibinfo{booktitle}{\emph{{IEEE} Big Data}}. \bibinfo{pages}{1028--1033}.
\newblock


\bibitem[\protect\citeauthoryear{Bojanowski, Grave, Joulin, and
  Mikolov}{Bojanowski et~al\mbox{.}}{2017}]%
        {DBLP:journals/tacl/BojanowskiGJM17}
\bibfield{author}{\bibinfo{person}{Piotr Bojanowski}, \bibinfo{person}{Edouard
  Grave}, \bibinfo{person}{Armand Joulin}, {and} \bibinfo{person}{Tom{\'{a}}s
  Mikolov}.} \bibinfo{year}{2017}\natexlab{}.
\newblock \showarticletitle{Enriching Word Vectors with Subword Information}.
\newblock \bibinfo{journal}{\emph{Trans. Assoc. Comput. Linguistics}}
  \bibinfo{volume}{5} (\bibinfo{year}{2017}), \bibinfo{pages}{135--146}.
\newblock


\bibitem[\protect\citeauthoryear{Brunner and Stockinger}{Brunner and
  Stockinger}{2020}]%
        {DBLP:conf/edbt/BrunnerS20}
\bibfield{author}{\bibinfo{person}{Ursin Brunner} {and} \bibinfo{person}{Kurt
  Stockinger}.} \bibinfo{year}{2020}\natexlab{}.
\newblock \showarticletitle{Entity Matching with Transformer Architectures -
  {A} Step Forward in Data Integration}. In \bibinfo{booktitle}{\emph{{EDBT}}}.
  \bibinfo{pages}{463--473}.
\newblock


\bibitem[\protect\citeauthoryear{Christen}{Christen}{2012}]%
        {DBLP:books/daglib/0030287}
\bibfield{author}{\bibinfo{person}{Peter Christen}.}
  \bibinfo{year}{2012}\natexlab{}.
\newblock \bibinfo{booktitle}{\emph{Data Matching - Concepts and Techniques for
  Record Linkage, Entity Resolution, and Duplicate Detection}}.
\newblock \bibinfo{publisher}{Springer}.
\newblock


\bibitem[\protect\citeauthoryear{Christophides, Efthymiou, Palpanas, Papadakis,
  and Stefanidis}{Christophides et~al\mbox{.}}{2021}]%
        {DBLP:journals/csur/ChristophidesEP21}
\bibfield{author}{\bibinfo{person}{Vassilis Christophides},
  \bibinfo{person}{Vasilis Efthymiou}, \bibinfo{person}{Themis Palpanas},
  \bibinfo{person}{George Papadakis}, {and} \bibinfo{person}{Kostas
  Stefanidis}.} \bibinfo{year}{2021}\natexlab{}.
\newblock \showarticletitle{An Overview of End-to-End Entity Resolution for Big
  Data}.
\newblock \bibinfo{journal}{\emph{{ACM} Comput. Surv.}} \bibinfo{volume}{53},
  \bibinfo{number}{6} (\bibinfo{year}{2021}), \bibinfo{pages}{127:1--127:42}.
\newblock


\bibitem[\protect\citeauthoryear{Dasgupta, Papadimitriou, and
  Vazirani}{Dasgupta et~al\mbox{.}}{2008}]%
        {DBLP:books/daglib/0017733}
\bibfield{author}{\bibinfo{person}{Sanjoy Dasgupta},
  \bibinfo{person}{Christos~H. Papadimitriou}, {and} \bibinfo{person}{Umesh~V.
  Vazirani}.} \bibinfo{year}{2008}\natexlab{}.
\newblock \bibinfo{booktitle}{\emph{Algorithms}}.
\newblock \bibinfo{publisher}{McGraw-Hill}.
\newblock


\bibitem[\protect\citeauthoryear{Demsar}{Demsar}{2006}]%
        {DBLP:journals/jmlr/Demsar06}
\bibfield{author}{\bibinfo{person}{Janez Demsar}.}
  \bibinfo{year}{2006}\natexlab{}.
\newblock \showarticletitle{Statistical Comparisons of Classifiers over
  Multiple Data Sets}.
\newblock \bibinfo{journal}{\emph{J. Mach. Learn. Res.}}  \bibinfo{volume}{7}
  (\bibinfo{year}{2006}), \bibinfo{pages}{1--30}.
\newblock


\bibitem[\protect\citeauthoryear{Devlin, Chang, Lee, and Toutanova}{Devlin
  et~al\mbox{.}}{2019}]%
        {DBLP:conf/naacl/DevlinCLT19}
\bibfield{author}{\bibinfo{person}{Jacob Devlin}, \bibinfo{person}{Ming{-}Wei
  Chang}, \bibinfo{person}{Kenton Lee}, {and} \bibinfo{person}{Kristina
  Toutanova}.} \bibinfo{year}{2019}\natexlab{}.
\newblock \showarticletitle{{BERT:} Pre-training of Deep Bidirectional
  Transformers for Language Understanding}. In
  \bibinfo{booktitle}{\emph{{NAACL-HLT}}}. \bibinfo{pages}{4171--4186}.
\newblock


\bibitem[\protect\citeauthoryear{Dong and Srivastava}{Dong and
  Srivastava}{2015}]%
        {DBLP:series/synthesis/2015Dong}
\bibfield{author}{\bibinfo{person}{Xin~Luna Dong} {and} \bibinfo{person}{Divesh
  Srivastava}.} \bibinfo{year}{2015}\natexlab{}.
\newblock \bibinfo{booktitle}{\emph{Big Data Integration}}.
\newblock \bibinfo{publisher}{Morgan {\&} Claypool Publishers}.
\newblock


\bibitem[\protect\citeauthoryear{Efthymiou, Papadakis, Stefanidis, and
  Christophides}{Efthymiou et~al\mbox{.}}{2019}]%
        {DBLP:conf/edbt/Efthymiou0SC19}
\bibfield{author}{\bibinfo{person}{Vasilis Efthymiou}, \bibinfo{person}{George
  Papadakis}, \bibinfo{person}{Kostas Stefanidis}, {and}
  \bibinfo{person}{Vassilis Christophides}.} \bibinfo{year}{2019}\natexlab{}.
\newblock \showarticletitle{MinoanER: Schema-Agnostic, Non-Iterative, Massively
  Parallel Resolution of Web Entities}. In \bibinfo{booktitle}{\emph{{EDBT}}}.
  \bibinfo{publisher}{OpenProceedings.org}, \bibinfo{pages}{373--384}.
\newblock


\bibitem[\protect\citeauthoryear{Fellegi and Sunter}{Fellegi and
  Sunter}{1969}]%
        {FellegiS69}
\bibfield{author}{\bibinfo{person}{I.~P. Fellegi} {and} \bibinfo{person}{A.~B.
  Sunter}.} \bibinfo{year}{1969}\natexlab{}.
\newblock \showarticletitle{{A Theory for Record Linkage}}.
\newblock \bibinfo{journal}{\emph{J. Amer. Statist. Assoc.}}
  \bibinfo{volume}{64}, \bibinfo{number}{328} (\bibinfo{year}{1969}),
  \bibinfo{pages}{1183--1210}.
\newblock


\bibitem[\protect\citeauthoryear{Fredman and Tarjan}{Fredman and
  Tarjan}{1987}]%
        {DBLP:journals/jacm/FredmanT87}
\bibfield{author}{\bibinfo{person}{Michael~L. Fredman} {and}
  \bibinfo{person}{Robert~Endre Tarjan}.} \bibinfo{year}{1987}\natexlab{}.
\newblock \showarticletitle{Fibonacci heaps and their uses in improved network
  optimization algorithms}.
\newblock \bibinfo{journal}{\emph{J. {ACM}}} \bibinfo{volume}{34},
  \bibinfo{number}{3} (\bibinfo{year}{1987}), \bibinfo{pages}{596--615}.
\newblock


\bibitem[\protect\citeauthoryear{Gale and Shapley}{Gale and Shapley}{1962}]%
        {DBLP:journals/tamm/GaleS62}
\bibfield{author}{\bibinfo{person}{D. Gale} {and} \bibinfo{person}{L.~S.
  Shapley}.} \bibinfo{year}{1962}\natexlab{}.
\newblock \showarticletitle{College Admissions and the Stability of Marriage}.
\newblock \bibinfo{journal}{\emph{Am. Math. Mon.}} \bibinfo{volume}{69},
  \bibinfo{number}{1} (\bibinfo{year}{1962}), \bibinfo{pages}{9--15}.
\newblock


\bibitem[\protect\citeauthoryear{Gemmell, Rubinstein, and Chandra}{Gemmell
  et~al\mbox{.}}{2011}]%
        {DBLP:journals/corr/abs-1108-6016}
\bibfield{author}{\bibinfo{person}{Jim Gemmell}, \bibinfo{person}{Benjamin
  I.~P. Rubinstein}, {and} \bibinfo{person}{Ashok~K. Chandra}.}
  \bibinfo{year}{2011}\natexlab{}.
\newblock \showarticletitle{Improving Entity Resolution with Global
  Constraints}.
\newblock \bibinfo{journal}{\emph{CoRR}}  \bibinfo{volume}{abs/1108.6016}
  (\bibinfo{year}{2011}).
\newblock


\bibitem[\protect\citeauthoryear{Giannakopoulos, Karkaletsis, Vouros, and
  Stamatopoulos}{Giannakopoulos et~al\mbox{.}}{2008}]%
        {DBLP:journals/tslp/GiannakopoulosKVS08}
\bibfield{author}{\bibinfo{person}{George Giannakopoulos},
  \bibinfo{person}{Vangelis Karkaletsis}, \bibinfo{person}{George~A. Vouros},
  {and} \bibinfo{person}{Panagiotis Stamatopoulos}.}
  \bibinfo{year}{2008}\natexlab{}.
\newblock \showarticletitle{Summarization system evaluation revisited: N-gram
  graphs}.
\newblock \bibinfo{journal}{\emph{{ACM} Trans. Speech Lang. Process.}}
  \bibinfo{volume}{5}, \bibinfo{number}{3} (\bibinfo{year}{2008}),
  \bibinfo{pages}{5:1--5:39}.
\newblock


\bibitem[\protect\citeauthoryear{Giannakopoulos and Palpanas}{Giannakopoulos
  and Palpanas}{2010}]%
        {giannakopoulos2010content}
\bibfield{author}{\bibinfo{person}{George Giannakopoulos} {and}
  \bibinfo{person}{Themis Palpanas}.} \bibinfo{year}{2010}\natexlab{}.
\newblock \showarticletitle{Content and type as orthogonal modeling features: a
  study on user interest awareness in entity subscription services}.
\newblock \bibinfo{journal}{\emph{International Journal of Advances on Networks
  and Services}} \bibinfo{volume}{3}, \bibinfo{number}{2}
  (\bibinfo{year}{2010}).
\newblock


\bibitem[\protect\citeauthoryear{Gotoh}{Gotoh}{1982}]%
        {gotoh1982improved}
\bibfield{author}{\bibinfo{person}{Osamu Gotoh}.}
  \bibinfo{year}{1982}\natexlab{}.
\newblock \showarticletitle{An improved algorithm for matching biological
  sequences}.
\newblock \bibinfo{journal}{\emph{Journal of molecular biology}}
  \bibinfo{volume}{162}, \bibinfo{number}{3} (\bibinfo{year}{1982}),
  \bibinfo{pages}{705--708}.
\newblock


\bibitem[\protect\citeauthoryear{Gutierrez and Sequeda}{Gutierrez and
  Sequeda}{2020}]%
        {Gutierrez20}
\bibfield{author}{\bibinfo{person}{Claudio Gutierrez} {and}
  \bibinfo{person}{Juan~F. Sequeda}.} \bibinfo{year}{2020}\natexlab{}.
\newblock \bibinfo{booktitle}{\emph{Knowledge Graphs: A Tutorial on the History
  of Knowledge Graph's Main Ideas}}.
\newblock \bibinfo{publisher}{Association for Computing Machinery},
  \bibinfo{pages}{3509–3510}.
\newblock
\urldef\tempurl%
\url{https://doi.org/10.1145/3340531.3412176}
\showURL{%
\tempurl}


\bibitem[\protect\citeauthoryear{Hassanzadeh, Chiang, Miller, and
  Lee}{Hassanzadeh et~al\mbox{.}}{2009}]%
        {DBLP:journals/pvldb/HassanzadehCML09}
\bibfield{author}{\bibinfo{person}{Oktie Hassanzadeh}, \bibinfo{person}{Fei
  Chiang}, \bibinfo{person}{Ren{\'{e}}e~J. Miller}, {and}
  \bibinfo{person}{Hyun~Chul Lee}.} \bibinfo{year}{2009}\natexlab{}.
\newblock \showarticletitle{Framework for Evaluating Clustering Algorithms in
  Duplicate Detection}.
\newblock \bibinfo{journal}{\emph{Proc. {VLDB} Endow.}} \bibinfo{volume}{2},
  \bibinfo{number}{1} (\bibinfo{year}{2009}), \bibinfo{pages}{1282--1293}.
\newblock


\bibitem[\protect\citeauthoryear{Herbold}{Herbold}{2020}]%
        {Herbold2020}
\bibfield{author}{\bibinfo{person}{Steffen Herbold}.}
  \bibinfo{year}{2020}\natexlab{}.
\newblock \showarticletitle{Autorank: A Python package for automated ranking of
  classifiers}.
\newblock \bibinfo{journal}{\emph{Journal of Open Source Software}}
  \bibinfo{volume}{5}, \bibinfo{number}{48} (\bibinfo{year}{2020}),
  \bibinfo{pages}{2173}.
\newblock


\bibitem[\protect\citeauthoryear{Kir{\'{a}}ly}{Kir{\'{a}}ly}{2013}]%
        {DBLP:journals/algorithms/Kiraly13}
\bibfield{author}{\bibinfo{person}{Zolt{\'{a}}n Kir{\'{a}}ly}.}
  \bibinfo{year}{2013}\natexlab{}.
\newblock \showarticletitle{Linear Time Local Approximation Algorithm for
  Maximum Stable Marriage}.
\newblock \bibinfo{journal}{\emph{Algorithms}} \bibinfo{volume}{6},
  \bibinfo{number}{3} (\bibinfo{year}{2013}), \bibinfo{pages}{471--484}.
\newblock


\bibitem[\protect\citeauthoryear{Konda, Das, C., Doan, Ardalan, Ballard, Li,
  Panahi, Zhang, Naughton, Prasad, Krishnan, Deep, and Raghavendra}{Konda
  et~al\mbox{.}}{2016}]%
        {DBLP:journals/pvldb/KondaDCDABLPZNP16}
\bibfield{author}{\bibinfo{person}{Pradap Konda}, \bibinfo{person}{Sanjib Das},
  \bibinfo{person}{Paul Suganthan~G. C.}, \bibinfo{person}{AnHai Doan},
  \bibinfo{person}{Adel Ardalan}, \bibinfo{person}{Jeffrey~R. Ballard},
  \bibinfo{person}{Han Li}, \bibinfo{person}{Fatemah Panahi},
  \bibinfo{person}{Haojun Zhang}, \bibinfo{person}{Jeffrey~F. Naughton},
  \bibinfo{person}{Shishir Prasad}, \bibinfo{person}{Ganesh Krishnan},
  \bibinfo{person}{Rohit Deep}, {and} \bibinfo{person}{Vijay Raghavendra}.}
  \bibinfo{year}{2016}\natexlab{}.
\newblock \showarticletitle{Magellan: Toward Building Entity Matching
  Management Systems}.
\newblock \bibinfo{journal}{\emph{Proc. {VLDB} Endow.}} \bibinfo{volume}{9},
  \bibinfo{number}{12} (\bibinfo{year}{2016}), \bibinfo{pages}{1197--1208}.
\newblock


\bibitem[\protect\citeauthoryear{K{\"{o}}pcke, Thor, and Rahm}{K{\"{o}}pcke
  et~al\mbox{.}}{2010}]%
        {DBLP:journals/pvldb/KopckeTR10}
\bibfield{author}{\bibinfo{person}{Hanna K{\"{o}}pcke},
  \bibinfo{person}{Andreas Thor}, {and} \bibinfo{person}{Erhard Rahm}.}
  \bibinfo{year}{2010}\natexlab{}.
\newblock \showarticletitle{Evaluation of entity resolution approaches on
  real-world match problems}.
\newblock \bibinfo{journal}{\emph{Proc. {VLDB} Endow.}} \bibinfo{volume}{3},
  \bibinfo{number}{1} (\bibinfo{year}{2010}), \bibinfo{pages}{484--493}.
\newblock


\bibitem[\protect\citeauthoryear{Kriege, Giscard, Bause, and Wilson}{Kriege
  et~al\mbox{.}}{2019}]%
        {DBLP:conf/icdm/KriegeGB019}
\bibfield{author}{\bibinfo{person}{Nils~M. Kriege},
  \bibinfo{person}{Pierre{-}Louis Giscard}, \bibinfo{person}{Franka Bause},
  {and} \bibinfo{person}{Richard~C. Wilson}.} \bibinfo{year}{2019}\natexlab{}.
\newblock \showarticletitle{Computing Optimal Assignments in Linear Time for
  Approximate Graph Matching}. In \bibinfo{booktitle}{\emph{{ICDM}}}.
  \bibinfo{pages}{349--358}.
\newblock


\bibitem[\protect\citeauthoryear{Kuhn and Yaw}{Kuhn and Yaw}{1955}]%
        {Kuhn55thehungarian}
\bibfield{author}{\bibinfo{person}{H.~W. Kuhn} {and} \bibinfo{person}{Bryn
  Yaw}.} \bibinfo{year}{1955}\natexlab{}.
\newblock \showarticletitle{The Hungarian method for the assignment problem}.
\newblock \bibinfo{journal}{\emph{Naval Res. Logist. Quart}}
  (\bibinfo{year}{1955}), \bibinfo{pages}{83--97}.
\newblock


\bibitem[\protect\citeauthoryear{Kurtzberg}{Kurtzberg}{1962}]%
        {kurtzberg1962approximation}
\bibfield{author}{\bibinfo{person}{Jerome~M Kurtzberg}.}
  \bibinfo{year}{1962}\natexlab{}.
\newblock \showarticletitle{On approximation methods for the assignment
  problem}.
\newblock \bibinfo{journal}{\emph{Journal of the ACM (JACM)}}
  \bibinfo{volume}{9}, \bibinfo{number}{4} (\bibinfo{year}{1962}),
  \bibinfo{pages}{419--439}.
\newblock


\bibitem[\protect\citeauthoryear{Lan, Chen, Goodman, Gimpel, Sharma, and
  Soricut}{Lan et~al\mbox{.}}{2020}]%
        {DBLP:conf/iclr/LanCGGSS20}
\bibfield{author}{\bibinfo{person}{Zhenzhong Lan}, \bibinfo{person}{Mingda
  Chen}, \bibinfo{person}{Sebastian Goodman}, \bibinfo{person}{Kevin Gimpel},
  \bibinfo{person}{Piyush Sharma}, {and} \bibinfo{person}{Radu Soricut}.}
  \bibinfo{year}{2020}\natexlab{}.
\newblock \showarticletitle{{ALBERT:} {A} Lite {BERT} for Self-supervised
  Learning of Language Representations}. In \bibinfo{booktitle}{\emph{{ICLR}}}.
\newblock


\bibitem[\protect\citeauthoryear{Li, Li, Suhara, Doan, and Tan}{Li
  et~al\mbox{.}}{2020}]%
        {DBLP:journals/pvldb/0001LSDT20}
\bibfield{author}{\bibinfo{person}{Yuliang Li}, \bibinfo{person}{Jinfeng Li},
  \bibinfo{person}{Yoshihiko Suhara}, \bibinfo{person}{AnHai Doan}, {and}
  \bibinfo{person}{Wang{-}Chiew Tan}.} \bibinfo{year}{2020}\natexlab{}.
\newblock \showarticletitle{Deep Entity Matching with Pre-Trained Language
  Models}.
\newblock \bibinfo{journal}{\emph{Proc. {VLDB} Endow.}} \bibinfo{volume}{14},
  \bibinfo{number}{1} (\bibinfo{year}{2020}), \bibinfo{pages}{50--60}.
\newblock


\bibitem[\protect\citeauthoryear{Li, Li, Suhara, Wang, Hirota, and Tan}{Li
  et~al\mbox{.}}{2021}]%
        {DBLP:journals/jdiq/LiLSWHT21}
\bibfield{author}{\bibinfo{person}{Yuliang Li}, \bibinfo{person}{Jinfeng Li},
  \bibinfo{person}{Yoshihiko Suhara}, \bibinfo{person}{Jin Wang},
  \bibinfo{person}{Wataru Hirota}, {and} \bibinfo{person}{Wang{-}Chiew Tan}.}
  \bibinfo{year}{2021}\natexlab{}.
\newblock \showarticletitle{Deep Entity Matching: Challenges and
  Opportunities}.
\newblock \bibinfo{journal}{\emph{{ACM} J. Data Inf. Qual.}}
  \bibinfo{volume}{13}, \bibinfo{number}{1} (\bibinfo{year}{2021}),
  \bibinfo{pages}{1:1--1:17}.
\newblock


\bibitem[\protect\citeauthoryear{Liu, Ott, Goyal, Du, Joshi, Chen, Levy, Lewis,
  Zettlemoyer, and Stoyanov}{Liu et~al\mbox{.}}{2019}]%
        {DBLP:journals/corr/abs-1907-11692}
\bibfield{author}{\bibinfo{person}{Yinhan Liu}, \bibinfo{person}{Myle Ott},
  \bibinfo{person}{Naman Goyal}, \bibinfo{person}{Jingfei Du},
  \bibinfo{person}{Mandar Joshi}, \bibinfo{person}{Danqi Chen},
  \bibinfo{person}{Omer Levy}, \bibinfo{person}{Mike Lewis},
  \bibinfo{person}{Luke Zettlemoyer}, {and} \bibinfo{person}{Veselin
  Stoyanov}.} \bibinfo{year}{2019}\natexlab{}.
\newblock \showarticletitle{RoBERTa: {A} Robustly Optimized {BERT} Pretraining
  Approach}.
\newblock \bibinfo{journal}{\emph{CoRR}}  \bibinfo{volume}{abs/1907.11692}
  (\bibinfo{year}{2019}).
\newblock


\bibitem[\protect\citeauthoryear{Lovasz and Plummer}{Lovasz and
  Plummer}{[n.d.]}]%
        {matchingTheoryBook}
\bibfield{author}{\bibinfo{person}{L. Lovasz} {and} \bibinfo{person}{M.~D.
  Plummer}.} \bibinfo{year}{[n.d.]}\natexlab{}.
\newblock \bibinfo{booktitle}{\emph{Matching theory}}.
\newblock


\bibitem[\protect\citeauthoryear{Malliaros, Meladianos, and
  Vazirgiannis}{Malliaros et~al\mbox{.}}{2018}]%
        {graphsTutorial}
\bibfield{author}{\bibinfo{person}{Fragkiskos~D. Malliaros},
  \bibinfo{person}{Polykarpos Meladianos}, {and} \bibinfo{person}{Michalis
  Vazirgiannis}.} \bibinfo{year}{2018}\natexlab{}.
\newblock \showarticletitle{Graph-based Text Representations: Boosting Text
  Mining, NLP and Information Retrieval with Graphs}. In
  \bibinfo{booktitle}{\emph{WWW Tutorials}}.
\newblock


\bibitem[\protect\citeauthoryear{Manning, Raghavan, and Sch{\"{u}}tze}{Manning
  et~al\mbox{.}}{2008}]%
        {DBLP:books/daglib/0021593}
\bibfield{author}{\bibinfo{person}{Christopher~D. Manning},
  \bibinfo{person}{Prabhakar Raghavan}, {and} \bibinfo{person}{Hinrich
  Sch{\"{u}}tze}.} \bibinfo{year}{2008}\natexlab{}.
\newblock \bibinfo{booktitle}{\emph{Introduction to information retrieval}}.
\newblock \bibinfo{publisher}{Cambridge University Press}.
\newblock


\bibitem[\protect\citeauthoryear{Melnik, Garcia{-}Molina, and Rahm}{Melnik
  et~al\mbox{.}}{2002}]%
        {DBLP:conf/icde/MelnikGR02}
\bibfield{author}{\bibinfo{person}{Sergey Melnik}, \bibinfo{person}{Hector
  Garcia{-}Molina}, {and} \bibinfo{person}{Erhard Rahm}.}
  \bibinfo{year}{2002}\natexlab{}.
\newblock \showarticletitle{Similarity Flooding: {A} Versatile Graph Matching
  Algorithm and Its Application to Schema Matching}. In
  \bibinfo{booktitle}{\emph{{ICDE}}}. \bibinfo{pages}{117--128}.
\newblock


\bibitem[\protect\citeauthoryear{Mikolov, Sutskever, Chen, Corrado, and
  Dean}{Mikolov et~al\mbox{.}}{2013}]%
        {DBLP:conf/nips/MikolovSCCD13}
\bibfield{author}{\bibinfo{person}{Tom{\'{a}}s Mikolov}, \bibinfo{person}{Ilya
  Sutskever}, \bibinfo{person}{Kai Chen}, \bibinfo{person}{Gregory~S. Corrado},
  {and} \bibinfo{person}{Jeffrey Dean}.} \bibinfo{year}{2013}\natexlab{}.
\newblock \showarticletitle{Distributed Representations of Words and Phrases
  and their Compositionality}. In \bibinfo{booktitle}{\emph{{NIPS}}}.
  \bibinfo{pages}{3111--3119}.
\newblock


\bibitem[\protect\citeauthoryear{Mudgal, Li, Rekatsinas, Doan, Park, Krishnan,
  Deep, Arcaute, and Raghavendra}{Mudgal et~al\mbox{.}}{2018}]%
        {DBLP:conf/sigmod/MudgalLRDPKDAR18}
\bibfield{author}{\bibinfo{person}{Sidharth Mudgal}, \bibinfo{person}{Han Li},
  \bibinfo{person}{Theodoros Rekatsinas}, \bibinfo{person}{AnHai Doan},
  \bibinfo{person}{Youngchoon Park}, \bibinfo{person}{Ganesh Krishnan},
  \bibinfo{person}{Rohit Deep}, \bibinfo{person}{Esteban Arcaute}, {and}
  \bibinfo{person}{Vijay Raghavendra}.} \bibinfo{year}{2018}\natexlab{}.
\newblock \showarticletitle{Deep Learning for Entity Matching: {A} Design Space
  Exploration}. In \bibinfo{booktitle}{\emph{{SIGMOD}}}.
  \bibinfo{pages}{19--34}.
\newblock


\bibitem[\protect\citeauthoryear{Nemenyi}{Nemenyi}{1963}]%
        {nemenyi1963distribution}
\bibfield{author}{\bibinfo{person}{P. Nemenyi}.}
  \bibinfo{year}{1963}\natexlab{}.
\newblock \bibinfo{booktitle}{\emph{Distribution-free Multiple Comparisons}}.
\newblock \bibinfo{publisher}{Princeton University}.
\newblock


\bibitem[\protect\citeauthoryear{Obraczka, Schuchart, and Rahm}{Obraczka
  et~al\mbox{.}}{2021}]%
        {DBLP:journals/corr/abs-2101-06126}
\bibfield{author}{\bibinfo{person}{Daniel Obraczka}, \bibinfo{person}{Jonathan
  Schuchart}, {and} \bibinfo{person}{Erhard Rahm}.}
  \bibinfo{year}{2021}\natexlab{}.
\newblock \showarticletitle{{EAGER:} Embedding-Assisted Entity Resolution for
  Knowledge Graphs}.
\newblock \bibinfo{journal}{\emph{CoRR}}  \bibinfo{volume}{abs/2101.06126}
  (\bibinfo{year}{2021}).
\newblock


\bibitem[\protect\citeauthoryear{Otto and Reichert}{Otto and Reichert}{2010}]%
        {OttoMdmSurvey10}
\bibfield{author}{\bibinfo{person}{Boris Otto} {and} \bibinfo{person}{Andreas
  Reichert}.} \bibinfo{year}{2010}\natexlab{}.
\newblock \showarticletitle{Organizing Master Data Management: Findings from an
  Expert Survey}. In \bibinfo{booktitle}{\emph{Proceedings of the 2010 ACM
  Symposium on Applied Computing (SAC)}}. \bibinfo{pages}{106–110}.
\newblock
\urldef\tempurl%
\url{https://doi.org/10.1145/1774088.1774111}
\showDOI{\tempurl}


\bibitem[\protect\citeauthoryear{Papadakis, Efthymiou, Thanos, and
  Hassanzadeh}{Papadakis et~al\mbox{.}}{2021}]%
        {paperExtendedVersion}
\bibfield{author}{\bibinfo{person}{George Papadakis}, \bibinfo{person}{Vasilis
  Efthymiou}, \bibinfo{person}{Emanouil Thanos}, {and} \bibinfo{person}{Oktie
  Hassanzadeh}.} \bibinfo{year}{2021}\natexlab{}.
\newblock \bibinfo{title}{Bipartite Graph Matching Algorithms for Entity
  Resolution: An Empirical Evaluation}.
\newblock
\newblock
\showeprint[arxiv]{2112.14030}
\urldef\tempurl%
\url{https://arxiv.org/abs/2112.14030}
\showURL{%
\tempurl}


\bibitem[\protect\citeauthoryear{Papadakis, Giannakopoulos, and
  Paliouras}{Papadakis et~al\mbox{.}}{2016}]%
        {DBLP:journals/www/0001GP16}
\bibfield{author}{\bibinfo{person}{George Papadakis}, \bibinfo{person}{George
  Giannakopoulos}, {and} \bibinfo{person}{Georgios Paliouras}.}
  \bibinfo{year}{2016}\natexlab{}.
\newblock \showarticletitle{Graph vs. bag representation models for the topic
  classification of web documents}.
\newblock \bibinfo{journal}{\emph{World Wide Web}} \bibinfo{volume}{19},
  \bibinfo{number}{5} (\bibinfo{year}{2016}), \bibinfo{pages}{887--920}.
\newblock


\bibitem[\protect\citeauthoryear{Papadakis, Mandilaras, Gagliardelli, Simonini,
  Thanos, Giannakopoulos, Bergamaschi, Palpanas, and Koubarakis}{Papadakis
  et~al\mbox{.}}{2020a}]%
        {DBLP:journals/is/PapadakisMGSTGB20}
\bibfield{author}{\bibinfo{person}{George Papadakis},
  \bibinfo{person}{Georgios~M. Mandilaras}, \bibinfo{person}{Luca
  Gagliardelli}, \bibinfo{person}{Giovanni Simonini},
  \bibinfo{person}{Emmanouil Thanos}, \bibinfo{person}{George Giannakopoulos},
  \bibinfo{person}{Sonia Bergamaschi}, \bibinfo{person}{Themis Palpanas}, {and}
  \bibinfo{person}{Manolis Koubarakis}.} \bibinfo{year}{2020}\natexlab{a}.
\newblock \showarticletitle{Three-dimensional Entity Resolution with JedAI}.
\newblock \bibinfo{journal}{\emph{Inf. Syst.}}  \bibinfo{volume}{93}
  (\bibinfo{year}{2020}), \bibinfo{pages}{101565}.
\newblock


\bibitem[\protect\citeauthoryear{Papadakis, Skoutas, Thanos, and
  Palpanas}{Papadakis et~al\mbox{.}}{2020b}]%
        {DBLP:journals/csur/PapadakisSTP20}
\bibfield{author}{\bibinfo{person}{George Papadakis},
  \bibinfo{person}{Dimitrios Skoutas}, \bibinfo{person}{Emmanouil Thanos},
  {and} \bibinfo{person}{Themis Palpanas}.} \bibinfo{year}{2020}\natexlab{b}.
\newblock \showarticletitle{Blocking and Filtering Techniques for Entity
  Resolution: {A} Survey}.
\newblock \bibinfo{journal}{\emph{{ACM} Comput. Surv.}} \bibinfo{volume}{53},
  \bibinfo{number}{2} (\bibinfo{year}{2020}), \bibinfo{pages}{31:1--31:42}.
\newblock
\urldef\tempurl%
\url{https://doi.org/10.1145/3377455}
\showDOI{\tempurl}


\bibitem[\protect\citeauthoryear{Pennington, Socher, and Manning}{Pennington
  et~al\mbox{.}}{2014}]%
        {DBLP:conf/emnlp/PenningtonSM14}
\bibfield{author}{\bibinfo{person}{Jeffrey Pennington},
  \bibinfo{person}{Richard Socher}, {and} \bibinfo{person}{Christopher~D.
  Manning}.} \bibinfo{year}{2014}\natexlab{}.
\newblock \showarticletitle{Glove: Global Vectors for Word Representation}. In
  \bibinfo{booktitle}{\emph{{EMNLP}}}. \bibinfo{pages}{1532--1543}.
\newblock


\bibitem[\protect\citeauthoryear{Rousseau and Vazirgiannis}{Rousseau and
  Vazirgiannis}{2013}]%
        {DBLP:conf/cikm/RousseauV13}
\bibfield{author}{\bibinfo{person}{Fran{\c{c}}ois Rousseau} {and}
  \bibinfo{person}{Michalis Vazirgiannis}.} \bibinfo{year}{2013}\natexlab{}.
\newblock \showarticletitle{Graph-of-word and {TW-IDF:} new approach to ad hoc
  {IR}}. In \bibinfo{booktitle}{\emph{CIKM}}. \bibinfo{pages}{59--68}.
\newblock


\bibitem[\protect\citeauthoryear{Saeedi, Nentwig, Peukert, and Rahm}{Saeedi
  et~al\mbox{.}}{2018a}]%
        {DBLP:journals/csimq/SaeediNPR18}
\bibfield{author}{\bibinfo{person}{Alieh Saeedi}, \bibinfo{person}{Markus
  Nentwig}, \bibinfo{person}{Eric Peukert}, {and} \bibinfo{person}{Erhard
  Rahm}.} \bibinfo{year}{2018}\natexlab{a}.
\newblock \showarticletitle{Scalable Matching and Clustering of Entities with
  {FAMER}}.
\newblock \bibinfo{journal}{\emph{Complex Syst. Informatics Model. Q.}}
  \bibinfo{volume}{16} (\bibinfo{year}{2018}), \bibinfo{pages}{61--83}.
\newblock


\bibitem[\protect\citeauthoryear{Saeedi, Peukert, and Rahm}{Saeedi
  et~al\mbox{.}}{2018b}]%
        {DBLP:conf/esws/SaeediPR18}
\bibfield{author}{\bibinfo{person}{Alieh Saeedi}, \bibinfo{person}{Eric
  Peukert}, {and} \bibinfo{person}{Erhard Rahm}.}
  \bibinfo{year}{2018}\natexlab{b}.
\newblock \showarticletitle{Using Link Features for Entity Clustering in
  Knowledge Graphs}. In \bibinfo{booktitle}{\emph{{ESWC}}}
  \emph{(\bibinfo{series}{Lecture Notes in Computer Science})},
  Vol.~\bibinfo{volume}{10843}. \bibinfo{publisher}{Springer},
  \bibinfo{pages}{576--592}.
\newblock


\bibitem[\protect\citeauthoryear{Schwartz, Steger, and Wei{\ss}l}{Schwartz
  et~al\mbox{.}}{2005}]%
        {DBLP:conf/wea/SchwartzSW05}
\bibfield{author}{\bibinfo{person}{Justus Schwartz}, \bibinfo{person}{Angelika
  Steger}, {and} \bibinfo{person}{Andreas Wei{\ss}l}.}
  \bibinfo{year}{2005}\natexlab{}.
\newblock \showarticletitle{Fast Algorithms for Weighted Bipartite Matching}.
  In \bibinfo{booktitle}{\emph{{WEA}}} \emph{(\bibinfo{series}{Lecture Notes in
  Computer Science})}, Vol.~\bibinfo{volume}{3503}. \bibinfo{pages}{476--487}.
\newblock


\bibitem[\protect\citeauthoryear{Wang, Tong, Long, Xu, Xu, and Lv}{Wang
  et~al\mbox{.}}{2019}]%
        {DBLP:conf/icde/WangTLXXL19}
\bibfield{author}{\bibinfo{person}{Yansheng Wang}, \bibinfo{person}{Yongxin
  Tong}, \bibinfo{person}{Cheng Long}, \bibinfo{person}{Pan Xu},
  \bibinfo{person}{Ke Xu}, {and} \bibinfo{person}{Weifeng Lv}.}
  \bibinfo{year}{2019}\natexlab{}.
\newblock \showarticletitle{Adaptive Dynamic Bipartite Graph Matching: {A}
  Reinforcement Learning Approach}. In \bibinfo{booktitle}{\emph{{ICDE}}}.
  \bibinfo{pages}{1478--1489}.
\newblock


\bibitem[\protect\citeauthoryear{Wang, Sisman, Wei, Dong, and Ji}{Wang
  et~al\mbox{.}}{2020}]%
        {DBLP:conf/icdm/WangSWDJ20}
\bibfield{author}{\bibinfo{person}{Zhengyang Wang}, \bibinfo{person}{Bunyamin
  Sisman}, \bibinfo{person}{Hao Wei}, \bibinfo{person}{Xin~Luna Dong}, {and}
  \bibinfo{person}{Shuiwang Ji}.} \bibinfo{year}{2020}\natexlab{}.
\newblock \showarticletitle{CorDEL: {A} Contrastive Deep Learning Approach for
  Entity Linkage}. In \bibinfo{booktitle}{\emph{{ICDM}}}.
\newblock


\bibitem[\protect\citeauthoryear{Watkins and Dayan}{Watkins and Dayan}{1992}]%
        {DBLP:journals/ml/WatkinsD92}
\bibfield{author}{\bibinfo{person}{Christopher J. C.~H. Watkins} {and}
  \bibinfo{person}{Peter Dayan}.} \bibinfo{year}{1992}\natexlab{}.
\newblock \showarticletitle{Technical Note Q-Learning}.
\newblock \bibinfo{journal}{\emph{Mach. Learn.}}  \bibinfo{volume}{8}
  (\bibinfo{year}{1992}), \bibinfo{pages}{279--292}.
\newblock


\bibitem[\protect\citeauthoryear{Wijaya and Bressan}{Wijaya and
  Bressan}{2009}]%
        {wijaya2009ricochet}
\bibfield{author}{\bibinfo{person}{Derry~Tanti Wijaya} {and}
  \bibinfo{person}{St{\'e}phane Bressan}.} \bibinfo{year}{2009}\natexlab{}.
\newblock \showarticletitle{Ricochet: A family of unconstrained algorithms for
  graph clustering}. In \bibinfo{booktitle}{\emph{International Conference on
  Database Systems for Advanced Applications}}. Springer,
  \bibinfo{pages}{153--167}.
\newblock


\bibitem[\protect\citeauthoryear{Winkler}{Winkler}{2006}]%
        {Winkler06}
\bibfield{author}{\bibinfo{person}{W.~E. Winkler}.}
  \bibinfo{year}{2006}\natexlab{}.
\newblock \bibinfo{booktitle}{\emph{Overview of Record Linkage and Current
  Research Directions}}.
\newblock \bibinfo{type}{{T}echnical {R}eport}. \bibinfo{institution}{Bureau of
  the Census}.
\newblock


\bibitem[\protect\citeauthoryear{Wu, Chaba, Sawlani, Chu, and
  Thirumuruganathan}{Wu et~al\mbox{.}}{2020}]%
        {DBLP:conf/sigmod/WuCSCT20}
\bibfield{author}{\bibinfo{person}{Renzhi Wu}, \bibinfo{person}{Sanya Chaba},
  \bibinfo{person}{Saurabh Sawlani}, \bibinfo{person}{Xu Chu}, {and}
  \bibinfo{person}{Saravanan Thirumuruganathan}.}
  \bibinfo{year}{2020}\natexlab{}.
\newblock \showarticletitle{ZeroER: Entity Resolution using Zero Labeled
  Examples}. In \bibinfo{booktitle}{\emph{{SIGMOD}}}.
  \bibinfo{pages}{1149--1164}.
\newblock


\end{thebibliography}

\pagebreak
\section*{Appendix}

\begin{appendix}\label{appendix}

\section{Algorithms}
\label{app:algorithms}

\vspace{4pt}
\noindent
\textbf{Ricochet Sequential Rippling Clustering (\textsf{RSR}).}
This algorithm, outlined in Algorithm \ref{algo:ricochetSRCC},  is an adaptation of the homonymous 
method for Dirty ER in \cite{DBLP:journals/pvldb/HassanzadehCML09} such that it exclusively considers partitions with just one entity from each input dataset. Initially, \textsf{RSR} sorts all nodes from both input datasets in descending order of the average weight of their adjacent edges (Line 7). Whenever a new seed is chosen from the sorted list, we consider all its adjacent edges with a weight higher than $t$ (Lines 8-11). The first adjacent vertex that is currently unassigned or is closer to the new seed than it is to the seed of its current partition is re-assigned to the new partition (Lines 14-16). If a partition is reduced to a singleton after a re-assignment, either because the chosen vertex (Line 17) or the seed (Line 24) was previously in it, it is placed in its nearest single-node partition (Lines 30-39). 

The algorithm stops when all nodes have been considered. In the worst case the algorithm has to iterate through $n$ vertices and each time reassign $n$ vertices to their most similar adjacent vertex, therefore its time complexity is $O(n \; m)$~\cite{wijaya2009ricochet}. 

\begin{algorithm2e}[tbh]
\DontPrintSemicolon
\small
\KwIn{Similarity Graph $G = (V_1, V_2, E)$, similarity threshold $t$}
\KwOut{A set of Partitions $C = \{c_1, c_2, \ldots, c_n\}$}
$C\gets \emptyset$\\
$Center \gets \emptyset$\\
\ForEach(\tcp*[h]{Initialization}){$v \in (V_1 \cup V_2)$}{
$simWithCenter(v) \gets 0$\\
$Partition(v) \gets \emptyset$\\
$centerOf(v) \gets v$
}
$Q \gets G.nodesInDecWeight(v, w(v))$ \tcp*{$w(v)=\sum_{e\in adj(v)}{e.sim}/|adj(v)|$}
\While{$Q \neq \emptyset$}{
    $v_i \gets Q.pop()$ \tcp*{the vertex with highest weight}
    $ToReassign \gets \emptyset$\\
    \ForEach(\tcp*[h]{for $v_i$'s adjacent edges}){
    $e = (v_i, v_j, sim) \in E :sim>t$}{
    \If{$v_j \in Center$}{
            \textbf{continue}\\
}
\If{$e.sim > simWithCenter(v_j)$}{
$Partition(centerOf(v_j)).remove(v_j)$\tcp*{remove $v_j$ from its previous partition}
$Partition(v_i) \gets Partition(v_i) \cup v_j$\\
$ToReassign \gets ToReassign \cup centerOf(v_j)$\tcp*{it is now a singleton}
$simWithCenter(v_j) \gets e.sim$\\
$centerOf(v_j) \gets v_i$\\
            \textbf{break}\\
}
}
    \If{$|Partition(v_i)| > 0$}{
\If(\tcp*{if $v_i$ was previously in another partition}){$centerOf(v_i) \neq v_i$}
{
 $Partition(centerOf(v_i)).remove(v_i)$\\
 $ToReassign \gets ToReassign \cup centerOf(v_i)$\\
}
$Center \gets Center \cup v_i$\\
            $Partition(v_i) \gets Partition(v_i) \cup v_i$\tcp*{put $v_i$ in its partition}
 $centerOf(v_i) \gets v_i$\\
 $simWithCenter(v_i) \gets 1$\\
}
\ForEach{$v_k \in ToReassign$}
{
$maxSim \gets 0$\\
$cMax \gets v_k$\\
\ForEach(\tcp*[h]{find singleton with the highest similarity with $v_k$ to reassign it}){$e = (v_k, v_{\ell}, sim) \in E :sim>t$}{
\If{$e.sim>maxSim$ \textbf{and} $|Partition(v_{\ell})|<2$}
{
$cMax \gets v_{\ell}$\\
$maxSim \gets e.sim$\\
}
}
\If{$maxSim>0$}
{
$Partition(v_k) \gets \emptyset$\\
$Partition(cMax) \gets Partition(cMax) \cup v_k$\\
}

}
}
\ForEach{$v_i \in (V_1 \cup V_2)$}{
\If{$|Partition(v_i)|>0$}
{
$C \gets C \cup  Partition(v_i)$
}
}
\textbf{return} $C$
\caption{Ricochet SR Clustering (\textsf{RSR})}
\label{algo:ricochetSRCC}
\end{algorithm2e}

\vspace{4pt}
\noindent
\textbf{Connected Components (\textsf{CNC}).}
This is the simplest algorithm for bipartite graph matching and is outlined in Algorithm \ref{algo:cccer}. First, it discards all edges with a weight lower than the given similarity threshold (Lines 1-3). Then, it computes the transitive closure of the pruned similarity graph (Line 4). In the output, it solely retains the partitions that contain two entities -- one from each input dataset (Lines 6-8). Using a simple depth-first approach, its time complexity is $O(n+m) \sim O(m)$, given that $m \gg n$  \cite{DBLP:books/daglib/0017733}.

\begin{algorithm2e}[tbh]
\DontPrintSemicolon
\small
\KwIn{Similarity Graph $G = (V_1, V_2, E)$, similarity threshold $t$}
\KwOut{A set of clusters $C = \{c_1, c_2, \ldots, c_n\}$}

\ForEach{$e = (v_i, v_j, sim) \in E$}{
    \If{$sim < t$}{
        $E \leftarrow E - (e)$\;
    }
}

$C_1 \leftarrow transitiveClosure(G)$\;

$C \gets \emptyset$\\
\ForEach{$c_i \in C_1$}{
    \If{$|c_i| == 2 \wedge c_i \cap V1 \neq \emptyset \wedge c_i \cap V2 \neq \emptyset$}{
        $C \gets C \cup \{c_i\}$ \\
    }
}
\textbf{return} $C$
\caption{Connected Components (\textsf{CNC})}
\label{algo:cccer}
\end{algorithm2e}

\vspace{4pt}
\noindent
\textbf{Row Column Assignment Clustering (\textsf{RCA}).}
This approach, outlined in Algorithm \ref{algo:row_column}, is based on the Row-Column Scan approximation method in
\cite{kurtzberg1962approximation} that solves the assignment problem. It requires two passes of the similarity graph, with each pass generating a candidate solution.
In the first pass, each entity from the source dataset creates a new partition, to which the most similar, currently unassigned entity from the target dataset is assigned (Lines 7-17). Note that, in principle, any pair of entities can be assigned to the same partition at this step even if their similarity is lower than $t$, since the assignment problem assumes that each vertex from $V1$ is connected to all vertices from $V2$ (any "job" can be performed by all "men"). 
In the second pass, the same procedure is applied to the entities/nodes of the target dataset (Lines 18-28). The value of each solution is the sum of the edge weights between the nodes assigned to the same (2-node) partition (Lines 17,28). The solution with the highest value is returned as output, after discarding the pairs with similarity less than $t$ (Lines 29-36).

At each pass the algorithm iterates over all nodes/entities of one of the entity collection searching for the node/entity with maximum similarity from the other entity collection. Therefore, its time complexity is $O(|V1|\;|V2| )$. 

\begin{algorithm2e}[t]
\DontPrintSemicolon
\small
\KwIn{Similarity Graph $G = (V_1, V_2, E)$, similarity threshold $t$}
\KwOut{A set of partitions $C = \{c_1, c_2, \ldots, c_n\}$}
$C_1\gets \emptyset$\\
$C_2 \gets \emptyset$\\
$M_1 \gets \emptyset$ \tcp*{matched nodes from $V_1$}
$M_2 \gets \emptyset$ \tcp*{matched nodes from $V_2$}
$D_1 \gets 0$ \tcp*{assignment value of $C_1$}
$D_2 \gets 0$ \tcp*{assignment value of $C_2$}
\ForEach{$v_i \in V_1$}{
    $c_i \gets \{v_i\}$ \tcp*{create a new partition containing $v_i$}
    $Q_i \gets V_2(sim(v_i))$ \tcp*{a priority queue of $V_2$'s nodes in decreasing sim with $v_i$}
    \While{$Q_i \neq \emptyset$}{
        $v_2 \gets Q_i.pop()$ \\
        \If(\tcp*[h]{if $v_2$ is not yet matched}){$v_2 \notin M_2$}{
            $c_i \gets c_i \cup \{v_2\}$ \tcp*{add $v_2$ to partition $c_i$}
            $M_2 \gets M_2 \cup \{v_2\}$ \\
            $D_1 \gets D_1 + \{sim(v_1,v_2)\}$ \\
            \Break
        }
    }
    $C_1 \gets C_1 \cup \{c_i\}$ \\
}
\ForEach{$v_j \in V_2$}{
    $c_j \gets \{v_j\}$ \tcp*{create a new partition containing $v_j$}
$Q_j \gets V_1(sim(v_j))$ \tcp*{a priority queue of $V_1$'s nodes in decreasing sim with $v_j$}    \While{$Q_j \neq \emptyset$}{
        $v_1 \gets Q_j.pop()$ \\
        \If(\tcp*[h]{if $v_1$ is not yet matched}){$v_1 \notin M_1$}{
            $c_j \gets c_j \cup \{v_1\}$ \tcp*{add $v_1$ to partition $c_j$}
            $M_1 \gets M_1 \cup \{v_1\}$ \\
            $D_2 \gets D_2 + \{sim(v_1,v_2)\}$ \\
\Break
        }
    }
    $C_2 \gets C_2 \cup \{c_j\}$ \\
}
\If(\tcp*[h]{get maximal assignment}){$D_1>D_2$}{
$C=C_1$\\
}
\Else {
$C=C_2$\\
}
\ForEach{$c=\{v_i,v_j\} \in C$}{
\If(\tcp*[h]{check similarities}){$sim(v_i,v_j)<t$}{
$C=C\setminus c$ \tcp{remove partition pairs with similarity less than $t$ }
}
}
\textbf{return} $C$
\caption{Row Column Clustering (\textsf{RCA})}
\label{algo:row_column}
\end{algorithm2e}

\vspace{4pt}
\noindent
\textbf{Best Assignment Heuristic (\textsf{BAH}).}
This algorithm applies a simple swap-based random-search algorithm to heuristically solve the Maximum Weight Bipartite Matching problem and uses the resulting solution to create the output partitions.
Its functionality is outlined in Algorithm \ref{algo:best_assignment}.
Initially, each entity from the smaller input dataset is connected to an entity from the larger input dataset (Line 9). In each iteration of the search process (Line 10), two entities from the larger dataset are randomly selected (Lines 12-13) in order to swap their current connections. 
If the sum of the edge weights of the new pairs is higher than the previous pairs (Line 15-19), the swap is accepted (Lines 20-24). The algorithm stops when a maximum number of search steps is reached or when a maximum run-time has been exceeded. In our case, the run-time has been set to 2 minutes.


\begin{algorithm2e}[tbh]
\DontPrintSemicolon
\small
\KwIn{Similarity Graph $G = (V_1, V_2, E): |V_1|>|V_2|$, similarity threshold $t$, max number of moves $maxNumMoves$}
\KwOut{A set of partitions $C = \{c_1, c_2, \ldots, c_n\}$}
$C\gets \emptyset$\\
$numMoves \gets 0$ \\
\ForEach{$(v^1_i, v^2_j) \in (V_1 \times V_2)$ }{
$d(v^1_i, v^2_j) \gets 0$ \tcp*{initialize pair contributions}
} 
\ForEach{$e = (v^1_i, v^2_j, sim) \in E$, with $e.sim$ > $t$}{
    $d(v^1_i, v^2_j) \gets sim$ \tcp*{initialize pair contributions}
}
\ForEach{$v^1_i \in V_1,  v^2_i \in V_2: i \leq |V_2|$}{
$c_i \gets \{v^1_i,v^2_i\}$ \tcp*{initialize partitions}
$p(v^1_i)=v^2_i$\\
} 

\While{$numMoves<maxNumMoves$}{
    $numMoves \gets numMoves + 1$\\
    $i=nextRand(|V_1|)$\\
    $j=nextRand(|V_1|):j\neq i$\\
    $D \gets 0 $\\
    \If(\tcp*[h]{check swaps}){$p(v^1_i) \neq null$} {
    $D \gets d(v^1_j,p(v^1_i))-d(v^1_i,p(v^1_i))$
    }
    \If(\tcp*[h]{check swaps}){$p(v^1_j) \neq null$} {
    $D \gets D+d(v^1_i,p(v^1_j))-d(v^1_j,p(v^1_j))$
    }
    \If(\tcp*[h]{if swaps increase assignment value}){$D \geq 0$}{
            $temp \gets p(v^1_j)$ \tcp*{perform swaps}
            $p(v^1_j) \gets p(v^1_i)$ \\
            $p(v^1_i) \gets temp$ \\
            $c_i \gets \{v^1_i,p(v^1_i)\}$ \\
            $c_j \gets \{v^1_j,p(v^1_j)\}$ \\
    }
}

\textbf{return} $C$
\caption{Best Assignment Heuristic (\textsf{BAH})}
\label{algo:best_assignment}
\end{algorithm2e}

\vspace{4pt}
\noindent
\textbf{Best Match Clustering (\textsf{BMC}).}
This algorithm is inspired from the Best Match strategy of~\cite{DBLP:conf/icde/MelnikGR02}, which solves the Stable Marriage problem~\cite{DBLP:journals/tamm/GaleS62}, as simplified in BigMat~\cite{DBLP:conf/bigdataconf/0002MD19}. 
Its functionality is outlined in Algorithm \ref{algo:best_match}. For each entity 
of the one dataset, this algorithm creates a new partition (Lines 4-5), 
in which the most similar, not-yet-clustered entity from the other dataset is also placed - provided that the corresponding edge weight is higher than $t$ (Lines 6-12). 
Note that the greedy heuristic for \textsf{BMC}
introduced in~\cite{DBLP:conf/icde/MelnikGR02} is the same, in principle, to Unique Mapping Clustering (see below). Note also that \textsf{BMC} is the only algorithm with an additional configuration parameter, apart from the similarity threshold: the input dataset that is used as the basis for creating partitions can be set to the source or the target dataset. In our experiments, we examine both options and retain the best one.

The algorithm iterates over the nodes of one of the datasets searching for its adjacent vertex with maximum similarity, therefore its time complexity is $O(m)$.

\begin{algorithm2e}[tbh]
\DontPrintSemicolon
\small
\KwIn{Similarity Graph $G = (V_1, V_2, E)$, similarity threshold $t$}
\KwOut{A set of clusters $C = \{c_1, c_2, \ldots, c_n\}$}
$C \gets \emptyset$\\
$M_2 \gets \emptyset$ \tcp*{matched nodes from $V_2$}

\ForEach{$v_i \in V_1$}{
    $c_i \gets \{v_i\}$ \tcp*{create a new cluster containing $v_i$}
    $Q_i \gets v_i.edgesDecOrder(t)$ \tcp*{edges in desc. sim > $t$}
    \While{$Q_i \neq \emptyset$}{
        $e \gets Q_i.pop()$ \\
        \If(\tcp*[h]{if $v_2$ is not yet matched}){$e.v_2 \notin M_2$}{
            $c_i \gets c_i \cup \{e.v_2\}$ \tcp*{add $v_2$ to cluster $c_i$}
            $M_2 \gets M_2 \cup \{e.v_2\}$ \\
            \Break
        }
    }
    $C \gets C \cup \{c_i\}$ \\
}
\textbf{return} $C$
\caption{Best Match Clustering (\textsf{BMC})}
\label{algo:best_match}
\end{algorithm2e}

\vspace{4pt}
\noindent
\textbf{Exact Clustering (\textsf{EXC}).}
This algorithm is inspired from the Exact strategy of~\cite{DBLP:conf/icde/MelnikGR02}. 
Its functionality is outlined in Algorithm \ref{algo:exact}.
Initially, it creates an empty priority queue for every vertex (Lines 2-5). Then, it populates the queue of every vertex $v_i$ with all its adjacent edges that exceed the given similarity threshold $t$, sorting them in decreasing weight (Lines 6-8). Subsequently,
\textsf{EXC} places two entities in the same partition (Lines 9-16) 
only if they are mutually the best matches, i.e., 
the most similar candidates of each other (Line 14).
This approach is basically a stricter, symmetric version of \textsf{BMC} and could also be conceived as a strict version of the reciprocity filter that was employed in~\cite{DBLP:conf/edbt/Efthymiou0SC19}.

Its time complexity 
is $O(n \; m)$, since the algorithm iterates over each vertex of one dataset searching for its adjacent vertex with maximum similarity and then performs the same search for the latter vertex. 

\vspace{4pt}
\noindent
\textbf{Unique Mapping Clustering (\textsf{UMC}).}
This algorithm is outlined in Algorithm \ref{algo:unique_mapping}. Initially, it iteraters over all edges and those with a weight higher than $t$ are placed in a priority queue that sorts them
in decreasing weight/similarity (Lines 5-7).
Subsequently, it iteratively forms a partition (Line 11) for the top-weighted pair (Line 9), 
as long as none of its entities has already been matched to some other (Line 10). 
This approach relies on the \textit{unique mapping constraint} of CCER, i.e., the restriction that 
each entity from the one input dataset
matches at most one entity from the other.
Note that the \textit{CLIP Clustering algorithm}, introduced for the multi-source 
ER problem in \cite{DBLP:conf/esws/SaeediPR18}, is equivalent to \textsf{UMC} when there are only two input datasets (i.e., in the CCER case that we study). 

Its time complexity 
is $O(m \; logm)$, due to the cost that is required for sorting all edges.

\vspace{4pt}
\noindent
\textbf{Kir\'{a}ly's Clustering (\textsf{KRC}).}
This algorithm is an adaptation of the linear time 3/2 approximation to the Maximum Stable Marriage problem, called ``New Algorithm'' in~\cite{DBLP:journals/algorithms/Kiraly13}. 
Intuitively, the entities of the source dataset (``men''~\cite{DBLP:journals/algorithms/Kiraly13}) propose to the entities (Line 16) 
from the target dataset with an edge weight higher than $t$ (``women''~\cite{DBLP:journals/algorithms/Kiraly13}) to form a partition (``get engaged''~\cite{DBLP:journals/algorithms/Kiraly13}). 
Its functionality is outlined in Algorithm \ref{algo:kiraly}. The entities of the target dataset accept a proposal under certain conditions (e.g., if it's the first proposal they receive - Line 17), 
and the partitions and preferences are updated accordingly (Lines 18, 22, 24). 
Entities from the source dataset get a second chance to make proposals (Lines 5, 27-30) 
and the algorithm terminates when all entities of the first dataset are in a partition (Line 13), 
or no more proposal chances are left (Line 27). 
We omit some of the details (e.g., the rare case of ``uncertain man''), due to space restrictions, and refer the reader to~\cite{,DBLP:journals/algorithms/Kiraly13} for more information (e.g., the acceptance criteria for proposals).
Its time complexity 
is $O(n + m \; logm)$ \cite{DBLP:journals/algorithms/Kiraly13}.


\begin{algorithm2e}[tbh]
\DontPrintSemicolon
\small
\KwIn{Similarity Graph $G = (V_1, V_2, E)$, similarity threshold $t$}
\KwOut{A set of clusters $C = \{c_1, c_2, \ldots, c_n\}$}
$C \gets \emptyset$\\

\ForEach{$v_i \in |V_1|$}{
    $Q1_i \gets \emptyset$ \tcp*{initialize a PQ in desc. sim}
}
\ForEach{$v_j \in |V_2|$}{
    $Q2_j \gets \emptyset$ \tcp*{initialize a PQ in desc. sim}
}

\ForEach{$e = (v_i, v_j, sim) \in E$, with $e.sim$ > $t$}{
    $Q1_i.push(e)$ \\
    $Q2_j.push(e)$ \\
}

\ForEach{$v_i \in V_1$}{
    $c_i \gets \{v_i\}$ \tcp*{create a new cluster containing $v_i$}
    $e \gets Q1_i.pop()$ \tcp*{the best edge for $v_i$}
    $v_j \gets e.v_j$ \tcp*{the best match for $v_i$ is $v_j$}
    $e_2 \gets Q2_j.pop()$ \tcp*{the best edge for $v_j$}
    
    \If(\tcp*[h]{if the best match for $v_j$ is $v_i$}){$e_2.v_i = v_i$}{
        $c_i \gets c_i \cup \{v_j\}$ \tcp*{add $v_j$ to cluster $c_i$}
        $C \gets C \cup \{c_i\}$ \\
    }
}
\textbf{return} $C$
\caption{Exact Clustering (\textsf{EXC})}
\label{algo:exact}
\end{algorithm2e}



\begin{algorithm2e}[t]
\DontPrintSemicolon
\small
\KwIn{Similarity Graph $G = (V_1, V_2, E)$, similarity threshold $t$}
\KwOut{A set of clusters $C = \{c_1, c_2, \ldots, c_n\}$}

$C \gets \emptyset$ \\
$freeM \gets \emptyset$ \tcp*{an initially empty linked list}

\ForEach(\tcp*[h]{$V_1$ corresponds to men in~\cite{DBLP:journals/algorithms/Kiraly13}}){$v_i \in V_1$}{
    $Q1_i \gets \emptyset$ \tcp*{$v_i$'s edges in desc. sim > $t$}
    $lastChance[i] \gets false$ \\
    $freeM.addLast(v_i)$ \tcp*{keeps insertion order}
}
\ForEach(\tcp*[h]{$V_2$ corresponds to women in~\cite{DBLP:journals/algorithms/Kiraly13}}){$v_j \in V_2$}{
    $Q2_j \gets \emptyset$ \tcp*{$v_j$'s edges in desc. sim > $t$}
    $fianc\acute{e}[j] \gets null$
}

\ForEach{$e = (v_i, v_j, sim) \in E$, with $e.sim$ > $t$}{
    $Q1_i.push(e)$ \\
    $Q2_j.push(e)$
}

\While{$freeM \neq \emptyset$}{
    $v_i \gets freeM.removeFirst()$ \tcp*{in insertion order}
    \uIf{$Q1_i \neq \emptyset$}{
         $v_j \gets Q1_i.pop()$ \tcp*{$v_i$'s preference is $v_j$}
        \uIf(\tcp*[h]{$v_j$ is free}){$fianc\acute{e}[j] = null$}{
            $C \gets C \cup \{\{v_i, v_j\}\}$ \tcp*{match $v_i$ to $v_j$}
        }
        \uElse{
            $v_i' \gets fianc\acute{e}[j]$\tcp*{$v_j$ was engaged to $v_i'$}
            \uIf(\tcp*[h]{refer to~\cite{DBLP:journals/algorithms/Kiraly13}}){acceptsProposal($v_j$, $v_i$)}{
                $C \gets C \setminus \{\{v_i', v_j\}\}$ \tcp*{$v_i'$ and $v_j$ break up} 
                $freeM.addLast(v_i')$ \tcp*{$v_i'$ is free again}
                $C \gets C \cup \{\{v_i, v_j\}\}$ \tcp*{match $v_i$ to $v_j$}
                $fianc\acute{e}[j] \gets v_i$ \tcp*{$v_j$ gets engaged to $v_i$}
            }
        }
    }
    \uElse{
        \If{$lastChance(v_i) = false$}{
            $lastChance[i] \gets true$ \tcp*{$2^{nd}$ chance for $v_i$}
            $Q1_i \gets recoverInitialQueue(v_i)$ \\
            $freeM.addLast(v_i)$ 
         }
    }
}


\textbf{return} $C$
\caption{Kir\'{a}ly's Clustering}
\label{algo:kiraly}
\end{algorithm2e}
\begin{algorithm2e}[t]
\DontPrintSemicolon
\small
\KwIn{Similarity Graph $G = (V_1, V_2, E)$, similarity threshold $t$}
\KwOut{A set of clusters $C = \{c_1, c_2, \ldots, c_n\}$}
$C \gets \emptyset$\\
$M_1 \gets \emptyset$ \tcp*{matched nodes from $V_1$}
$M_2 \gets \emptyset$ \tcp*{matched nodes from $V_2$}
$Q \gets \emptyset$ \\
\ForEach{$e = (v_i, v_j, sim) \in E$}{
    \If{e.sim > t}{
        $Q.put(e)$  \tcp*{a PQ of edges in desc. sim > $t$}
    }
}
\While{$Q \neq \emptyset$}{
    $e \gets Q.pop()$ \tcp*{the entity pair with highest sim}
    \If(\tcp*[h]{$v_i$, $v_j$ not matched}){$e.v_i \notin M_1$ \textbf{and} $e.v_j \notin M_2$}{
        $C \gets C \cup \{\{e.v_i, e.v_j\}\}$ \\
        $M_1 \gets M_1 \cup \{e.v_i\}$ \\
        $M_2 \gets M_2 \cup \{e.v_j\}$ \\
    }
}
\textbf{return} $C$
\caption{Unique Mapping Clustering}
\label{algo:unique_mapping}
\end{algorithm2e}

\section{Similarity Functions}
\label{app:similarity}

\begin{figure*}[th!]
\centering
\includegraphics[width=0.95\textwidth]{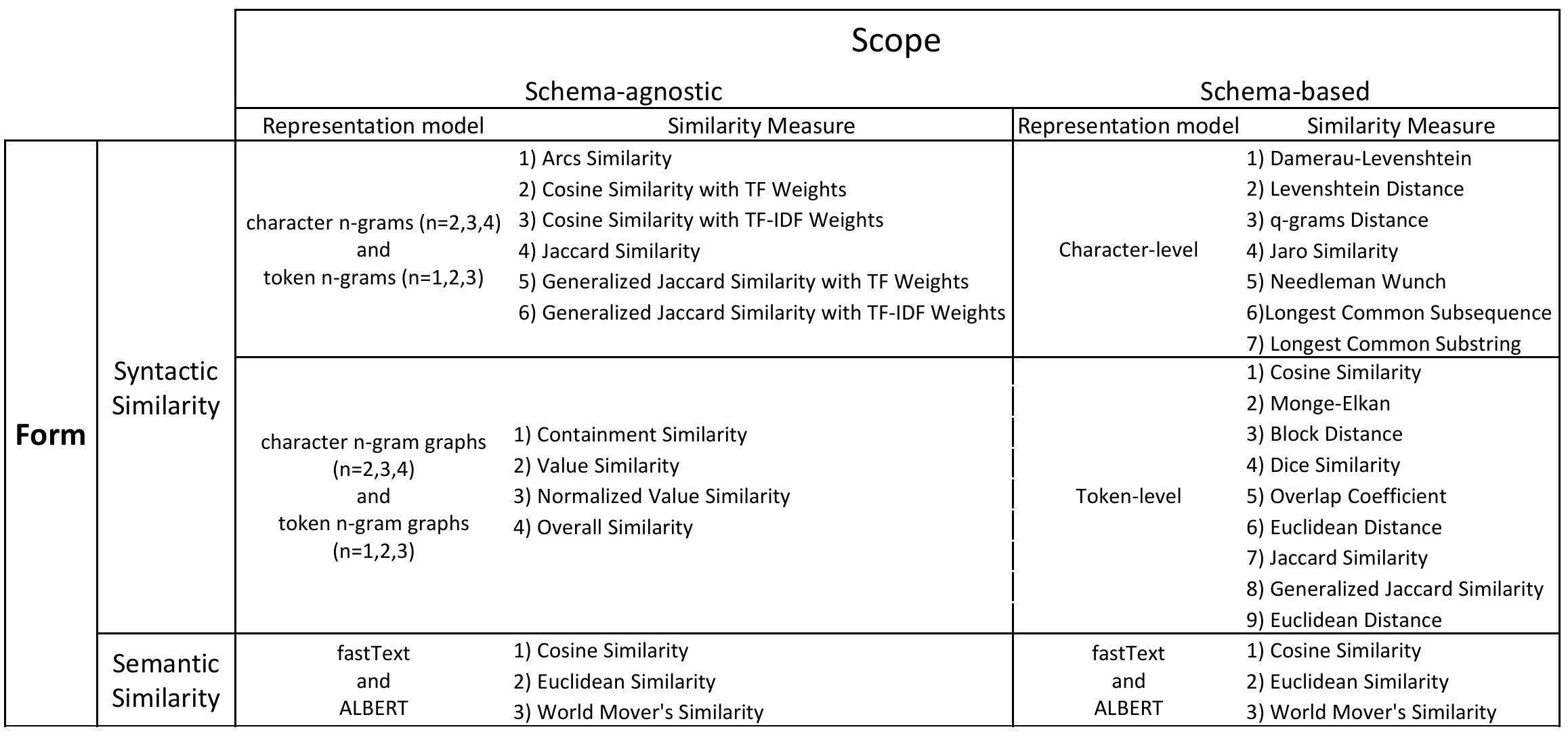}
\vspace{-10pt}
\caption{Taxonomy of the similarity functions we used to generate the similarity graphs. We use $n \in \{2,3,4\}$ for character and $n \in \{1, 2, 3\}$ for token n-grams for both vector and graph models, as in \cite{DBLP:journals/www/0001GP16}. The graph similarities are defined in \cite{DBLP:journals/tslp/GiannakopoulosKVS08}. }
\vspace{-10pt}
\label{fig:einputTaxonomy}
\end{figure*}

In this section, we provide more details about the similarity measures mentioned in Figure~\ref{fig:einputTaxonomy}.

\begin{enumerate}
    \item For the schema-based syntactic representations, which involve short textual values, we considered 16 established similarity measures: 
    Cosine Similarity, Block Distance, Levenshtein Distance, Damerau-Levenshtein Distance, Euclidean Distance, Jaccard Similarity, Generalized Jaccard Similarity, Dice Similarity, Overlap Coefficient, Jaro Similarity, Longest Common Subsequence, Longest Common Substring, Monge-Elkan Similarity, Needleman-Wunch, q-grams Distance, and Simon White Similarity.
    
    \item For the n-gram vectors, we used six similarity measures: Cosine and Generalized Jaccard Similarity with both TF and TF-IDF weights, Enhanced Jaccard Similarity with TF weights and ARCS Similarity with TF-IDF weights.
    
    \item For the n-gram graphs, we used four graph similarity measures: Containment, Value, Normalized Value and Overall Similarity (i.e., the average of the three measures) \cite{DBLP:journals/tslp/GiannakopoulosKVS08}. 
    
    \item For the semantic models, we consider Cosine Similarity, Euclidean Similarity (=1/(1+Euclidean distance)) and World Mover's Similarity (=1/(1+World Mover's Distance)).
\end{enumerate}

Each category of similarity functions is described in more detail in the following.

\subsection{Schema-based syntactic functions}

For this category, we use the following similarity and distance measures, as defined and implemented in Simmetrics.

\subsubsection{Character-level measures.} The following similarity measures are applied to two strings $s_1$ and $s_2$ at character level.

\textit{Levenshtein Distance:} 
Counts the (minimum) number of insert, delete and substitute operations required to transform one string into the other. 

\textit{Damerau-Levenshtein Distance:} Demerau-Levenshtein Distance only differs to Levenshtein Distance by including transpositions among the operations allowed.

\textit{Jaro Similarity:} The Jaro Similarity of two strings $s_1$ and $s_2$ is given by the formula: 
$$
similarity(s_1, s_2) = 
\begin{cases} 
      0 & \text{, if } m=0 \\
      \frac{1}{3}\left(\frac{m}{|s_1|} + \frac{m}{|s_2|} + \frac{m-t}{m}\right) & \text{, else,}
   \end{cases}
$$
where $m$ is the number of common characters, and $t$ is the number of transpositions.

\textit{Needleman-Wunch:} This similarity measure is the result of applying an algorithm that assigns three scores (seen as parameters) to two sequences of characters $s_1$ and $s_2$, depending on whether aligned characters are a match, a mismatch, or a gap. A match occurs when the two aligned characters are the same, a mismatch when they are not the same, and a gap when for the aligned we need an insert or delete operation. The match, mismatch, gap scores used in this study, as in Simmetrics, are 0, -1, and -2, respectively.

\textit{q-grams Distance}: It applies a Block Distance (see below) similarity metric over all tri-grams in a string.

\textit{Longest Common Substring Similarity:} As the name suggests, this measure counts the size of the longest common substring ($lcs_{str}$) between two strings, divided by the size of the longest string:
$similarity(s_1,s_2) = |lcs_{str}(s_1,s_2)| / max(|s_1|, |s_2|)$.

\textit{Longest Common Subsequence Similarity:} The difference between this measure and the previous is that a subsequence does not need to consist of consecutive characters:
$similarity(s_1,s_2) = |lcs_{seq}(s_1,s_2)| / max(|s_1|, |s_2|)$.

\subsubsection{Word-level measures} The following similarity measures are applied to two strings $a$ and $b$ that are treated as sets or multisets (bags) of words.

\textit{Cosine Similarity}: The similarity is defined as the cosine of the angle between the multisets (bags) of words $a$ and $b$ expressed as sparse vectors.
$similarity(a,b) = a \cdot b / (||a|| \; ||b||)$. 

\textit{Euclidean Distance}: Compares the frequency of occurrence of each word $w$ in two strings $a$ and $b$
$distance(a,b) = ||a - b|| = \sqrt{\sum_w{(freqA(w)-freqB(w))^2}}$

\textit{Block Distance}: Also known as \textit{L1 Distance}, \textit{City Block Distance} and \textit{Manhattan Distance} between two multisets (bags) of words $a$ and $b$ is the sum of the absolute differences of the frequency of each word in $a$ vs in $b$: $distance(a,b) = ||a - b||_1$.

\textit{Overlap Coefficient}: The size of the intersection divided by the smaller of the size of the two sets of words: 
$similarity(a,b) = |a \cap b| / min(|a|,|b|)$.

\textit{Dice Similarity}: The Dice Similarity is defined as twice the shared information (intersection) divided by sum of cardinalities of the two sets of words: 
$similarity(a,b) = 2 |a \cap b| / (|a| + |b|)$.

\textit{Simon White Similarity}: This similarity is the same as Dice Similarity, with the only difference being that it considers $a$ and $b$ as multisets (bags) of words. 

\textit{Jaccard Similarity}: Computes the size of the intersection divided by the size of the union for two sets of words
$similarity(a,b) = |a \cap b| / |a \cup b|$.

\textit{Generalized Jaccard Similarity}: Same as the Jaccard Similarity, except that the Generalized Jaccard Similarity considers multisets (bags) of words, instead of sets. 

\textit{Monge-Elkan Similarity}: This similarity is the average similarity of the most similar words between two sets of words $a$ and $b$:
$similarity(a,b) = \frac{1}{|a|} \sum_{w_i \in a}{max_{w_j \in b}\left(sim(w_i,w_j)\right)}$, 
where $sim$ is the optimized Smith-Waterman algorithm~\cite{gotoh1982improved} that operates as the secondary character-level similarity to compute the similarity of individual words.

\subsection{Schema-agnostic syntactic functions}

For this category, we use the following similarity functions, as defined and implemented in JedAI.

\subsubsection{Bag Models \cite{DBLP:books/daglib/0021593}}

There are two types of n-grams, the character and the token
ones. These give rise to two types of bag models: the \textit{character n-grams
model} and the \textit{token n-grams model}.
Collectively, they are called \textit{bag} or \textit{vector space models},
because they model an entity $e_i$ as a vector with one dimension
for every distinct n-gram in an entity collection $E$: $BM(e_i)=(w_{i1},\dots,w_{im})$, 
where $m$ stands for the \textit{dimensionality} of $E$ (i.e., the number of distinct
n-grams in it), while $w_{ij}$ is the weight of the $j^{th}$ dimension that
quantifies the importance of the corresponding n-gram for $e_i$. 

The most common weighting schemes are:

\emph{(i)} \textit{Term Frequency} (\textsf{TF}) sets weights in proportion
  to the number of times the corresponding n-grams appear in the values of entity $e_i$. 
More formally, $TF(t_j, e_i)$=$f_j/N_{e_i}$, where $f_j$ stands for the
occurrence frequency of $t_j$ in $e_i$, while $N_{e_i}$ is the number of n-grams in $e_i$,
normalizing \textsf{TF} so as to mitigate the effect of different lengths on the weights.

\emph{(ii)} \textit{Term Frequency-Inverse Document Frequency}
(\textsf{TF-IDF}) discounts the \textsf{TF} weight for the most common
tokens in the entire entity collection $E$, as they typically correspond to noise (i.e.,
stop words). Formally, $TF$-$IDF(t_j, e_i)=TF(t_j, e_i)\cdot IDF(t_j)$,
where $IDF(t_j)$ is the inverse document frequency of the n-gram $t_j$, i.e.,
$IDF(t_j)=\log{|E|/(|\{ e_k \in E : t_j \in e_k \}|+1)}$. In this way, high
weights are given to n-grams with high frequency in $e_i$, but low frequency in $E$.

To construct the bag model for a specific entity, we aggregate the vectors
corresponding to each one of its attribute values. The end result is
a weighted vector $(a_i(w_{1}),....,a(w_{m}))$, where $a_i(w_{j})$ is the sum of weights, i.e., $a_i(w_{j})=\sum_{a_k \in A_i} w_{ij}$, where $a_k$ stands for an individual attribute value in the set of values $A_i$ of entity $e_i$.

To compare two bag models, $BM(e_i)$ and $BM(e_j)$, one of the following similarity
measures is typically used:

\emph{(i)} \textit{ARCS Similarity} (\textsf{ARCS}) sums the inverse Document Frequency of the common n-grams in two bag models. That is, the rarer the common n-grams are, the higher gets the overall similarity. Formally: $ARCS(BM(e_i),BM(e_j))=\sum_{k \in BM(e_i){\cap}BM(e_j)}{\log 2 /\log (DF_1(k) \cdot DF_2(k))}$.

\emph{(ii)} \textit{Cosine Similarity} (\textsf{CS}) measures the cosine of the angle
 of the weighted vectors. Formally, it is equal to their dot product
 similarity, normalized by the product of their magnitudes:\\
$CS(BM(e_i),BM(e_j))=\sum_{k=1}^{m}{w_{ik}w_{jk}}/||BM(e_i)||/||BM(e_j)||$.
 
\emph{(iii)} \textit{Jaccard Similarity} (\textsf{JS}) treats the
document vectors as sets, with weights higher than (equal to) 0 indicating the
presence (absence) of the corresponding n-gram. On this basis, it defines as
similarity the ratio between the sizes of set intersection and union:
$JS(BM(e_i),BM(e_j)){=}|BM(e_i){\cap}BM(e_j)|/|BM(e_i){\cup}BM(e_j)|$.

\emph{(iv)} \textit{Generalized Jaccard Similarity} (\textsf{GJS}) extends
\textsf{JS} so that it takes into account the weights associated with every
n-gram:\\
{\small$GJS(BM(e_i),BM(e_j)){=}\sum_{k=1}^{m}min(w_{ik},w_{jk})/\sum_{k=1}^{m}max(w_{ik},w_{jk})$}. 

Both CS and GJS apply seamlessly to both TF and TF-IDF weights.

\subsubsection{Graph Models \cite{DBLP:journals/tslp/GiannakopoulosKVS08,DBLP:conf/cikm/RousseauV13}.}
Recent works suggest that graph models outperform the bag ones in various tasks \cite{graphsTutorial}, from Information Retrieval \cite{DBLP:conf/cikm/RousseauV13} to Document Classification \cite{DBLP:journals/www/0001GP16}. 
There are two graph models, one for each type of n-grams, i.e.,
\textit{token n-gram graphs} \cite{DBLP:conf/cikm/RousseauV13} and
\textit{character n-gram graphs} \cite{DBLP:journals/tslp/GiannakopoulosKVS08}. Both models represent each entity $e_i$ as an undirected graph
$G_{i}$ that contains one vertex for each n-gram in the attribute values of $e_i$.
An edge connects every pair of vertices/n-grams that
co-occur within a window of size $n$ in the values of $e_i$. Every 
edge is weighted according to the co-occurrence frequency of the
corresponding n-grams. Thus, the graphs
incorporate \textit{contextual information} in the form of n-grams'
closeness. 

To construct the model for an entity, we merge the graphs of its attribute values using the update
operator, which is described in~\cite{DBLP:journals/tslp/GiannakopoulosKVS08,giannakopoulos2010content}.
To compare graph models, we can use the following graph similarity
measures~\cite{DBLP:journals/tslp/GiannakopoulosKVS08}:

\emph{(i)} \textit{Containment Similarity} (\textsf{CoS}) estimates the number
of edges shared by two graph models, $G_i$ and $G_j$, regardless of the
corresponding weights (i.e., it merely estimates the portion of common n-grams in the original
texts). 
Formally:\\ $CoS(G_i,G_j) = \sum_{e\in G_i}{\mu(e,G_j)}/min(|G_i|,|G_j|)$,
where $|G|$ is the size of graph G, and $\mu(e,G)=1$ if $e \in G$, or 0
otherwise.

\emph{(ii)}  \textit{Value Similarity} (\textsf{VS}) 
extends \textsf{CoS} by considering the weights of common edges. Formally, using $w_e^k$ for the weight of edge $e$ in $G_k$:
{\small
$VS(G_i,G_j)=\sum_{e\in (G_i\cap G_j)}{\frac{min(w_e^i,w_e^j)}{max(w_e^i,w_e^j)\cdot max(|G_i|,|G_j|)}}$}.

\emph{(iii)} \textit{Normalized Value Similarity} (\textsf{NS}) extends
\textsf{VS} by mitigating the impact of imbalanced graphs, i.e., the cases where
the comparison between a large graph with a much smaller one yields
similarities close to 0. Formally:\\
{\small
$NS(G_i,G_j){=}\sum_{e\in
(G_i\cap G_j)}{min(w_e^i,w_e^j)/max(w_e^i,w_e^j)}/min(|G_i|,|G_j|)$}.

\emph{(iv)} \textit{Overall Similarity} (\textsf{OS}) constitutes the average of the above graph similarity measures, which are all defined $[0,1]$. Formally: 
$OS(G_i,G_j){=}(CoS(G_i,G_j)+VS(G_i,G_j)+NS(G_i,G_j))/3$.
\end{appendix}

\section{Additional Experiments}

\subsection{Critical Difference Analysis} 

\begin{figure}[t]
    \centering
    \includegraphics[width=\columnwidth]{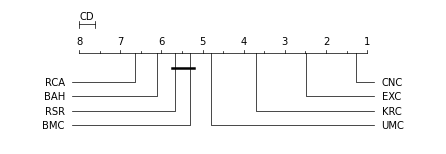}
    \vspace{-25pt}
    \caption{Nemenyi diagram based on Precision.}
    \vspace{-10pt}
    \label{fig:nemenyiPr}
\end{figure}

\begin{figure}[t]
    \centering
    \includegraphics[width=\columnwidth]{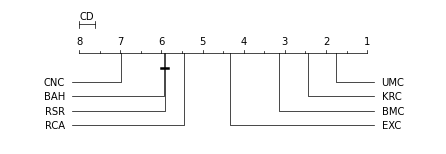}
    \vspace{-25pt}
    \caption{Nemenyi diagram based on Recall.}
    \vspace{-10pt}
    \label{fig:nemenyiRe}
\end{figure}

In addition to F-Measure, which is examined in Section \ref{sec:expAnalysis} and Figure \ref{fig:nemenyi}, we performed a post-hoc Nemenyi test to identify the critical difference of the eight algorithms with respect to Precision and Recall. The corresponding Nemenyi diagrams appear in Figures \ref{fig:nemenyiPr} and \ref{fig:nemenyiRe}, respectively. In both cases, the critical distance is the same as that of F-Measure, namely 0.37.
We observe that for Precision, only the difference between \textsf{RSR} and \textsf{BMC} is insignificant, while for Recall, the only insignificant difference pertains to \textsf{BAH} and \textsf{RSR}. The best performing algorithm in terms of Precision is \textsf{CNC}, followed by \textsf{EXC}, \textsf{KRC} and \textsf{UMC}, while the best recall is achieved by \textsf{UMC}, with \textsf{KRC} in the second place. These results verify the patterns in Section \ref{sec:expAnalysis} (Table \ref{tb:avPer} and Figure \ref{fig:effectiveness} in particular), which highlight the excellent balance between Precision and Recall that is achieved by \textsf{UMC} and \textsf{KRC}.

\subsection{Threshold Analysis} 


As explained in Section \ref{sec:expAnalysis}, the similarity threshold constitutes the most important configuration parameter for the effectiveness and time efficiency of all bipartite graph matching algorithms. It is crucial, therefore, to understand how easily this parameter can be fine-tuned a-priori, examining the main factors that determine its optimal value. To this end, Table \ref{tb:thrDistribution} presents the descriptive statistics of similarity threshold per algorithm and type of edge weights. These statistics include the average value along with the corresponding standard deviation as well as the minimum and maximum values together with the first, second and third quartile ($Q_1$, $Q_2$ and $Q_3$, respectively). We also report the Pearson correlation of similarity thresholds with the normalized size of similarity graphs, i.e., the number of their edges divided by the Cartesian product.

\begin{table}[t]\centering
    \caption{The distribution of similarity thresholds per algorithm and type of input.}
    \vspace{-10pt}
    {\small
	\begin{tabular}{ | l | r | r | r | r | r | r | c |}
		\cline{2-8}
		\multicolumn{1}{c|}{}&
		\multicolumn{1}{c|}{\textbf{mean$\pm$std}} &
		\multicolumn{1}{c|}{\textbf{min.}} &
		\multicolumn{1}{c|}{$\mathbf{Q_1}$} &
		\multicolumn{1}{c|}{$\mathbf{Q_2}$} &
		\multicolumn{1}{c|}{$\mathbf{Q_3}$} &
		\multicolumn{1}{c|}{\textbf{max.}}  & 
		\multicolumn{1}{c|}{$\mathbf{\rho(t, \frac{|E|}{||V_1 \times V_2||})}$}  \\
		\hline
	    \textsf{CNC} & 0.76$\pm$0.16 & 0.30 & 0.65 & 0.80 & 0.90 & 0.95 & -0.09\\
        \textsf{RSR} & 0.76$\pm$0.16 & 0.20 & 0.65 & 0.80 & 0.90 & 0.95 & -0.14\\
        \textsf{RCA} & 0.66$\pm$0.21 & 0.05 & 0.50 & 0.70 & 0.80 & 0.95 & -0.22\\
        \textsf{BAH} & 0.66$\pm$0.25 & 0.05 & 0.50 & 0.70 & 0.85 & 0.95 & -0.31\\
        \textsf{BMC} & 0.67$\pm$0.20 & 0.20 & 0.53 & 0.70 & 0.80 & 0.95 & -0.16\\
        \textsf{EXC} & 0.63$\pm$0.23 & 0.05 & 0.45 & 0.65 & 0.80 & 0.95 & -0.19\\
        \textsf{KRC} & 0.61$\pm$0.25 & 0.05 & 0.40 & 0.65 & 0.80 & 0.95 & -0.17\\
        \textsf{UMC} & 0.63$\pm$0.23 & 0.05 & 0.45 & 0.65 & 0.80 & 0.95 & -0.16 \\
        \hline
        \multicolumn{8}{c}{\textbf{(a) Schema-based syntactic inputs}}\\
        \hline
        \textsf{CNC} & 0.41$\pm$0.23 & 0.05 & 0.20 & 0.40 & 0.60 & 0.95 & 0.43\\
        \textsf{RSR} & 0.41$\pm$0.24 & 0.05 & 0.20 & 0.40 & 0.60 & 0.95 & 0.39\\
        \textsf{RCA} & 0.31$\pm$0.24 & 0.05 & 0.10 & 0.25 & 0.50 & 0.95 & 0.35\\
        \textsf{BAH} & 0.30$\pm$0.24 & 0.05 & 0.05 & 0.25 & 0.45 & 0.95 & 0.35\\
        \textsf{BMC} & 0.33$\pm$0.24 & 0.05 & 0.10 & 0.30 & 0.50 & 0.95 & 0.37\\
        \textsf{EXC} & 0.29$\pm$0.23 & 0.05 & 0.10 & 0.25 & 0.45 & 0.90 & 
        0.38\\
        \textsf{KRC} & 0.27$\pm$0.24 & 0.05 & 0.05 & 0.20 & 0.45 & 0.90 & 
        0.31\\
        \textsf{UMC} & 0.30$\pm$0.25 & 0.05 & 0.05 & 0.20 & 0.50 & 0.95 & 
        0.33\\
        \hline
        \multicolumn{8}{c}{\textbf{(b) Schema-agnostic syntactic inputs}}\\
        \hline
        \textsf{CNC} & 0.69$\pm$0.36 & 0.00	& 0.48 & 0.95 & 0.95 & 0.95 & -0.13 \\
        \textsf{RSR} & 0.80$\pm$0.24 & 0.15 & 0.70 & 0.95 & 0.95 & 0.95 & -0.29\\
        \textsf{RCA} & 0.77$\pm$0.27 & 0.10 & 0.55 & 0.95 & 0.95 & 0.95 & -0.32\\
        \textsf{BAH} & 0.66$\pm$0.30 & 0.05 & 0.41 & 0.80 & 0.95 & 0.95 & -0.25\\
        \textsf{BMC} & 0.78$\pm$0.27 & 0.10 & 0.56 & 0.95 & 0.95 & 0.95 & -0.26\\
        \textsf{EXC} & 0.77$\pm$0.26 & 0.10 & 0.56 & 0.95 & 0.95 & 0.95 & -0.24\\
        \textsf{KRC} & 0.75$\pm$0.29 & 0.05 & 0.51 & 0.95 & 0.95 & 0.95 & -0.25\\
        \textsf{UMC} & 0.76$\pm$0.28 & 0.10 & 0.51 & 0.93 & 0.95 & 0.95 & -0.22\\
        \hline
        \multicolumn{8}{c}{\textbf{(c) Schema-based semantic inputs}}\\
        \hline
        \textsf{CNC} & 0.68$\pm$0.29 & 0.00 & 0.55 & 0.75 & 0.95 & 0.95 & -0.31\\
       \textsf{RSR} & 0.74$\pm$0.24 & 0.10 & 0.55 & 0.85 & 0.95 & 0.95 &       -0.27\\
        \textsf{RCA} & 0.66$\pm$0.28 & 0.05 & 0.45 & 0.80 & 0.85 & 0.95 & -0.34 \\
        \textsf{BAH} & 0.60$\pm$0.33 & 0.05 & 0.30 & 0.75 & 0.85 & 0.95 & -0.37\\
        \textsf{BMC} & 0.71$\pm$0.26 & 0.10 & 0.50 & 0.85 & 0.95 & 0.95 & -0.30 \\
        \textsf{EXC} & 0.67$\pm$0.27 & 0.05 & 0.45 & 0.80 & 0.90 & 0.95 & -0.28\\
        \textsf{KRC} & 0.64$\pm$0.25 & 0.05 & 0.45 & 0.68 & 0.84 & 0.95 & -0.31\\
        \textsf{UMC} & 0.67$\pm$0.27 & 0.05 & 0.46 & 0.80 & 0.85 & 0.95 & -0.33\\
        \hline
        \multicolumn{8}{c}{\textbf{(d) Schema-agnostic semantic inputs}}\\
	\end{tabular}
	\vspace{-10pt}
}
	\label{tb:thrDistribution}
\end{table}

We observe different patterns for each type of edge weights. More specifically, the schema-agnostic syntactic inputs yield relatively high thresholds: the average values of all algorithms fluctuate between 0.61 and 0.76, while the median ($Q_2$) is slightly higher, between 0.65 and 0.8. Yet, the variance is also high, as indicated by the standard deviation, which is consistently higher than 0.16. The reason is that the similarity thresholds of all algorithms cover the entire space in $[0.05, 0.95]$, as suggested by the minimum and maximum values. For most algorithms, though, the first ($Q_1$) and third ($Q_3$) quartiles significantly restrict the optimal values to the range 0.4/0.5-0.8 or 0.65-0.90. The slightly negative correlation with the normalized graph size suggests that in datasets where the portion of edges is high, the optimal threshold should be slightly lower. The reason is that the schema-based syntactic weights provide higher confidence in datasets with relatively small graphs (e.g., $D_1$), while large graphs indicate non-zero similarities with too many non-matching entities. These conditions occur in datasets with lower confidence in the similarity values, such as the bibliographic ones $D_4$ and $D_9$, which convey a limited vocabulary. In the former cases, high thresholds suffice for achieving high performance, but in the latter ones, slightly lower thresholds are required in order to restricting the impact on recall. 

Regarding the schema-agnostic syntactic weights, we observe that the standard deviation is higher than the schema-based counterparts, even though the average and median values are much lower, even by 50\%. The reason is that the similarity thresholds still cover the entire space in $[0.05, 0.95]$, with the interquartile range ($Q_3$ - $Q_1$) increasing from 0.25 or 0.3 to 0.4 for practically all algorithms. This means that this type of weights yields a larger diversity of graphs, which hinders the configuration of the similarity thresholds. Note that the correlation with the normalized graph size is significantly positive for all algorithms. The reason is that large graphs emanate from similarity functions that associate many non-matching entities, albeit with low weights. These settings call for higher thresholds in order to prune the non-matching edges and increase precision. In other words, smaller thresholds should be used in smaller graphs in order to maintain a high recall. 

\begin{figure*}[t]
    \centering
    \includegraphics[width=0.245\textwidth]{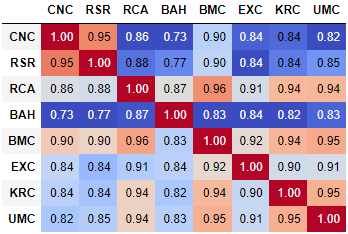}
    \includegraphics[width=0.245\textwidth]{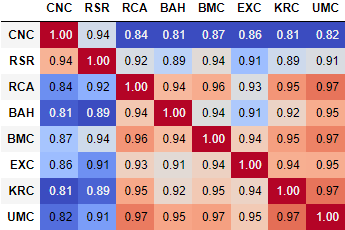}
    \includegraphics[width=0.245\textwidth]{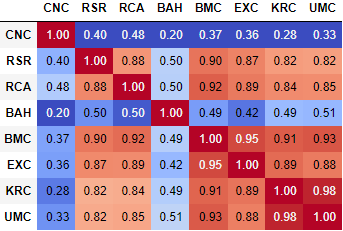}
    \includegraphics[width=0.245\textwidth]{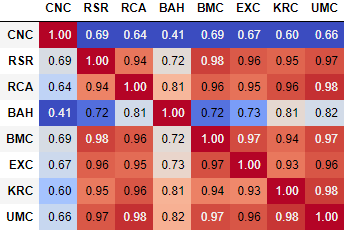}
    \vspace{-12pt}
    \caption{Pearson correlation between the similarity thresholds of every algorithm for schema-based syntactic, schema-agnostic syntactic inputs, schema-based semantic and schema-agnostic semantic inputs, from left to right.}
    \label{fig:pearsonCorrelations}
\end{figure*}

Among the semantic inputs, the schema-based ones share similar patterns with their syntactic counterparts, as the average thresholds are high for all algorithms. Compared to syntactic weights, though, the median and the third quartile get higher, but the first one gets lower, thus increasing the interquartile range and the standard deviation. This means that threshold fine-tuning gets harder. Yet, the normalized graph size provide stronger indications for the best thresholds, as suggested by the significantly higher negative correlations. Similar patterns apply to the schema-agnostic semantic inputs, albeit with slightly lower thresholds.

Note that the relative order of the algorithms with respect to the average thresholds remains the same in both types of syntactic weights: \textsf{CNC} and \textsf{RSR} exhibit the highest thresholds, with \textsf{KRC} lying at the other extreme. followed in close distance by \textsf{EXC} and \textsf{UMC}. These patterns do not hold in the case of semantic inputs, except for \textsf{RSR}, which consistently yields the highest average threshold. In fact, \textsf{CNC} uses lower thresholds than \textsf{KRC}, \textsf{EXC} and \textsf{UMC} for schema-based weights. 

These patterns can be verified at a finer granularity through Table \ref{tb:datasetThresholds}, which provides the average threshold per dataset for each algorithm and type of edge weights. Another pattern that emerges from this table is that every algorithm exhibits highly similar thresholds for every row (i.e., dataset). This means that knowing the optimal threshold for a particular algorithm over a specific dataset provides strong indications for fine-tuning the rest of the algorithms over the same dataset. The high correlation between the optimal thresholds of the eight algorithms is verified by the diagrams in Figure \ref{fig:pearsonCorrelations}, which present the Pearson correlations for each pair across the four types of edge weights. We observe that this correlation is highly positive, taking values well above 0.8 in the vast majority of cases, especially for the syntactic weights. The only exceptions are \textsf{BAH} and \textsf{CNC} for semantic weights, which fluctuate from 0.2 to 0.8. 

Overall, the similarity thresholds used by bipartite graph matching algorithms depend on the type of edge weights and the normalized graph size. With the exception of schema-agnostic syntactic weights, relatively high thresholds are used on average. Their optimal value is more stable in the case of schema-based syntactic weights, with the interquartile range limited to 0.3 over even 0.25 for most algorithms. Most importantly, the optimal threshold for a particular similarity graph is relatively stable across different algorithms. In other words, it depends more on the characteristics of the input, than the functionality of the graph matching algorithm.

\begin{table}[t]\centering
 \setlength{\tabcolsep}{2.5pt}
    \caption{The average similarity threshold and its standard deviation per algorithm, dataset and type of edge weights.}
    \vspace{-10pt}
    {\small
	\begin{tabular}{ | l | r | r | r | r | r | r | r | r | }
		\cline{2-9}
		\multicolumn{1}{c|}{}&
		\multicolumn{1}{c|}{\textbf{CNC}} &
		\multicolumn{1}{c|}{\textbf{RSR}} &
		\multicolumn{1}{c|}{\textbf{RCA}} &
		\multicolumn{1}{c|}{\textbf{BAH}} &
		\multicolumn{1}{c|}{\textbf{BMC}} &
		\multicolumn{1}{c|}{\textbf{EXC}} &
		\multicolumn{1}{c|}{\textbf{KRC}} &
		\multicolumn{1}{c|}{\textbf{UMC}} \\
		\hline
        \hline
        $\mathbf{D_1}$ & .81$\pm$.15 & .85$\pm$.12 & .75$\pm$.14 & .83$\pm$.13 & .76$\pm$.14 & .69$\pm$.21 & .72$\pm$.17 & .71$\pm$.17 \\
        $\mathbf{D_2}$ & .63$\pm$.15 & .60$\pm$.14 & .42$\pm$.20 & .16$\pm$.17 & .48$\pm$.20 & .34$\pm$.23 & .39$\pm$.22 & .38$\pm$.20 \\
        $\mathbf{D_3}$ & .61$\pm$.14 & .60$\pm$.15 & .45$\pm$.17 & .43$\pm$.19 & .50$\pm$.17 & .40$\pm$.20 & .37$\pm$.25 & .41$\pm$.21\\
        $\mathbf{D_4}$ & .73$\pm$.12 & .71$\pm$.13 & .49$\pm$.22 & .49$\pm$.23 & .54$\pm$.17 & .53$\pm$.22 & .40$\pm$.23 & .43$\pm$.21 \\
        $\mathbf{D_5}$ & .86$\pm$.10 & .86$\pm$.10 & .80$\pm$.09 & .82$\pm$.10 & .79$\pm$.12 & .75$\pm$.13 & .73$\pm$.12 & .73$\pm$.14\\
        $\mathbf{D_6}$ & .86$\pm$.07 & .88$\pm$.07 & .84$\pm$.08 & .85$\pm$.11 & .83$\pm$.09 & .82$\pm$.09 & .82$\pm$.09 & .85$\pm$.10\\
        $\mathbf{D_7}$ & .77$\pm$.19 & .76$\pm$.21 & .72$\pm$.21 & .77$\pm$.19 & .70$\pm$.22 & .69$\pm$.22 & .68$\pm$.23 & .69$\pm$.23\\
        $\mathbf{D_8}$ & .75$\pm$.16 & .77$\pm$.15 & .73$\pm$.16 & .72$\pm$.17 & .73$\pm$.16 & .71$\pm$.18 & .72$\pm$.18 & .72$\pm$.17\\
        $\mathbf{D_9}$ & .71$\pm$.15 & .72$\pm$.15 & .59$\pm$.18 & .61$\pm$.18 & .59$\pm$.18 & .49$\pm$.23 & .55$\pm$.22 & .58$\pm$.19\\
        $\mathbf{D_{10}}$ & .64$\pm$.13 & .63$\pm$.12 & .56$\pm$.13 & .55$\pm$.12 & .55$\pm$.12 & .53$\pm$.16 & .38$\pm$.23 & .48$\pm$.13\\
        \hline
        \multicolumn{9}{c}{(a) \textbf{Schema-based, syntactic inputs}} \\
        \hline
        $\mathbf{D_1}$ & .60$\pm$.24 & .67$\pm$.20 & .64$\pm$.21 & .62$\pm$.23 & .64$\pm$.21 & .58$\pm$.21 & .60$\pm$.21 & .63$\pm$.22 \\
        $\mathbf{D_2}$ & .33$\pm$.15 & .28$\pm$.14 & .14$\pm$.10 & .07$\pm$.04 & .19$\pm$.11 & .13$\pm$.10 & .12$\pm$.10 & .11$\pm$.09 \\
        $\mathbf{D_3}$ & .34$\pm$.22 & .28$\pm$.19 & .15$\pm$.12 & .14$\pm$.13 & .18$\pm$.14 & .17$\pm$.14 & .10$\pm$.08 & .45$\pm$.08 \\
        $\mathbf{D_4}$ & .47$\pm$.21 & .43$\pm$.22 & .25$\pm$.18 & .31$\pm$.20 & .31$\pm$.19 & .34$\pm$.20 & .24$\pm$.17 & .28$\pm$.17 \\
        $\mathbf{D_5}$ & .42$\pm$.19 & .39$\pm$.19 & .29$\pm$.18 & .26$\pm$.19 & .29$\pm$.20 & .25$\pm$.20 & .23$\pm$.18 & .24$\pm$.18 \\
        $\mathbf{D_6}$ & .33$\pm$.17 & .32$\pm$.18 & .29$\pm$.18 & .26$\pm$.17 & .29$\pm$.18 & .24$\pm$.17 & .24$\pm$.16 & .27$\pm$.17 \\
        $\mathbf{D_7}$ & .55$\pm$.28 & .64$\pm$.31 & .51$\pm$.29 & .50$\pm$.27 & .57$\pm$.30 & .50$\pm$.31 & .53$\pm$.32 & .55$\pm$.31 \\
        $\mathbf{D_8}$ & .48$\pm$.17 & .51$\pm$.17 & .43$\pm$.16 & .43$\pm$.17 & .43$\pm$.16 & .39$\pm$.16 & .41$\pm$.15 & .43$\pm$.16 \\
        $\mathbf{D_9}$ & .38$\pm$.18 & .39$\pm$.18 & .26$\pm$.16 & .30$\pm$.16 & .27$\pm$.17 & .22$\pm$.15 & .22$\pm$.15 & .26$\pm$.16 \\
        $\mathbf{D_{10}}$ & .24$\pm$.15 & .19$\pm$.14 & .12$\pm$.10 & .13$\pm$.12 & .15$\pm$.11 & .15$\pm$.12 & .07$\pm$.03 & .11$\pm$.09 \\
        \hline
        \multicolumn{9}{c}{(b) \textbf{Schema-agnostic, syntactic inputs}} \\
        \hline
        $\mathbf{D_1}$ & .70$\pm$.33 & .79$\pm$.23 & .79$\pm$.23 & .69$\pm$.34 & .77$\pm$.25 & .71$\pm$.25 & .74$\pm$.25 & .77$\pm$.26\\
        $\mathbf{D_2}$ & .72$\pm$.25 & .62$\pm$.32 & .60$\pm$.28 & .17$\pm$.04 & .65$\pm$.28 & .60$\pm$.28 & .57$\pm$.25 & .57$\pm$.25 \\
        $\mathbf{D_3}$ & .65$\pm$.28 & .62$\pm$.32 & .27$\pm$.18 & .17$\pm$.11 & .27$\pm$.18 & .45$\pm$.28 & .27$\pm$.18 & .30$\pm$.21 \\
        $\mathbf{D_4}$ & .60$\pm$.41 & .73$\pm$.33 & .72$\pm$.33 & .61$\pm$.30 & .72$\pm$.33 & .72$\pm$.33 & .66$\pm$.37 & .70$\pm$.33\\
        $\mathbf{D_5}$ & .72$\pm$.38 & .87$\pm$.17 & .83$\pm$.20 & .70$\pm$.30 & .86$\pm$.19 & .85$\pm$.20 & .79$\pm$.29 & .78$\pm$.28 \\
        $\mathbf{D_6}$ & .75$\pm$.38 & .89$\pm$.15 & .87$\pm$.19 & .73$\pm$.26 & .87$\pm$.17 & .87$\pm$.18 & .87$\pm$.18 & .86$\pm$.17\\
        $\mathbf{D_7}$ & .71$\pm$.40 & .85$\pm$.23 & .86$\pm$.23 & .79$\pm$.24 & .86$\pm$.23 & .86$\pm$.24 & .86$\pm$.23 & .86$\pm$.23\\
        $\mathbf{D_8}$ & .75$\pm$.28 & .72$\pm$.32 & .67$\pm$.32 & .65$\pm$.35 & .72$\pm$.25 & .62$\pm$.39 & .70$\pm$.28 & .72$\pm$.25 \\
        $\mathbf{D_9}$ & .72$\pm$.35 & .73$\pm$.33 & .69$\pm$.36 & .74$\pm$.27 & .69$\pm$.36 & .67$\pm$.36 & .69$\pm$.36 & .69$\pm$.36\\
        $\mathbf{D_{10}}$ & .70$\pm$.21 & .65$\pm$.28 & .57$\pm$.25 & .42$\pm$.46 & .45$\pm$.42 & .57$\pm$.25 & .32$\pm$.39 & .37$\pm$.39\\
        \hline
        \multicolumn{9}{c}{(c) \textbf{Schema-based, semantic inputs}}\\
        \hline
        $\mathbf{D_1}$ & .80$\pm$.21 & .87$\pm$.11 & .80$\pm$.21 & .87$\pm$.11 & .80$\pm$.21 & .80$\pm$.21 & .80$\pm$.21 & .80$\pm$.21\\
        $\mathbf{D_2}$ & .72$\pm$.25 & .72$\pm$.18 & .52$\pm$.39 & .25$\pm$.00 & .62$\pm$.25 & .62$\pm$.25 & .62$\pm$.25 & .60$\pm$.28\\
        $\mathbf{D_3}$ & .80$\pm$.21 & .80$\pm$.21 & .60$\pm$.28 & .12$\pm$.11 & .80$\pm$.21 & .70$\pm$.28 & .52$\pm$.18 & .60$\pm$.28\\
        $\mathbf{D_4}$ & .71$\pm$.29 & .67$\pm$.30 & .62$\pm$.29 & .64$\pm$.31 & .69$\pm$.29 & .66$\pm$.33 & .55$\pm$.25 & .64$\pm$.32\\
        $\mathbf{D_5}$ & .67$\pm$.25 & .70$\pm$.21 & .65$\pm$.28 & .67$\pm$.25 & .65$\pm$.28 & .62$\pm$.32 & .60$\pm$.28 & .65$\pm$.28\\
        $\mathbf{D_6}$ & .62$\pm$.32 & .65$\pm$.28 & .60$\pm$.35 & .57$\pm$.39 & .60$\pm$.35 & .60$\pm$.35 & .60$\pm$.35 & .60$\pm$.35\\
        $\mathbf{D_7}$ & .55$\pm$.47 & .76$\pm$.37 & .75$\pm$.39 & .73$\pm$.39 & .76$\pm$.37 & .68$\pm$.38 & .73$\pm$.38 & .75$\pm$.39\\
        $\mathbf{D_8}$ & - & - & - & - & - & - & - & -\\
        $\mathbf{D_9}$ & .72$\pm$.25 & .72$\pm$.25 & .67$\pm$.25 & .65$\pm$.28 & .67$\pm$.25 & .62$\pm$.25 & .62$\pm$.25 & .67$\pm$.25 \\
        $\mathbf{D_{10}}$ & - & - & - & - & - & - & - & -\\
        \hline
        \multicolumn{9}{c}{(d) \textbf{Schema-agnostic, semantic inputs}}
	\end{tabular}
	\vspace{-10pt}
}
	\label{tb:datasetThresholds}
\end{table}

\subsection{Trade-off between F-Measure and Run-time}

\begin{figure*}[ht!]
\centering
\includegraphics[width=0.57\textwidth]{figures/f1RtLegend.png}\\
\includegraphics[width=0.33\textwidth]{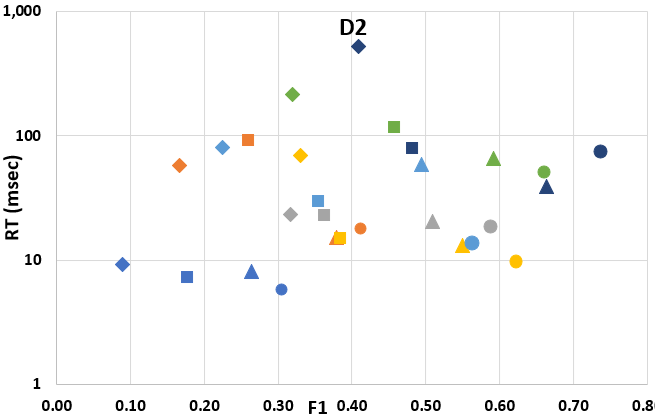}
\includegraphics[width=0.33\textwidth]{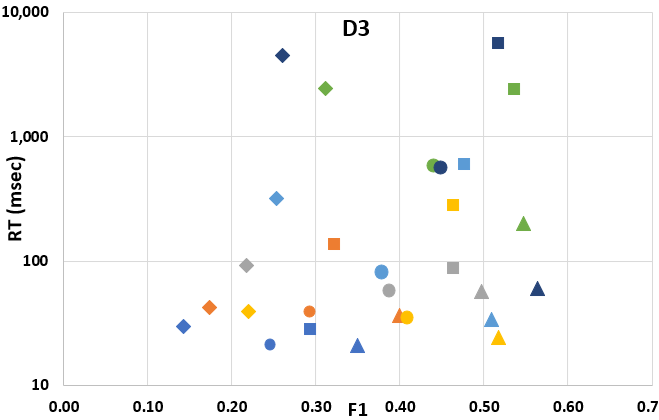}
\includegraphics[width=0.33\textwidth]{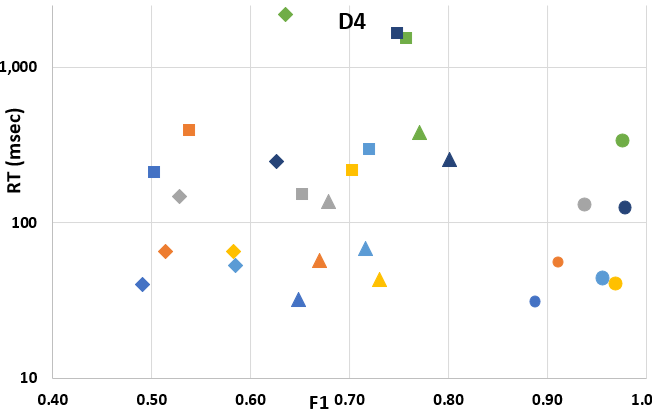}
\includegraphics[width=0.33\textwidth]{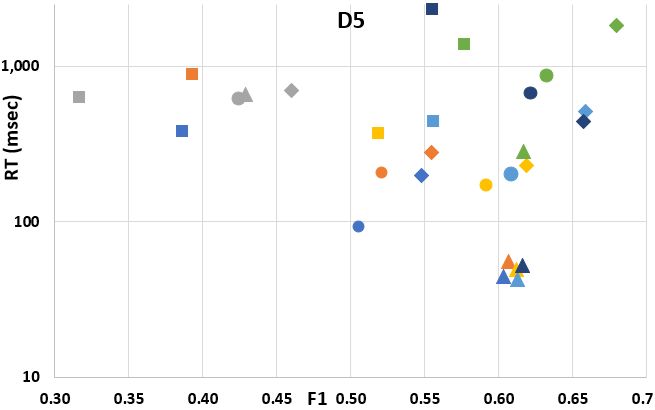}
\includegraphics[width=0.33\textwidth]{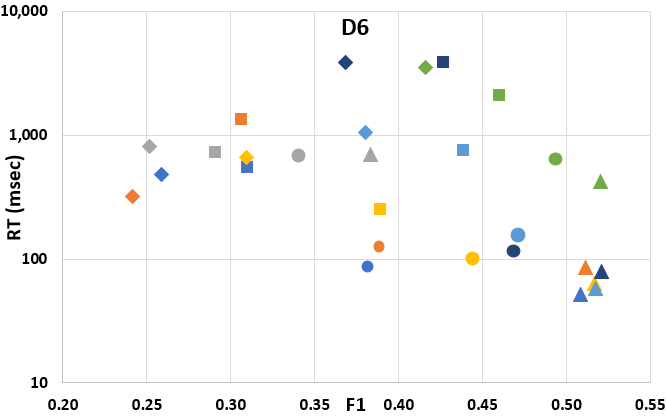}
\includegraphics[width=0.33\textwidth]{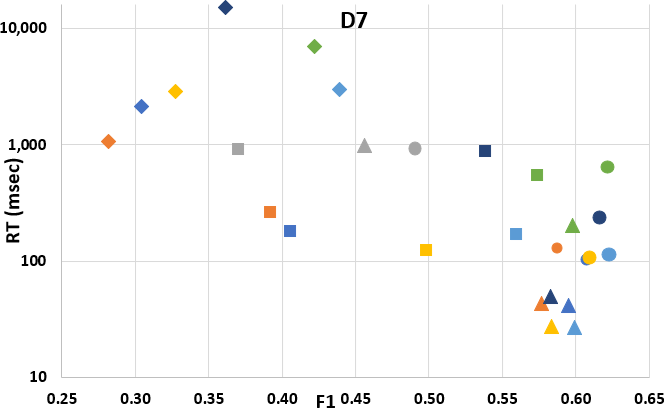}
\includegraphics[width=0.33\textwidth]{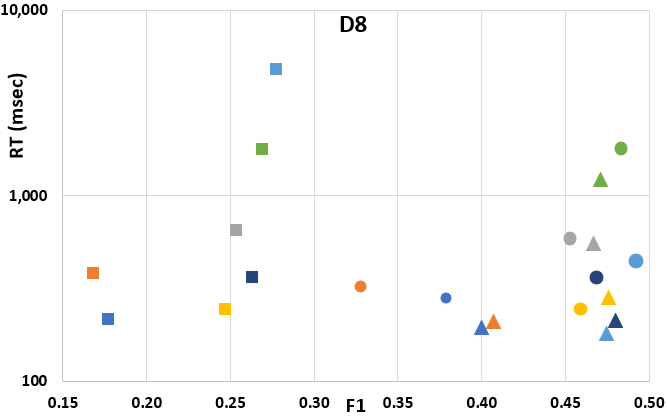}
\includegraphics[width=0.33\textwidth]{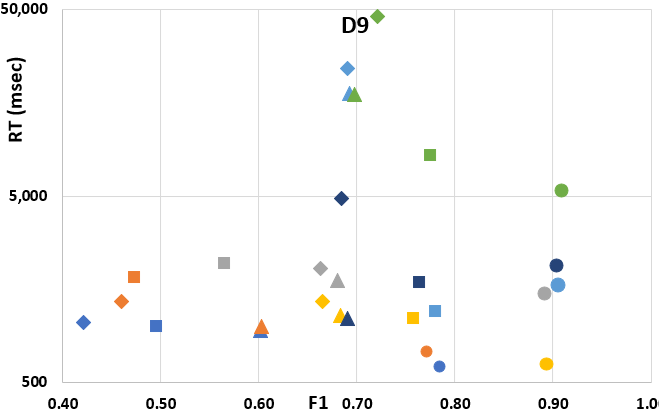}
\includegraphics[width=0.33\textwidth]{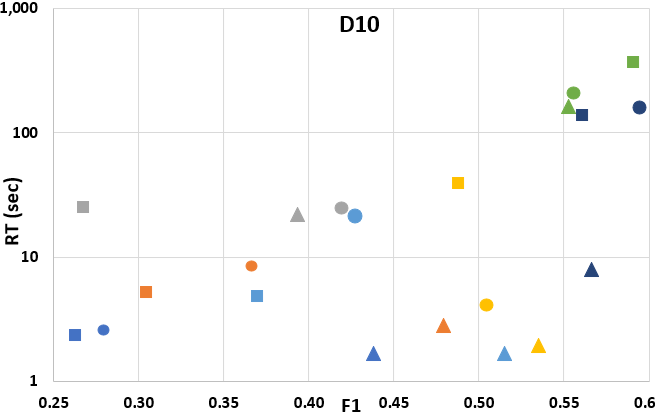}
\vspace{-10pt}
\caption{Scatter plots of the average F-measure (on the horizontal axis) and the average run-time (on the vertical axis) per algorithm and type of input across $D_2$-$D_{10}$. Every algorithm corresponds to a different color (see the legend) and every type of input to a different shape: 
circle stands for the schema-agnostic syntactic inputs, triangle for the schema-based syntactic ones, rhombus for the schema-agnostic semantic ones and rectangle for the schema-based semantic ones. Note that logarithmic scale is used in all datasets and that $D_{10}$ reports seconds instead of milliseconds, as in all other cases. }
\vspace{-10pt}
\label{fig:f1Rt9Datasets}
\end{figure*}

In this section, we examine the best trade-off that is achieved on average by all combinations of algorithms and types of edge weights across the datasets $D_2$-$D_{10}$. To this end, Figure \ref{fig:f1Rt9Datasets} contains one diagram per dataset, excluding the combinations including \textsf{BAH}, as their average performance consistently underperforms with respect to both F-Measure and run-time.

Starting with $D_2$, we observe the best performance clearly corresponds to \textsf{UMC} coupled with schema-agnostic syntactic weights, which achieves the highest macro-averaged F-Measure (0.738). The next best combinations reduce the average F-Measure by 10\% for a similar run-time. Note in Table \ref{tb:datasetThresholds} the average similarity threshold of \textsf{UMC} is just 0.11, due to the very small size of the input graph, as shown in Table \ref{tb:inputs}.

In $D_3$, the best F-Measure is achieved by schema-based weights, with the syntactic ones outperforming the semantic ones in terms of run-time, due to the lower normalized graph size and the higher similarity thresholds, on average. Among the algorithms, \textsf{UMC} is again the best option, dominating the second-best approach, \textsf{KRC}, with respect to both effectiveness and time efficiency. \textsf{BMC} and \textsf{EXC} reduce the run-time almost by 50\%, though at the cost of much lower F-Measure (<10\%).

$D_4$ is dominated by schema-agnostic syntactic inputs, due to the noise in the form of misplaced attribute values (e.g., the author of a publication is added in its title). This type of error cannot be addressed by schema-based weights, thus reducing significantly their effectiveness. In contrast, schema-agnostic weights consider address this noise inherently, as they take into account the entire textual information per entity profile. Among the algorithms, \textsf{UMC} and \textsf{KRC} exhibit the highest average F-Measure (0.98) at the cost of high run-time, due to the low similarity thresholds they employ. The best trade-off is achieved by \textsf{BMC}, which significantly increases the similarity threshold, reducing the run-time by at least 2/3 for a negligible reduction to F-Measure (1\%).

$D_5$ constitutes a highly noisy dataset, with many missing values in all attributes. As a result, the effectiveness of both types of schema-based semantic weights is rather low. Yet, the schema-agnostic ones consider all attribute values per entity, with the semantic weights leveraging the contextual knowledge offered by fastText and Albert pre-trained embeddings. As a result, the schema-agnostic semantic weights achieve the highest F-Measure, followed in close distance by their syntactic counterparts. Note that the difference in the run-time between the two types of edges is in favor of the semantic ones in most cases, despite the much larger similarity graphs they provide as input. This should be attributed to the much higher similarity thresholds they employ, on average. Among the algorithms, \textsf{KRC} is the top performing one in terms of effectiveness (F1=0.68), but \textsf{UMC} and \textsf{EXC} offer a better trade-off: F1 drops to 0.66, but the run-time is reduced by 2/3.

$D_6$ and $D_7$ share similar levels and forms of noise with $D_5$. However, the performance of schema-agnostic weights, especially the semantic ones, is much lower in both cases, probably due to the TVDB entities that are included in both datasets, but not in $D_5$. As a result, in $D_6$ the schema-based syntactic weights dominate all others, with the F-Measure of all algorithms (except \textsf{RCA}) confined in 0.508 and 0.521. The best trade-off is thus achieved by the fastest ones, namely  \textsf{CNC} (F1=0.508, run-time=52ms) and \textsf{EXC} (F1=0.517, run-time=58ms). Note that the very low run-time should be attribute to the very small graph sizes (see Table \ref{tb:inputs}) and the very high similarity thresholds (see Table \ref{tb:datasetThresholds}).

Similarly, the schema-based syntactic weights achieve relatively high F-Measure ($\sim$0.6) for very low run-time ($\ll$100 ms), due to very small input graphs and the relatively high similarity thresholds. However, the schema-agnostic syntactic weights offer slightly higher F-Measure ($\sim$0.62) for significantly higher run-time ($\sim$100ms), due to the significantly larger graphs and the smaller similarity thresholds. The optimal choice is actually \textsf{EXC}, which achieves the highest macro-average F-Measure among all combinations (F1=0.623) for an acceptable run-time (112ms).

$D_8$ constitutes a highly noisy dataset, which restricts the F-Measure of all combinations below 0.5. The two types of syntactic weights are competing for the best performance. The schema-based ones trade slightly lower effectiveness for significantly lower run-times and vice versa for the schema-agnostic ones. Among the former, \textsf{UMC} offers the best balance (F1=0.480, run-time=213ms), while the latter are dominated by \textsf{EXC} (F1=0.492, run-time=443ms). The final choice depends on the requirements of the application at hand.

$D_9$ is another bibliographic dataset with noise in the form of misplaced values, similar to $D_4$. As a result, the best performance is achieved by schema-agnostic syntactic weights, with \textsf{BMC} dominating all other combinations. Its F-Measure is 1.5\% lower than the maximum one (0.894 vs 0.909 for \textsf{KRC}), while achieving the second lowest run-time (627 vs 611 for \textsf{CNC}). This is because the schema-agnostic syntactic graphs are by far the smallest ones (see Table \ref{tb:inputs}), while \textsf{BMC} uses the highest average similarity threshold among all top-performing algorithms.

Finally, $D_{10}$ constitutes a rather noisy dataset with the highest portion of missing values. The F-Measure remains below 0.6 in all cases, with the top performance corresponding to two algorithms, \textsf{KRC} and \textsf{UMC}, in combination with any type of edges. This should be attributed to the relatively balanced number of entities in the two constituent entity collections. Among the two algorithms, \textsf{UMC} clearly outperforms \textsf{KRC} in terms of run-time. The best trade-off is actually achieved by its combination with schema-based syntactic weights (F1=0.57, run-time=8sec). 

Overall, the best combination of bipartite graph matching algorithms and type of edge weights depends on the datasets at hand and the type of noise it incorporates. On average, though, \textsf{UMC} in combination with syntactic similarities (mostly schema-agnostic ones) is consistently close to the best trade-off between F-Measure and run-time, if not the best one.

\balance

\end{document}